\documentclass[prb,10pt,twocolumn,aps,superscriptaddress,longbibliography,nofootinbib,floatfix]{revtex4-2}
\usepackage{amsmath}
\usepackage{amssymb}
\usepackage{txfonts}
\usepackage{siunitx}
\usepackage{graphicx}
\usepackage{bm}
\usepackage{hyperref}
\usepackage{float}
\usepackage{braket}
\usepackage[T1]{fontenc}
\usepackage[latin1]{inputenc}
\usepackage[table,dvipsnames]{xcolor}
\usepackage{lipsum}     
\usepackage[most]{tcolorbox}
\usepackage{blindtext}

\definecolor{NCSBlue}{rgb}{0,0.5137,0.788} 
\hypersetup{
  colorlinks=true,
  citecolor=NCSBlue,
  linkcolor=NCSBlue,
  urlcolor=NCSBlue}

\newcommand{\refcite}[1]{Ref.~\cite{#1}}

\newcommand{\refeq}[1]{Eq.~(\ref{#1})}

\newcommand{\refeqand}[2]{Eqs.~(\ref{#1}) and (\ref{#2})}
\newcommand{\reffig}[1]{Fig.~\ref{#1}}
\newcommand{\reffigand}[2]{Figs.~\ref{#1} and \ref{#2}}
\newcommand{\refsec}[1]{Sec.~\ref{#1}}

\newcommand{\punc}[1]{\,{\text{#1}}}
\newcommand{\sub}[1]{_{\text{#1}}}

\newcommand{\super}[1]{^{\text{#1}}}
\newcommand{\Chi}{\mathrm{X}}

\DeclareMathOperator{\Res}{Res}

\DeclareMathOperator{\erf}{erf}
\DeclareMathOperator{\erfc}{erfc}

\newcommand{\DTO}{Dy$_2$Ti$_2$O$_7$}
\newcommand{\HTO}{Ho$_2$Ti$_2$O$_7$}

\begin{document}
	
	\title{Dynamic scaling near the Kasteleyn transition in spin ice: critical relaxation of monopoles and strings following a field quench} 
	\author{Sukla Pal}
          \affiliation{School of Physics and Astronomy, The University of Nottingham, Nottingham, NG7 2RD, United Kingdom}
          \affiliation{Pitaevskii BEC Center, CNR-INO and Dipartimento di Fisica, Universit\`a di Trento, I-38123 Trento, Italy}
	   \author{Stephen Powell}
	\affiliation{School of Physics and Astronomy, The University of Nottingham, Nottingham, NG7 2RD, United Kingdom}

	\date{\today}
	
	  \begin{abstract}
       We study dynamics in classical spin ice following a magnetic field quench to close to the Kasteleyn transition, using Monte Carlo simulations and dynamic scaling theory to characterize the relaxation of the magnetization and the density of magnetic monopoles. We have previously argued that this dynamics can be described in terms of seeding and growth of strings of flipped spins, and our results here demonstrate that a solvable stochastic model based on independent strings correctly describes the relaxation as well as the distribution of string lengths within the critical scaling regime near the transition. We also show how generalized scaling forms capture the behavior over a broader range of monopole densities and provide a clear understanding of the breakdown of the scaling picture further from the critical point.
	\end{abstract}
	\maketitle
\section{Introduction}

Condensed matter systems subject to geometrical frustration \cite{Moessner2006}, which can combine local constraints with large fluctuations, have been of profound theoretical and experimental interest in the last few decades. Foundational examples include models for geometrically frustrated antiferromagnets \cite{Anderson1956}, proton disorder in crystalline ice \cite{Pauling1935}, and dimer-tiling problems \cite{Fowler1937}. More recently, the notion of geometric frustration has come to encompass a much broader domain, from cold atom physics \cite{Glaetzle2014, Zhu2016, Menu2024}, liquid crystals \cite{Takeaki2013}, and nanophotonics \cite{Jacqmin2014, Schmidt2016, Koraltan2021} to microwave cavity and circuit quantum electrodynamics \cite{Petrescu2012, Biondi2015}.

Phase transitions in frustrated systems provide a particularly striking example of the novel phenomena that they can host. While the standard Landau--Ginzburg--Wilson (LGW) theory \cite{Ma2018, Wilson1974}, based on the concepts of symmetry breaking and long-range order, provides a powerful description of universal behavior near critical points, a broad family of transitions outside the LGW paradigm are now known to exist. These include so-called deconfined critical points in quantum magnets \cite{Senthil2004a, Senthil2004b, Senthil2006, Metlitski2018}, as well driven dissipative quantum systems \cite{Ferioli2023} and fractional quantum Hall systems \cite{Song2024}. A variety of such transitions have been proposed in frustrated and constrained classical systems, including geometrically frustrated magnets \cite{Jaubert2008,Powell2015,Szabo2025} and dimer models \cite{Alet2006,Wilkins2019,Desai2021}.

While the concepts of scaling and universality were originally introduced for equilibrium phase transitions, they can also be applied to systems out of equilibrium \cite{Hohenberg1977, Polkovnikov2005, Zurek2005, Polkovnikov2011, Kolodrubetz2012}. Dynamic scaling theory describes dynamics close to a critical point, by incorporating scaling with time. Important examples of this include the behavior following a quench \cite{Liu2014}, a sudden change in one of the system parameters, and the Kibble--Zurek mechanism \cite{Kibble1980, Zurek1985, Hindmarsh2000, Chandran2012}, where a parameter is changed at a constant rate.

In this work, we apply scaling theory to describe the dynamics following a quench in a model of spin ice, a geometrically frustrated magnetic pyrochlore oxide. The low-temperature dynamics of frustrated magnetic systems is interesting in many ways \cite{FmagSpringer2011}, and this is particularly true for the spin ice compounds, locally constrained spin systems \cite{Sakakibara2003, Castelnovo2008, Henley2010, SIceSpringer2011} that host spin liquids \cite{Gingras2014, Savary2016, BPowell2020}. In these phases, characterized by an emergent gauge field and fractionalized excitations, the system remains disordered at lowest temperatures while exhibiting strong correlations \cite{Ramirez1999, Pauling1935}.
\begin{figure}
        \centering
        \includegraphics[width=1.0\linewidth]{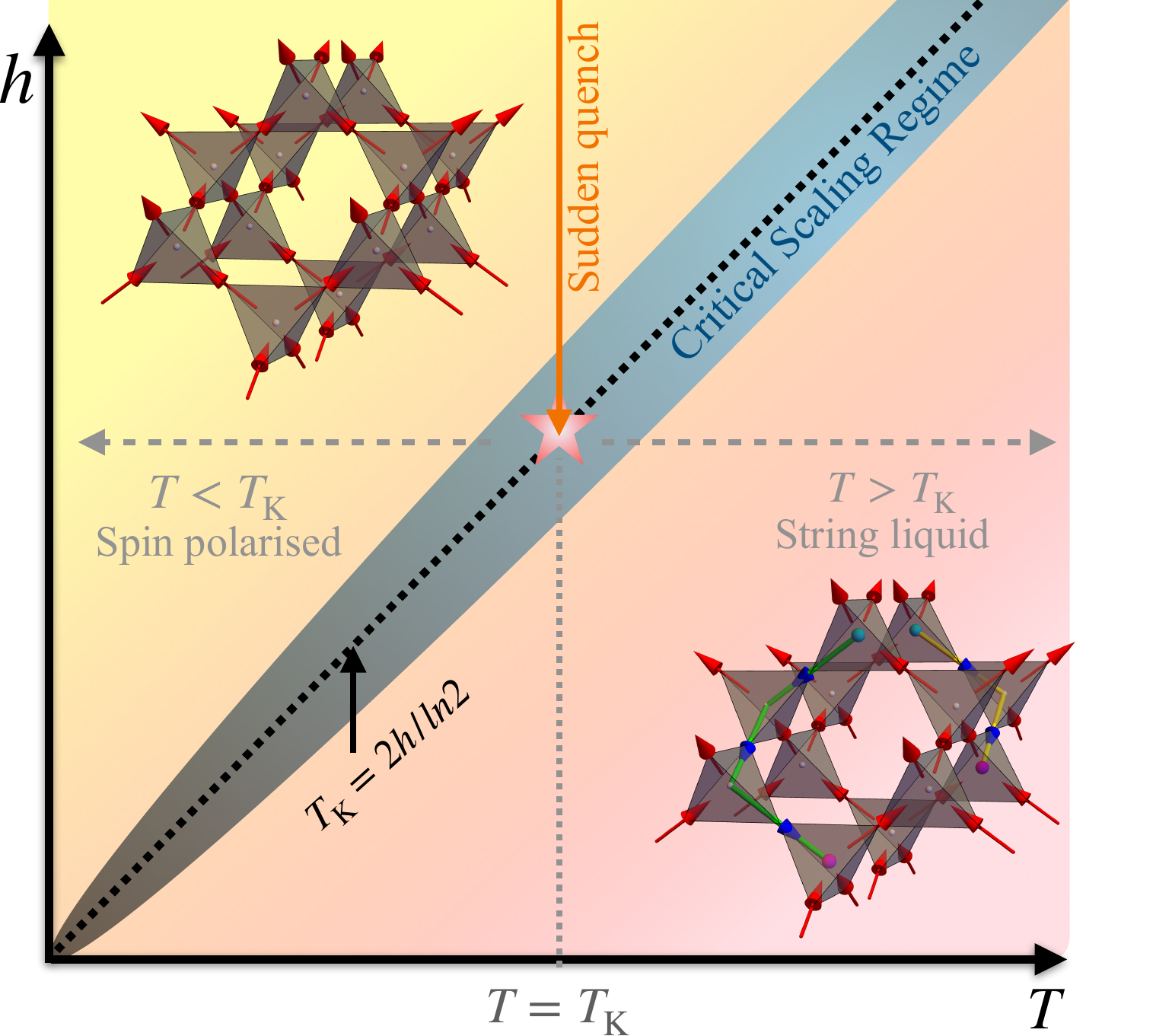}
        \caption{Schematic phase diagram describing the Kasteleyn transition of spin ice in an applied magnetic field \(h\) along the \([100]\) crystal direction. In equilibrium, there is a crossover at temperature \(T = T\sub{K} \propto h\) between a spin-polarized regime and a string liquid (illustrated in top-left and bottom-right insets, respectively). This becomes a sharp transition (black dotted line) when the energy \(\Delta\) of a magnetic monopole (see main text) is much larger than \(T\). We consider a sudden field quench (solid orange arrow) from \(h=+\infty\) to the critical scaling regime near the transition.}
        \label{fig:schematic}
\end{figure}

In the ``classical'' spin ices  \DTO\ \cite{Paulsen2014, Raban2022, Hallen2022} and \HTO\ \cite{Ehlers2004}, the spin liquid is a so-called Coulomb phase \cite{Henley2010, Fenell2009, Powell2015, Bramwell2001}, exhibiting dipolar correlations \cite{Henley2010, Morris2009} and deconfined magnetic monopoles \cite{Castelnovo2008}, which carry a fraction of the (dipolar) spin degree of freedom and are topological in character. Considerable attention has been paid to spin dynamics in spin ice \cite{Jaubert2009,Castelnovo2010,Jaubert2011,Mostame2014b,Grams2014,Jackson2014}, which can be understood in terms of motion of these quasiparticles, and is therefore described as magnetic transport \cite{Tang2025}.

The quench protocol that we consider involves starting with a large magnetic field along a \(\langle 100 \rangle\) crystal direction which is suddenly reduced to smaller but nonzero value. A strong field along this direction brings the system into a fully magnetized state, which is separated from the Coulomb phase by a Kasteleyn transition \cite{Jaubert2008, Powell2008, Powell2013} at a critical ratio of temperature to field. This unconventional phase transition, first recognized in a 2D classical dimer model \cite{Kasteleyn1963} and subsequently proposed to occur in various other classical \cite{Grigera2018,Potts2022} and quantum \cite{Yuan2012,Powell2022} systems, is characterized by topological order and involves no symmetry breaking. It has an asymmetric character, being apparently first order on one side but continuous on the other, and displays anisotropic scaling in real space.

By reducing the applied field, one can drive the system into the scaling regime near the Kasteleyn transition, as illustrated in Fig.~\ref{fig:schematic}. We have previously argued \cite{Pal2024} that the relaxation of the magnetization and monopole density is driven by the nucleation and growth of strings of flipped spins \cite{Morris2009, Zapke2025}, which subsequently form a dense network of clusters. Here, we employ dynamic scaling arguments to describe how this relaxation occurs for final parameter values that are near criticality. We show that not only physical observables but also the distribution of string lengths is governed by universal scaling functions, and calculate explicit forms for these functions in the limit of low monopole density, based on an effective stochastic model of a single string. A brief outline of this theoretical description and some of our numerical results have been presented previously \cite{Powell2025}.

The paper is organized as follows. In Sec.~\ref{Sec:theory}, we introduce the model that we use to describe classical spin ice and review the equilibrium Kasteleyn transition. We then turn to the dynamics in Sec.~\ref{Sec:DynamicsTheory}, where we derive dynamic scaling forms for the magnetization, monopole density, and string length distribution.  In Sec.~\ref{Sec:Res}, we present Monte Carlo simulation results for these quantities and compare them with the theory. We conclude in \refsec{conclu}, which includes a discussion of relevance to experiments. Additional data for other parameter values are shown in the appendix.

\section{Model and equilibrium properties}
\label{Sec:theory}

\subsection{Hamiltonian}
\label{Sec:Hamiltonian}

We use a minimal model of classical spin ice with vector spins \(\boldsymbol{S}_i\) on the sites \(i\) of a pyrochlore lattice, with ferromagnetic nearest-neighbor interactions. Crystal-field effects give a strong easy-axis anisotropy, effectively constraining the spins to \(\boldsymbol{S}_i = \sigma_i\hat{\boldsymbol{n}}_i\), where \(\sigma_i = \pm 1\) is an effective Ising degree of freedom and the fixed unit vector \(\hat{\boldsymbol{n}}_i\) points along the local \(\langle 111 \rangle\) axis between the centers of the two tetrahedra to which each site belongs \cite{Bramwell2001}.

Within the dumbbell model \cite{Castelnovo2008}, the energy can be expressed in terms of the ``magnetic charge'' on tetrahedron \(\alpha\),
\begin{equation}
n_\alpha = -\frac{1}{2}\eta_\alpha \sum_{i \in \alpha} \sigma_i\punc,
\end{equation}
which takes values \(0\), \(\pm 1\), and \(\pm 2\). (Here, \(\eta_\alpha = \pm 1\) for the two tetrahedron orientations in the pyrochlore lattice.) Including exchange \(J\) and dipolar interactions \(D\) between nearest-neighbor spins only, the Hamiltonian can be written as
\begin{align}
\label{eq:H}
H &= \frac{\Delta}{2} \sum_{\langle i j \rangle} \sigma_i \sigma_j\\
&= H\sub{gs} + \Delta\sum_\alpha n_\alpha^2
\punc,
\end{align}
where
\( \Delta = \frac{2}{3}J + \frac{8}{3}\left(1 + \sqrt{\frac{2}{3}}\right)D\) \cite{Castelnovo2008} and \(H\sub{gs}\) is a constant. In our numerical results, we set \(\Delta = \qty{2.8}{\kelvin}\), corresponding to \DTO\ \cite{Gardiner2010,Pal2024}.

The Hamiltonian \(H\) is minimized, \(H = H\sub{gs}\), in any state with \(n_\alpha = 0\) for all \(\alpha\), which requires that two spins point into and two spins point out of every tetrahedron. The number of such configurations, which are said to obey the ``ice rules'', grows exponentially with the number of lattice sites \(N\sub{s}\) \cite{Ramirez1999}. The minimal excitations are tetrahedra with \(n_\alpha = \pm 1\), which are referred to as magnetic monopoles \cite{Castelnovo2008}. While monopoles are created in charge-neutral pairs by flipping a spin, they are deconfined in the Coulomb phase of spin ice \cite{Henley2010} and can be treated as fractionalized quasiparticles with energy \(\Delta\). In the classical spin ice materials, dipolar interactions between spins in fact lead to effective magnetic Coulomb interactions between the monopoles \cite{Castelnovo2008}. As we argue in the following, we expect these to have only quantitative effects, and not to modify the universal scaling behavior of interest here.

The total magnetization is
\begin{equation}
\boldsymbol{M} = \sum_i \boldsymbol{S}_i\punc,
\end{equation}
in units of the magnetic moment \(\mu\approx 10\mu\sub{B}\) of each spin. An external magnetic field \(\boldsymbol{h}\) is applied along a \(\langle 100 \rangle\) crystal direction, which we take as the \(+z\) axis and refer to as ``upwards''. It couples to the spins through a Zeeman term
\begin{equation}
\label{eq:Hh}
H_h = -\mu \sum_i \boldsymbol{h}\cdot\boldsymbol{S}_i
\punc.
\end{equation}

\subsection{Equilibrium Kasteleyn transition}
\label{Sec:res_monopole}

We first briefly review the theory of the Kasteleyn transition in equilibrium in the nearest-neighbor model of spin ice \cite{Jaubert2008,Powell2008,Powell2013}.

For \(h > 0\), the full Hamiltonian \(H + H_h\) has a unique ground state, illustrated in the top-left inset of \reffig{fig:schematic}. In this configuration, all spins have positive projection along \(+z\) (i.e., each points upwards to the extent permitted by the Ising anisotropy), and so \(M_z = N\sub{s}/\sqrt{3}\). This minimizes \(H_h\), while also satisfying the ice rule on every tetrahedron, giving \(H=H\sub{gs}\).

Starting from this state and flipping any spin downwards reduces \(M_z\) by \(2/\sqrt{3}\) and therefore increases \(H_h\) by \(+2 h\), where we define\footnote{Note that Refs.~\cite{Jaubert2008} and \cite{Powell2013} define \(h\) as \(h_{\text{\cite{Jaubert2008}}} = \sqrt{3}h\).}  \(h = \mu\lvert \boldsymbol{h}\rvert/\sqrt{3}\). Doing so also breaks the ice rule on each of the two tetrahedra to which the spin belongs, producing monopoles with charges \(n_\alpha = \pm 1\), and so increases \(H\) by \(2\Delta\). The important excitations above the ground state are in fact strings of such flipped spins oriented along the \(z\) direction, as shown in the bottom-right inset of \reffig{fig:schematic}, for which the ice rule is obeyed on all tetrahedra except those at its ends. (These include isolated spin flips, treated as strings of length \(\ell = 1\).) Each string contributes entropy due to the number of possible paths, which is \(2^\ell\) for an isolated string, and hence free energy \(F_\ell = 2\Delta + (2 h - T \ln 2) \ell\). (We set \(k\sub{B} = 1\) here and throughout.)

This implies that, for \(T < T\sub{K} = \frac{2}{\ln 2}h\), strings of any length increase the free energy and appear at a density \(e^{-F_\ell/T}\) that vanishes in the limit \(\Delta/T \rightarrow \infty\). By contrast, for \(T > T\sub{K}\) and any finite \(\Delta/T\), sufficiently long strings always have \(F_\ell < 0\) and hence proliferate, giving a ``liquid'' of strings. (The string density in equilibrium is set by the effective hard-core interactions between strings, which reduce the entropic contributions due to the excluded volume \cite{Jaubert2008}.) The resulting crossover of the string density becomes a phase transition, referred to as a Kasteleyn transition \cite{Kasteleyn1963}, at \(T = T\sub{K}\) in the limit \(\Delta/T\rightarrow \infty\).

A quantity that plays the role of the order parameter, though not associated with any broken symmetry, is the deviation from saturation magnetization
\begin{equation}
\sigma = 1 - \frac{\sqrt{3}}{N\sub{s}} M_z\punc,
\end{equation}
which is zero in the ground state with all spins up. Inserting a string of length \(\ell\) increases \(\sigma\) by \(2\ell / N\sub{s}\), and so \(\sigma\) can be viewed as the areal density of strings. It therefore gives a signature of the Kasteleyn transition, showing a crossover from a small value for \(T < T\sub{K}\) when strings are suppressed (strictly zero in the limit \(\Delta/T \rightarrow \infty\)) to a larger value for \(T > T\sub{K}\) when they proliferate.

At the equilibrium Kasteleyn transition, the relevant scaling variables are reduced temperature \(\theta = (T - T\sub{K})/{T\sub{K}}\) and monopole fugacity \(z = e^{-\Delta/T}\), which both vanish at the critical point. The monopole density and magnetization are described by scaling forms \cite{Powell2013}
\begin{align}
\label{eq:StaticScaling}
\rho_1 &\approx \lvert\theta\rvert^{2}\Psi\super{eq}_\pm(z\lvert\theta\rvert^{-3/2})\\
\sigma &\approx \lvert\theta\rvert\Phi\super{eq}_\pm(z\lvert\theta\rvert^{-3/2})
\punc,
\end{align}
up to logarithmic corrections. (The subscript \(\pm\) indicates that the functions also depend on the sign of \(\theta\). The rational exponents and logarithmic corrections are due to the fact that the equilibrium transition is at its upper critical dimension \cite{Jaubert2008}.) The scaling functions, calculated using mean-field theory \cite{Jaubert2008,Powell2013}, are
\begin{align}
\Psi_\pm\super{eq}(v)&=4v^{4/3}\psi\left(\pm \frac{2\ln2}{3} v^{-2/3}\right)\\
\Phi_\pm\super{eq}(v) &= 4v^{2/3}\left[\psi\left(\pm \frac{2\ln2}{3} v^{-2/3}\right)\right]^2
\punc,
\end{align}
where \(\psi(x)\) is the positive solution of \(\psi^3-x \psi - \frac{1}{3}=0\).

\begin{figure}
    \includegraphics[width=0.95\linewidth]{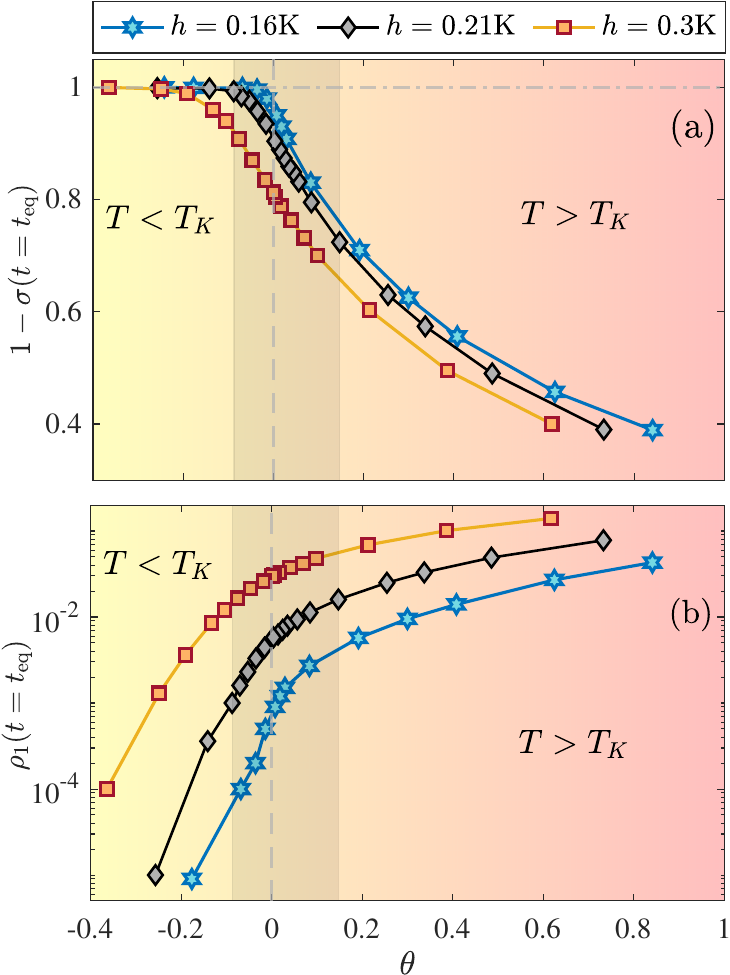}
    \caption{Monte Carlo results for (a) magnetization and (b) monopole density at equilibrium as a function of reduced temperature $\theta = (T - T\sub{K})/T\sub{K}$ for several values of the field \(h\) after the quench. The system contains \(N\sub{s}=128000\) spins and error bars are smaller than symbol sizes.}
    \label{fig:KT_varyH}
\end{figure}

As a preliminary check on our simulations of the dynamics, we show long-time results in \reffigand{fig:KT_varyH}{fig:KT_FiniteSize}, which are consistent with the established equilibrium behavior near the transition.
In \reffig{fig:KT_varyH}, we show the dimensionless magnetization \(m = 1 - \sigma\) and monopole density \(\rho_1\) as a function of reduced temperature \(\theta\), for several values of field \(h\). In each case \(m\) [panel (a)] saturates at \(1\) at low temperature and decreases continuously as \(\theta\) increases through zero. Because the Kasteleyn temperature \(T\sub{K}\) is proportional to \(h\), smaller \(h\) values correspond to lower \(T\), and hence smaller \(z\), at given \(\theta\). The transition is therefore sharpest for the smallest field, \(h = \qty{0.16}{\kelvin}\), and becomes increasingly rounded as \(h\) increases.

\begin{figure}
        \includegraphics[width=0.9\linewidth]{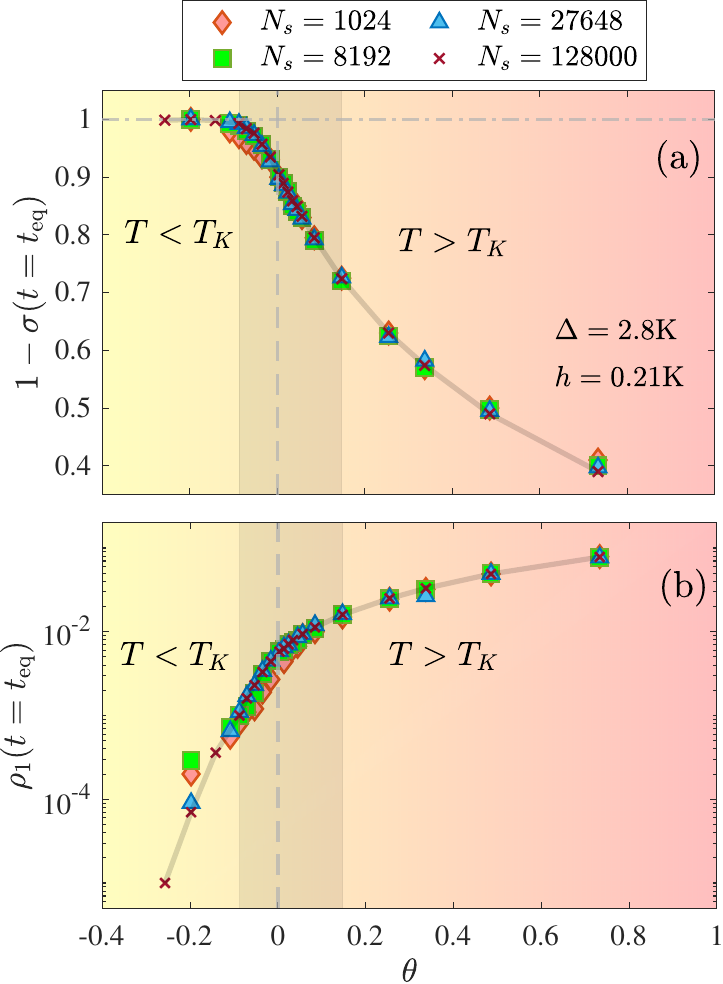}
        \caption{Monte Carlo results for (a) magnetization and (b) monopole density at equilibrium as a function of reduced temperature \(\theta\) for several system sizes \(N\sub{s}\). In each case, the magnetic field after the quench is $h=\qty{0.21}{\kelvin}$ corresponding to Kasteleyn temperature $T\sub{K} = \qty{0.606}{\kelvin}$. Finite-size effects are prominent only for the smallest \(N\sub{s}\) shown. The solid gray lines connecting the points is a guide for the eye. Error bars are smaller than symbol sizes.}
\label{fig:KT_FiniteSize}
\end{figure}

The same quantities are plotted in \reffig{fig:KT_FiniteSize} for four different system sizes \(N\sub{s}\) and fixed \(h = \qty{0.21}{\kelvin}\). Finite system size can also cause rounding of the transition, because of finite-length strings spanning the system \cite{Jaubert2008}, an effect that can be incorporated in scaling theory \cite{Powell2013}. This rounding is visible for the smallest \(N\sub{s} = 1024\), but is generally quite small for larger sizes, especially close to the transition. The results for the dynamics that we present below are mainly for the largest size shown here, \(N\sub{s} = 128000\).

\section{Dynamics: Theory}
\label{Sec:DynamicsTheory}

We use the ``standard model''  of the dynamics of classical spin ice \cite{Hallen2022}, which is believed to provide a good description for the relevant temperatures \cite{Ryzhkin2005,Jaubert2009,Pal2024}. Spin flips are attempted at randomly chosen sites at a temperature-independent rate \(N\sub{s}\tau^{-1}\), and accepted with a probability given by the Glauber function \cite{Glauber1963}
\begin{equation}
\label{eq:Glauber}
P\sub{G}(\delta E)= \frac{1}{e^{\delta E/T} + 1}
\end{equation}
for a spin flip with energy change \(\delta E\). This probability satisfies detailed balance and so gives a Boltzmann distribution at temperature \(T\) in the long-time limit.

Suppose that for time \(t < 0\), a large field \(h_u \gg T\) has been applied, and so the configuration at \(t=0\) has all spins pointing upwards (but still subject to the local easy-axis constraint), with magnetization \(M_z = \frac{1}{\sqrt{3}} N\sub{s}\) and hence \(\sigma = 0\). The field is subsequently reduced, allowing some spins to flip and \(\sigma\) to increase.

The relaxation of the magnetization can be understood in terms of the seeding and growth of strings of flipped spins \cite{Pal2024}. Creating a string requires creating a pair of monopoles on adjacent tetrahedra, and therefore costs an energy \(2\Delta\), whereas growth or contraction of a string occurs by moving one of the monopoles and hence changes the energy only by the Zeeman term. Near the critical point at \(z = 0\), there is a separation of timescales for these two processes, and one can approximate the behavior in terms of a low density of independent strings.

\subsection{Single-string model}

To describe this regime, and as a first step towards the full scaling behavior, we therefore define a single-string model, which is expected to apply only in the limit of vanishing \(z\). We will show that this model leads to dynamical-scaling forms for the string-length distribution, magnetization, and monopole density that are valid for long time and temperature \(T\) close to \(T\sub{K}\). (This single-string model is explained in more detail in \refcite{Pal2024} for the case \(h\sub{f} = 0\).)

\begin{figure}
\includegraphics[width=\columnwidth]{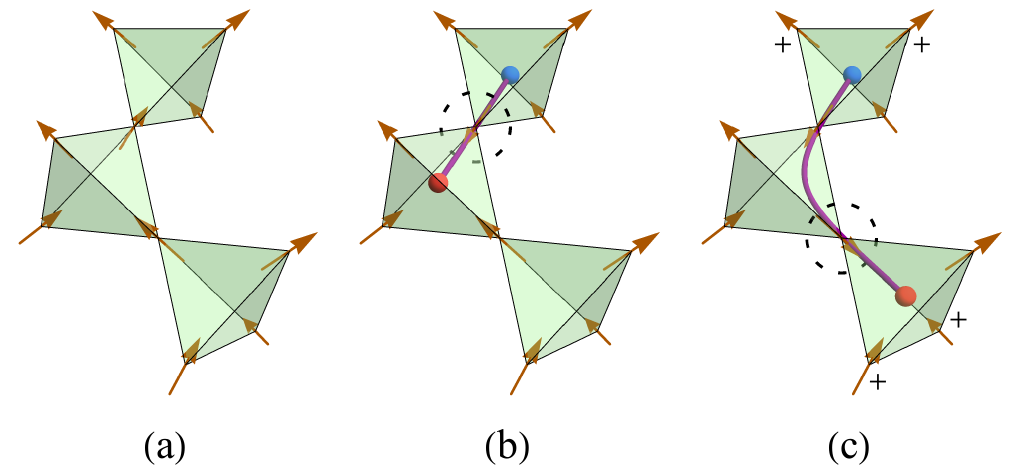}
\caption{Illustration of string creation and growth. (a) In the starting configuration, all spins point upwards. (b) Flipping any spin (circled) creates a pair of monopoles on the adjacent tetrahedra, which can be interpreted as a string of length \(\ell = 1\). Flipping this spin back upward would cause the two monopoles to annihilate and the string to disappear. (c) The monopoles can be separated by flipping other spins downwards. In this case, a second spin (circled) has been flipped, increasing the string length to \(\ell=2\). At this point, flipping any of the four spins labeled with \(+\) would extend the string to length \(\ell = 3\), while flipping the down spin adjacent to either monopole would shrink the string to length \(\ell = 1\).}
\label{FigString}
\end{figure}

Within the approximation of independent strings, we describe the configuration at time \(t\) after the quench only in terms of the number of strings \(N(\ell,t)\) of each length, neglecting all information about the positions and shapes of the strings. The spin-flip dynamics leads to an effective stochastic process for \(N(\ell,t)\), which involves four processes, illustrated in \reffig{FigString}:
\begin{enumerate}
\item If a spin is flipped downward in a region containing only up spins, a pair of monopoles are created on the adjacent tetrahedra. In terms of strings, this corresponds to creation of a string of length \(\ell = 1\). The energy change for this process is \(\delta E = 2\Delta + 2h\sub{u}\), including \(\Delta\) for each monopole plus the Zeeman energy of the flipped spin, and so its acceptance probability is \(r_* = P\sub{G}(2\Delta + 2h_u)=(e^{2(\Delta + h_u)/T} + 1)^{-1}\). For low string density, the number of spins available to flip can be approximated by the total number of spins, \(N\sub{s}\), and so the rate of string creation is \(N\sub{s}r_*\).

\item A string grows when one of its terminating monopoles moves in a way that increases their separation. For an isolated string, this happens by flipping downward any of \(4\) spins in the planes above and below, as shown in \reffig{FigString}(c). The number of monopoles is unchanged, and so the associated energy cost, \(\delta E = 2 h_u\), is entirely due to the Zeeman energy of the flipped spin.\footnote{As noted in \refcite{Pal2024}, Coulomb interactions between monopoles, which we do not include here, would modify the energy cost, reducing the rate at which short strings grow.} Because we neglect the spatial structure of the string, all \(4\) processes contribute to the effective rate \(r_+ = 4(e^{2h_u/T} + 1)^{-1}\).

\item Similarly, a string shrinks if one of \(2\) downward-pointing spins flip up, moving one of the two monopoles closer to the other. The Zeeman energy in this case is \(\delta E = -2h_u\), and so the rate is \(r_- = 2(e^{-2h_u/T} + 1)^{-1}\).

\item A string of length \(\ell = 1\) disappears if the single down spin is flipped up, causing its two terminating monopoles to annihilate. This process changes the energy by \(\delta E = -2\Delta - 2h\sub{u}\), and so occurs with rate
\(r_0 = (e^{-2(\Delta + h_u)/T} + 1)^{-1}\).

\end{enumerate}

\begin{figure}
\includegraphics{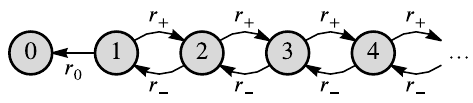}
\caption{Graph representing the dynamical model for a single string with rate matrix \(W_{\ell\ell'}\). Vertices (circles) denote strings of length \(\ell\), with \(\ell=0\) representing a string that has shrunk to zero length and hence disappeared. Edges (arrows) are transitions with associated rates for growth \(r_+\), contraction \(r_-\), and annihilation \(r_0\). To describe the evolution of the string population, we also include string creation, which occurs with rate \(r_*\).}
\label{FigGraph}
\end{figure}

Taking into account these four processes, the population of strings evolves according to
\begin{equation}
\label{eq:NelltODE}
\frac{\partial}{\partial t}N(\ell, t) = \delta_{\ell,1}N\sub{s}r_* + \sum_{\ell'=1}^{\infty} W_{\ell \ell'}N(\ell',t)
\punc,
\end{equation}
where \(W_{\ell\ell'} = \langle \ell \rvert W \lvert \ell' \rangle\) is an element of the rate matrix for the stochastic process governing a single string,
\begin{multline}
W = -(r_0 - r_-)\lvert 1\rangle\langle 1 \rvert \\+ \sum_{\ell=1}^{\infty}\left[-(r_-+r_+)\lvert \ell\rangle\langle \ell\rvert + r_+\lvert \ell+1 \rangle \langle \ell \rvert + r_- \lvert \ell \rangle \langle \ell+1 \rvert\right]\punc,
\end{multline}
which is illustrated in \reffig{FigGraph}. (We omit from \(W\) the state \(\lvert 0 \rangle\), representing a string that has shrunk to zero length and hence disappeared. This does not affect the dynamics, but effectively projects the probability vector into the orthogonal subspace, \(1 - \lvert 0 \rangle\langle 0 \rvert\).) To describe the string population at time \(t\) following the quench, we solve this model with the initial condition \(N(\ell, 0) = 0\) for all \(\ell \ge 1\).

In our simulations, we study the distribution of string lengths as a function of time, which can be compared directly to \(N(\ell,t)\). We also measure the density of monopoles \(\rho_1(t)\) and the magnetization \(1 - \sigma(t)\), and so we would like to express these observables in terms of \(N(\ell,t)\). Because each string is terminated by a pair of monopoles, the total number \(n_1\) of monopoles is simply given by double the number of strings. The density of monopoles is \(\rho_1 = n_1/N\sub{d}\), where \(N\sub{d} = \frac{1}{2}N\sub{s}\) is the number of tetrahedra, and so we can write \(\rho_1 = 4r_*M_0(t)\), where
\begin{equation}
\label{eq:DefineMnt}
M_n(t) = \frac{1}{N\sub{s}r_*}\sum_{\ell = 1}^{\infty} \ell^n N(\ell,t)
\end{equation}
is the \(n\)th moment of the length distribution, excluding \(\ell = 0\). As explained in \refsec{Sec:res_monopole}, a string of length \(\ell\) contains \(\ell\) spins flipped from up to down, and so
\(\sigma = 2r_*M_1(t)\).

Because the equilibrium Kasteleyn transition can be understood on the level of a single string, it must be consistent with the single-string model of the dynamics. Comparing the rates of growth \(r_+\) and shrinkage \(r_-\), we indeed find that strings tend to grow, \(r_+ > r_-\), when \(e^{2h_u/T} < 2\), or \(T > T\sub{K}\), and to shrink when \(T < T\sub{K}\). For the latter, this gives a long-time steady state with a low density of short strings and \(\sigma \rightarrow 0\) as \(z \rightarrow 0\), consistent with the equilibrium picture. For \(T > T\sub{K}\), strings (and hence \(\sigma\)) grow without limit and the single-string picture fails at long times. This is also consistent with the string liquid, with density determined by interactions, in the high-temperature phase.

To describe scaling near the transition, we can express the rates in terms of the scaling variables \(\theta\) and \(z\). Near the critical point, i.e., for \(T\simeq T\sub{K}\) and small \(z\), we find \(r_* = \frac{1}{2}z^2 + O(z^4, \theta z^2)\) and \(r_0 = 1 + O(z^2)\) for the creation and annihilation rates. For growth and shrinkage,
\begin{align}
r_+ &= \frac{4}{3} + \frac{8}{9}\theta \ln 2 + O(\theta^2)\\
r_- &= \frac{4}{3} - \frac{4}{9}\theta \ln 2 + O(\theta^2)\punc,
\end{align}
which imply
\begin{equation}
\frac{r_+}{r_-} = 1 + \theta \ln 2 + O(\theta^2)\punc.
\end{equation}

We first solve this model exactly, finding the leading-order behavior as the critical point and long-time limit are approached, before showing how the same results can be derived by using a continuum-\(\ell\) description.

\subsubsection{Exact asymptotic solution}

The solution of the master equation, \refeq{eq:NelltODE}, for \(\ell \ge 1\) can be written as
\begin{equation}
\label{eq:Nellt}
N(\ell, t) = N\sub{s}r_* \int_0^t dt'\, \langle \ell \rvert e^{Wt'} \lvert 1 \rangle\punc,
\end{equation}
where \(e^{W t'}\) is an operator exponential. To calculate this, we use the eigenvectors of \(W\),
\begin{equation}
\label{eq:eigenvector}
\lvert \psi(q) \rangle = \sum_{\ell=1}^{\infty} \lvert \ell \rangle \left[ q^\ell - \frac{r_+ + (r_0 - r_-)q}{r_0 - r_- + r_-q}\left(\frac{r_+}{r_-q}\right)^{\ell-1} \right]
\end{equation}
with \(q\) a complex number, to write
\begin{equation}
\lvert 1 \rangle = -\oint_{\mathfrak{C}} \frac{d q}{2\pi i} \frac{1}{q}\frac{r_0 - r_- + r_-q}{r_+ + (r_0 - r_-)q}\lvert\psi(q)\rangle\punc,
\end{equation}
where the contour \(\mathfrak{C}\) encloses the origin (in a counterclockwise direction) but not the pole at \(q = r_+/(r_- - r_0)\). The required inner product can then be simplified to
\begin{equation}
\langle \ell \rvert e^{Wt} \lvert 1 \rangle = \oint_\mathfrak{C} \frac{d q}{2\pi i} q^{\ell-2}\frac{r_+ - r_- q^2}{r_+ + (r_0 - r_-)q}e^{t\lambda(q)}\punc,
\label{EqPrltintegral}
\end{equation}
where \(\lambda(q) = r_+(q^{-1}-1) + r_-(q - 1)\) is the eigenvalue \cite{Pal2024}.

Performing the integral in \refeq{eq:Nellt} gives
\begin{equation}
\label{eq:Ncontourintegral}
N(\ell,t) = N\sub{s}r_* \oint_{\mathfrak{C}} \frac{d q}{2\pi i}
\left[p_\ell(q) - p_\ell(q)\big\rvert_{t=0}\right]
\punc,
\end{equation}
where
\begin{equation}
p_\ell(q) = q^{\ell-2} \frac{r_+ - r_- q^2}{r_+ + (r_0 - r_-)q}\frac{e^{t\lambda(q)}}{\lambda(q)}\punc.
\end{equation}
Each term in the integrand of \refeq{eq:Ncontourintegral} has poles where \(\lambda(q) = 0\), i.e., at \(q=1\) and \(q = r_+/r_-\). If we choose \(\mathfrak{C}\) so that it does not enclose either, e.g., a counterclockwise circle of radius \(< \min(1, r_+/r_-)\), then the second term is analytic within the contour (for \(\ell \ge 1\)), and so Cauchy's theorem gives
\begin{equation}
\oint_{\mathfrak{C}} \frac{d q}{2\pi i}
p_\ell(q)\big\rvert_{t=0} = 0\punc.
\end{equation}

The remaining integral can be evaluated for \(t\gg \tau\equiv(r_+r_-)^{-1/2}\) by deforming \(\mathfrak{C}\) to pass through the saddle point, \(\lambda'(q_0)=0\), at \(q_0 = \sqrt{r_+/r_-}\). This involves crossing the pole at \(q = r_+/r_-\) when \(\theta < 0\) (and hence \(r_+ < r_-\)), and the pole at \(q = 1\) when \(\theta > 0\), adding a contribution given by the residue in either case. One therefore has
\begin{equation}
\label{eq:NltIntegral}
\frac{N(\ell,t)}{N\sub{s}r_*} = \oint_{\mathfrak{C}'} \frac{d q}{2\pi i}
p_\ell(q)
- \Theta(\theta)\Res(p_\ell,1) - \Theta(-\theta)\Res(p_\ell,q_0^2)
\end{equation}
where \(\mathfrak{C}'\) is the deformed contour passing through \(q_0\), \(\Theta\) is the unit step function, and \(\Res\) denotes the residue.

We expand near the saddle point by writing \(q = q_0 + i\xi \sqrt{\tau/t}\), in terms of which \(t\lambda(q)\approx -x^2 - \xi^2\) to leading order in \(\theta\), where 
\begin{equation}
x = \frac{\ln 2}{\sqrt{3}}\theta t^{1/2}
\punc.
\end{equation}
The integral is therefore dominated by \(\xi = O(1)\) and can be extended to \(-\infty < \xi <\infty\). For small \(\theta\), the poles \(q=1\) and \(q=r_+/r_-\) are at \(\xi = +i x\) and \(\xi = -i x\) respectively, to the same order. We therefore find the asymptotic expression
\begin{equation}
\frac{N(\ell,t)}{N\sub{s}} \approx \frac{1}{2}z^2 \left[\int_{-\infty}^{\infty}\frac{d\xi}{2\pi}\tilde{p}_\ell(\xi) - i \Res(\tilde{p}_\ell,i\lvert x \rvert)\right]
\punc,
\end{equation}
where
\begin{equation}
\tilde{p}_\ell(\xi) = \frac{2i \xi e^{-\xi^2-x^2} e^{2y(x+i\xi)}}{x^2+\xi^2}
\end{equation}
is given by expanding \(p_\ell(q)\) around \(q = q_0\), for large \(t\) and \(\ell\) and small \(\lvert \theta\rvert\), with \(x\) and
\begin{equation}
y=\frac{\sqrt{3}}{4}\ell t^{-1/2}
\end{equation}
both \(O(1)\).

\begin{figure}
\includegraphics{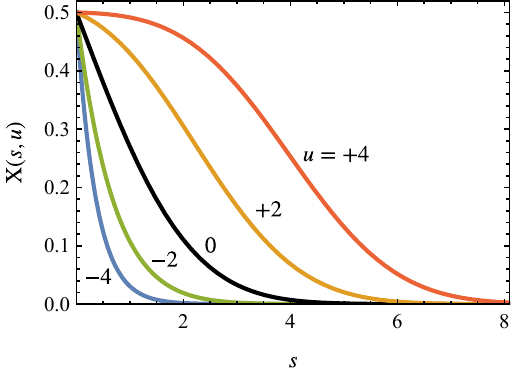}
\caption{Scaling function \(\Chi(s,u)\), defined by \refeq{eq:Chi}, plotted as a function of \(s\) with each curve labeled by its value of \(u\). Within the single-string model, the number of strings of length \(\ell\) at time \(t\) after the quench is given by \(N(\ell,t)\approx N\sub{s} z^2 \Chi(\ell t^{-1/2},\theta t^{1/2})\), for large \(\ell\) and \(t\) and small reduced temperature \(\theta = (T-T\sub{K})/T\sub{K}\).}
\label{FigXplot}
\end{figure}

The integral over \(\xi\) can be evaluated in terms of the complementary error function \(\erfc x = 1 - \erf x\), where
\begin{equation}
    \erf x = \frac{2}{\sqrt{\pi}}\int_0^x d\xi\,e^{-\xi^2}\punc.
\end{equation}
The result is \(N(\ell,t)/N\sub{s} \approx z^2 \Chi(\ell t^{-1/2},\theta t^{1/2})\), where
\begin{equation}
\label{eq:Chi}
\Chi\left(\!\frac{4y}{\sqrt{3}},\frac{\sqrt{3}x}{\ln 2}\!\right) = \frac{1}{4}\left[\erfc(y-x)+ e^{4xy}\erfc(y+x)\right]
\end{equation}
is plotted in \reffig{FigXplot}. This gives a dynamic scaling form for the distribution of string lengths, depending on \(\ell\), \(t\), and \(\theta\) only through the combinations \(\ell t^{-1/2}\) and \(\theta t^{1/2}\) for long times close to the critical point.

\begin{figure}
\includegraphics{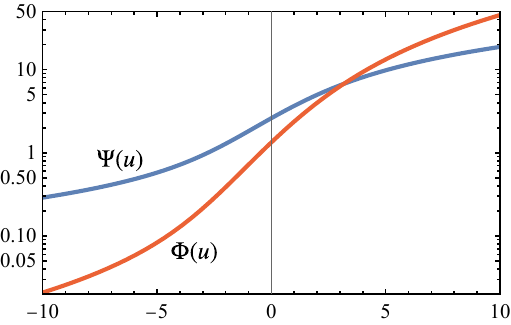}
\caption{Scaling functions \(\Psi\) and \(\Phi\), defined in \refeqand{eq:Psi}{eq:Phi}, describing the monopole density and deviation from saturation magnetization respectively, at long times for temperatures close to the Kasteleyn transition [see \refeq{eq:ScalingForms1}].}
\label{FigScalingFunction}
\end{figure}

To calculate the moments \(M_n(t)\) of the string length distribution, one can calculate the sum in \refeq{eq:DefineMnt} using the exact expression in \refeq{eq:Ncontourintegral} and then use the same saddle-point approximation for the resulting contour integral \cite{Powell2025}, or simply replace the sum over \(\ell\) by an integral over \(y\) and use the asymptotic expression for \(N(\ell, t)\). Either gives \(M_0(t) \approx \frac{1}{2}\sqrt{t}\Psi(\theta t^{1/2})\) and \(M_1(t) \approx t\Phi(\theta t^{1/2})\), where
\begin{align}\label{eq:Psi}
\Psi\left(\!\frac{\sqrt{3}x}{\ln 2}\!\right) &= \frac{4}{\sqrt{3}}\left[x + \frac{e^{-x^2}}{\sqrt{\pi}} + \left(x + \frac{1}{2x}\right)\erf x\right]\\
\label{eq:Phi}
\Phi\left(\!\frac{\sqrt{3}x}{\ln 2}\!\right) &= \frac{4}{3}\left[1 + x^2 + \frac{e^{-x^2}}{\sqrt{\pi}}\left(x + \frac{1}{2x}\right) + \left(1 - \frac{1}{4x^2}+x^2\right)\erf x\right]
\end{align}
are defined such that
\begin{equation}
\begin{aligned}
    \rho_1 &\approx z^2 t^{1/2}\Psi(\theta t^{1/2})\\
    \sigma &\approx z^2 t\Phi(\theta t^{1/2})\punc.
\end{aligned}
\label{eq:ScalingForms1}
\end{equation}
These expressions are dynamic scaling forms for the monopole density and magnetization, depending on \(\theta\) only through the combination \(\theta t^{1/2}\) for long times close to the critical point. The scaling functions \(\Psi\) and \(\Phi\) are plotted in \reffig{FigScalingFunction}.

\subsubsection{Continuum string distribution}

We now show that identical scaling results can be found starting from \refeq{eq:NelltODE} by using a continuum-\(\ell\) description. This is justified at long times near the critical point because the integral giving \(N(\ell,t)\), \refeq{eq:NltIntegral}, is then dominated by a small region near \(q=1\), where the coefficients of \(\lvert \ell\rangle\) in the eigenvector \(\lvert\psi(q)\rangle\), \refeq{eq:eigenvector}, are slowly varying functions of \(\ell\).

We therefore write \(N(\ell,t)\approx N\sub{s}\varphi(\ell,t)\) for \(\ell \ge 1\) where \(\varphi\) is, by assumption, a smooth function of \(\ell\) and \(t\). Expanding \refeq{eq:NelltODE} and keeping derivatives up to second order in \(\ell\) gives
\begin{equation}
\label{eq:NelltPDE}
\frac{\partial}{\partial t}\varphi(\ell, t) = \left[ -(r_+ - r_-) \frac{\partial}{\partial \ell} + \frac{1}{2}(r_++r_-)\frac{\partial^2}{\partial\ell^2}\right]\varphi(\ell,t)\punc,
\end{equation}
where the differential operator on the right-hand side is the continuum version of the rate matrix \(W_{\ell\ell'}\). The source term in \refeq{eq:NelltODE} corresponds to the inhomogeneous boundary condition \(\varphi(0, t) = r_*/(r_0 + r_+ - r_-)\). This equation describes an advection--diffusion process with a source at \(\ell = 0\) and a drift velocity \(r_+ - r_-\) that changes sign at \(T = T\sub{K}\).

Near the critical point, expanding to leading order in \(\theta\) and \(z\) gives
\begin{equation}
\label{eq:NelltPDE2}
\frac{\partial}{\partial t}\varphi(\ell, t) = -\frac{4\ln2}{3}\theta \frac{\partial}{\partial \ell}\varphi(\ell,t) + \frac{4}{3}\frac{\partial^2}{\partial\ell^2}
\varphi(\ell,t)
\end{equation}
with boundary condition \(\varphi(0,t) = \frac{1}{2}z^2\). The general solution can be written as
\begin{equation}
\varphi(\ell,t) = z^2 \Chi(\ell t^{-1/2},\theta t^{1/2}) + \varphi\sub{c}(\ell, t)\punc,
\end{equation}
with \(\Chi\) as in \refeq{eq:Chi} and the complementary function \(\varphi\sub{c}\) given by any solution of \refeq{eq:NelltPDE2} with homogeneous boundary condition \(\varphi\sub{c}(0,t) = 0\). All dependence on initial conditions is included in the latter, which gives a transient correction that decays to zero at long times for any initial distribution with finite width. The solution at long times is therefore simply \(\varphi(\ell, t) = z^2\Chi(\ell t^{-1/2}, \theta t^{1/2})\). Integrating over \(\ell\) then reproduces the scaling forms of \refeq{eq:ScalingForms1} for \(\rho_1\) and \(\sigma\) with the same functions \(\Psi\) and \(\Phi\).

\subsection{Generalized scaling}
\label{SecGeneralizedScaling}

The single-string model treats strings as sparse enough that interactions between them can be neglected. Since new strings are created with rate \(\propto r_* \sim z^2\), this assumption is justified in the limit of vanishing \(z\), but will clearly fail for larger \(z\), at least beyond the shortest times. Including \(z\) within dynamic scaling therefore requires going between the single-string picture.

\begin{figure}
\includegraphics[width=\columnwidth]{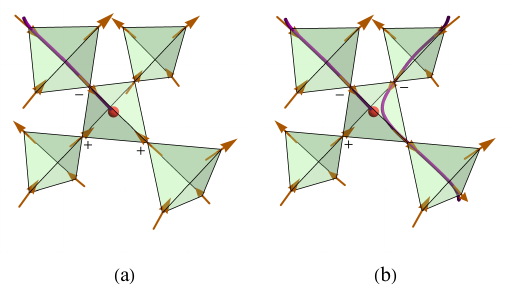}
\caption{Illustration of excluded-volume effect for strings. (a) A string of flipped spins in an otherwise fully polarized region is terminated by a monopole on the central tetrahedron. As in \reffig{FigString}, there are two spins (labeled \(+\)) on this tetrahedron that could be flipped to extend the string and one (labeled \(-\)) to shrink it. (b) When a second string passes through the same tetrahedron (i.e., another two spins are flipped down on the tetrahedron), only one up spin remains that can be flipped to extend the string.}
\label{FigString2}
\end{figure}

To do so, we will include the effect of nonzero string density (i.e., \(\sigma>0\)) at the mean-field level. Consider the situation, shown in \reffig{FigString2}, where one of the two monopoles at either end of a string occupies a tetrahedron through which another string passes. (In other words, another two spins are flipped downwards on the tetrahedron.) In this case, only one up spin remains that can be flipped to extend the string, and so the rate for this process is decreased from \(r_+\). In addition, there are two down spins, rather than one, that can be flipped back up, and so the rate of shrinking the string is greater than \(r_-\). This excluded-volume effect is equivalent to the hard-core repulsion between strings that reduces their entropy at finite density and allows for a liquid of strings in equilibrium \cite{Jaubert2008}.

To include this effect within a mean-field approximation, we reduce the drift coefficient \(r_+ - r_-\) in \refeq{eq:NelltPDE} by an amount proportional to the areal string density \(\sigma\). Equivalently, we modify \refeq{eq:NelltPDE2} to
\begin{equation}
\frac{\partial\varphi}{\partial t} = -\frac{4\ln2}{3}\left[\theta - b\int_0^\infty d\ell\,\ell \varphi(\ell,t) \right] \frac{\partial\varphi}{\partial \ell}
+ \frac{4}{3}\frac{\partial^2\varphi}{\partial\ell^2}\punc,
\end{equation}
where \(b\) is a positive constant, independent of \(\theta\) and \(z\).

By rescaling the variables \(t\), \(\ell\) and \(z\), we can write the solution of this nonlinear integro-differential equation as
\begin{equation}
\label{eq:ScalingForms2Chi}
\varphi(\ell, t) = z^2\Chi(\ell t^{-1/2}, \theta t^{1/2}, z\lvert\theta\rvert^{-3/2})\punc,
\end{equation}where \(\Chi(s,u,v)\) is a function that depends on the phenomenological parameter \(b\) and that we expect to agree with \(\Chi(s,u)\) in the limit \(v \rightarrow 0\). On physical grounds, we assume the existence of a unique solution for an initial condition peaked near \(\ell = 0\). Integrating over \(\ell\) then gives the generalized scaling forms
\begin{equation}
\begin{aligned}
\rho_1 &\approx z^2 t^{1/2} \Psi(\theta t^{1/2},z\lvert\theta\rvert^{-3/2})\\
\sigma &\approx z^2 t \Phi(\theta t^{1/2},z\lvert\theta\rvert^{-3/2})\punc,
\end{aligned}
\label{eq:ScalingForms2}
\end{equation}
which, along with \refeq{eq:ScalingForms2Chi}, generalize the single-string scaling behavior to include the full dependence on \(z\). For \(z\rightarrow 0\), these relations reduce to \refeq{eq:ScalingForms1}, meaning that \(\Psi(u,v)\rightarrow \Psi(u)\) and \(\Phi(u,v)\rightarrow\Phi(u)\) for \(v\rightarrow0\).

In the long-time limit \(t\rightarrow\infty\), they should instead reduce to the equilibrium scaling forms in \refeq{eq:StaticScaling}. Our generalized scaling forms are indeed compatible with these equilibrium results, despite being derived using mean-field theories that are formulated quite differently. By matching the functions, we infer the limits \(\Psi(u,v)\rightarrow \lvert u\rvert^{-1}v^{-2}\Psi\super{eq}_\pm(v)\) and \(\Phi(u,v)\rightarrow u^{-2}v^{-2}\Phi\super{eq}_\pm(v)\) for \(u\rightarrow\pm\infty\).

Note that \(\Phi(u) \rightarrow (u \ln 2)^{-2}\) as \(u \rightarrow -\infty\) and \(\Phi\super{eq}_-(v)\rightarrow (v/\ln 2)^2\) as \(v\rightarrow0\). The two limits \(t\rightarrow\infty\) and \(z\rightarrow 0\) therefore commute when \(\theta < 0\), both giving asymptotic behavior \(\sigma \approx (\ln 2)^{-2}z^2\theta^{-2}\). We similarly find the consistent result \(\rho_1 \approx (2/\ln2)z^2\lvert \theta\rvert^{-1}\) for \(\theta < 0\), taking the two limits in either order. By contrast, the dynamical scaling forms, \refeq{eq:ScalingForms1}, do not have finite long-time limits for \(\theta > 0\), and so \(t\rightarrow\infty\) and \(z\rightarrow 0\) do not commute in this case. We interpret this as due to the fact that the long-time limit for fixed \(z\) and \(\theta > 0\) is dominated by interactions, which set the equilibrium string density \cite{Jaubert2008}, whereas taking \(z\rightarrow 0\) first and using the single-string model neglects interactions between strings.

Returning briefly to finite-size scaling, the reciprocal of the linear system size is also a relevant scaling variable \cite{Cardy1996}. For the Kasteleyn transition, system sizes parallel, \(L_z\), and perpendicular, \(L_\perp\), to the field in fact have different scaling dimensions and need to be treated separately \cite{Jaubert2008,Powell2013}. This would further generalize the scaling forms in \refeq{eq:ScalingForms2} to involve scaling functions of four variables, but we do not attempt to confirm this behavior quantitatively here.

\section{Dynamics: Results}
\label{Sec:Res}

To assess the validity of our predicted dynamic scaling laws near the Kasteleyn transition, we perform Monte Carlo (MC) simulations on the microscopic model defined in \refsec{Sec:Hamiltonian}. As explained in \refsec{Sec:DynamicsTheory}, we use dynamics where independent spin flips are attempted at a fixed rate \(N\sub{s}\tau\sub{flip}^{-1}\) and accepted with probability \(P\sub{G}\) given in \refeq{eq:Glauber}. We use parameters for \DTO, for which a MC time step corresponds to \(\tau\sub{flip} \simeq \qty{3}{\milli\second}\) \cite{Snyder2004} and the activation energy is \(\Delta = \qty{2.8}{\kelvin}\) \cite{Gardiner2010,Pal2024}.

For each magnetic field \(h\), our simulations are performed for a range of $T$ near the transition temperature $T\sub{K} \propto h$, for which the reduced temperature \(\theta\) is small. This corresponds to a range of values for \(z = e^{-\Delta/T}\) around \(e^{-\Delta/T\sub{K}}\). We present data for fields \(h = \qty{0.16}{\kelvin}\), \(\qty{0.21}{\kelvin}\), and \(\qty{0.3}{\kelvin}\), which therefore allow us to access the critical behavior at small \(z\). To establish the significance of finite system size, we use systems with $N\sub{s} = 1024$, $8192$, $27648$, and $128000$ spins and periodic boundary conditions. Finite-size effects are generally quite small, as seen at equilibrium in \refsec{Sec:res_monopole} and noted previously in the dynamics \cite{Pal2024}.

We first present results for the two thermodynamic quantities, monopole density and magnetization. We then consider the distribution of string lengths, which is likely to be less accessible in experiment but provides a more direct test of the validity of the picture of dynamics based on isolated strings.

\subsection{Dynamic scaling: monopoles and magnetization density}
\label{sec:res_Dscaling}

In this section, we compare MC simulations with our theoretical predictions for monopole density and magnetization following a quench. These include both the predictions of the single-string model, \refeq{eq:ScalingForms1}---which applies only in the limit \(z \rightarrow 0\) but for which the explicit scaling functions are given in \refeqand{eq:Psi}{eq:Phi}---and the generalized scaling expressions, \refeq{eq:ScalingForms2}.

\begin{figure}[!] 
\includegraphics[width=0.9\linewidth]{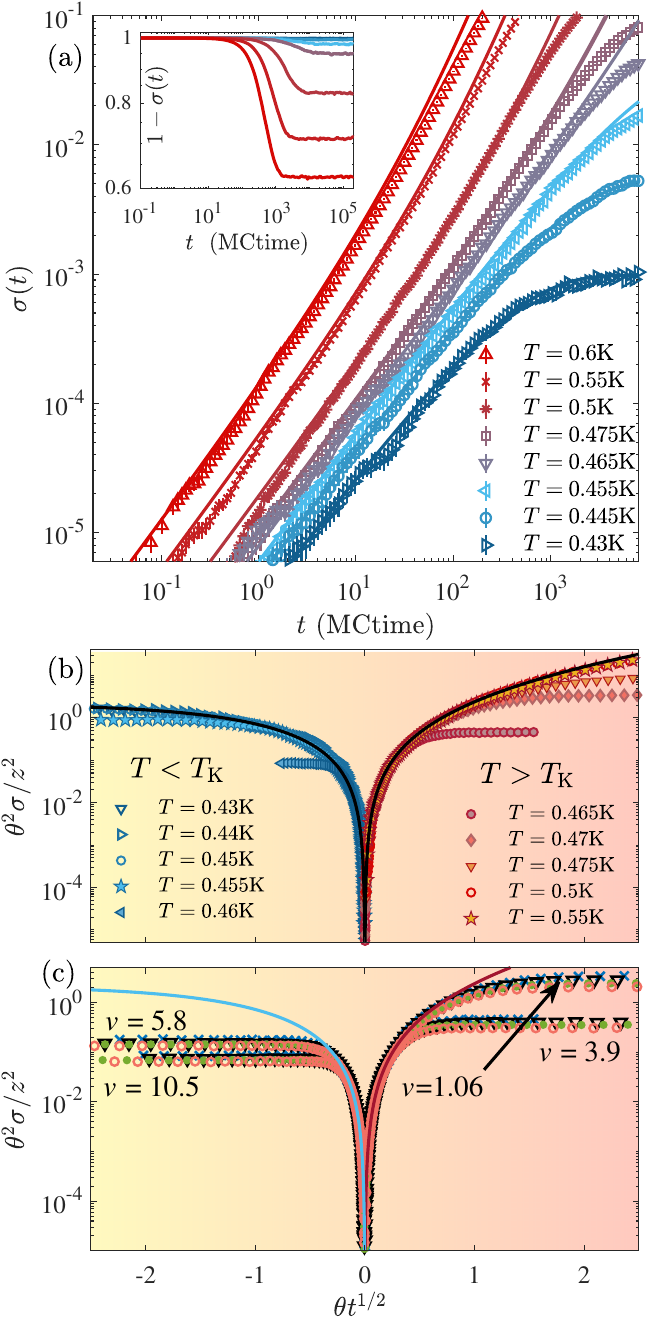}
\caption{Time evolution of \(\sigma\), the deviation from saturation magnetization, for $N_s = 128000$. (a) MC data (symbols) for $h=\qty{0.16}{\kelvin}$ at a set of temperatures close to $T_K = \qty{0.462}{\kelvin}$, compared with the scaling function (solid lines) in Eq.~(\ref{eq:Phi}). The inset shows the magnetization density \(1 - \sigma\). (b) Same data and scaling function plotted in terms of dynamic scaling variables. (c) Illustration of generalized scaling behavior given in Eq.~(\ref{eq:ScalingForms2}), which predicts that data for each value of $v = z\lvert \theta \rvert^{-3/2}$ should collapse onto a single curve. The plot includes data for $h=\qty{0.16}{\kelvin}$ (blue cross), $h=\qty{0.21}{\kelvin}$ (black triangle), $h=\qty{0.25}{\kelvin}$ (green filled circle) and $h=\qty{0.3}{\kelvin}$ (coral empty circle), chosen to give four different values of $v = 10.5$, $5.8$, $3.9$ and $1.06$.
In all panels, the MC data are averaged over 100 to 200 independent runs and error bars are smaller than or similar to symbol sizes.}
\label{fig:MzScaling128000h0p16}
\end{figure}

\begin{figure}[!] 
\centering
\includegraphics[width=0.9\linewidth]{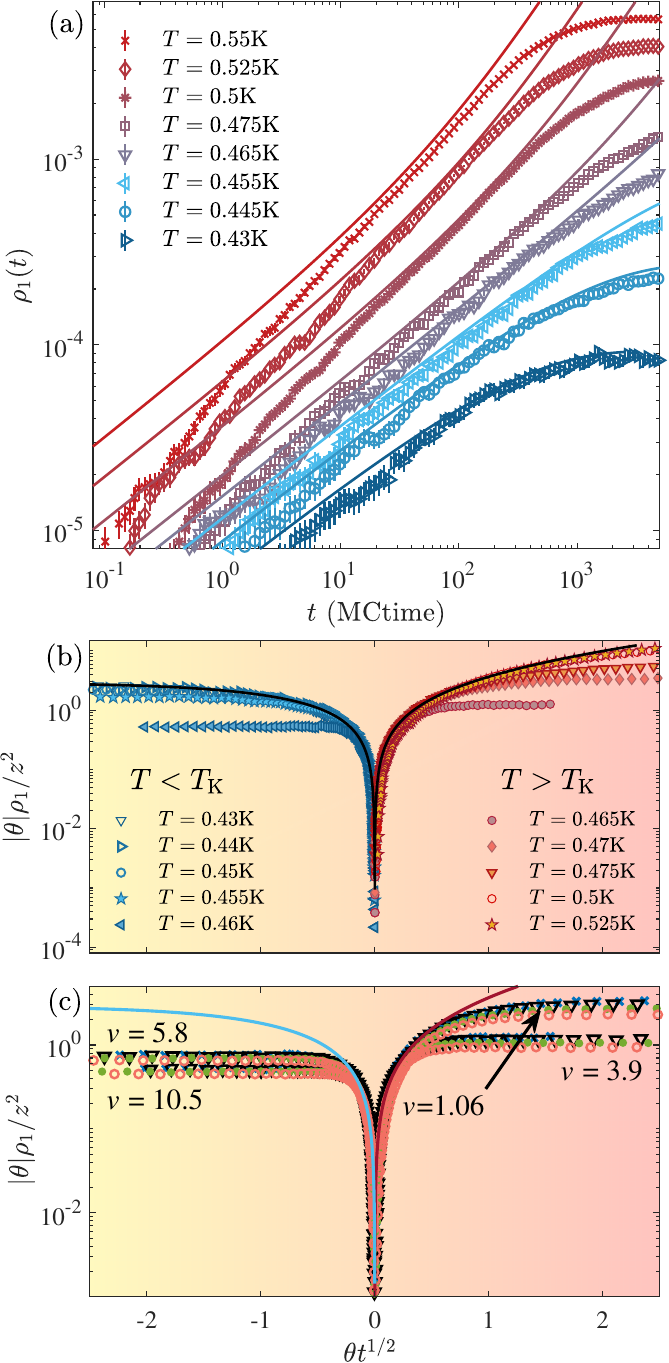}
\caption{Time evolution of the monopole density $\rho_1$ for $N_s = 128000$, with panels corresponding to those in Fig.~\ref{fig:MzScaling128000h0p16}. (a) MC data (symbols) for $h=\qty{0.16}{\kelvin}$ and the corresponding scaling function (solid lines) from Eq.~(\ref{eq:Psi}). (b) Same data and scaling function plotted in terms of dynamic scaling variables. (c) Illustration of generalized scaling behavior given in Eq.~(\ref{eq:ScalingForms2}). The plot includes data for the same values of \(h\), and hence \(v\), as in Fig.~\ref{fig:MzScaling128000h0p16}(c). 
In all panels, the MC data are averaged over 100 to 200 independent runs and error bars are smaller than or similar to symbol sizes.}
\label{fig:MonScaling128000h0p16}
\end{figure} 

Results for magnetization and monopole density are given in Figures~\ref{fig:MzScaling128000h0p16} and \ref{fig:MonScaling128000h0p16} respectively, for \(N\sub{s} = 128000\), \(h = \qty{0.16}{\kelvin}\), and a range of temperatures \(T\) around \(T\sub{K} = \qty{0.462}{\kelvin}\).

In each figure, panel (a) shows the observable, \(\sigma\) and \(\rho_1\) respectively, versus time for each temperature, along with the theoretical result from the single-string model. For these temperatures, \(0.0015 \lesssim z \lesssim 0.01\), and quantitative agreement is expected as long as corrections due to nonzero \(z\) are small. Based on standard scaling arguments \cite{Goldenfeld}, this occurs when the other relevant variables \(\lvert\theta\rvert\) and \(t^{-1}\) are large enough to dominate the effects of \(z\), but still small enough that scaling applies. The generalized scaling forms of \refeq{eq:ScalingForms2} can be expressed in terms of the combinations \(z\lvert\theta\rvert^{-3/2}\) and \(z t^{3/4}\), which implies that the small-\(z\) forms should apply for reduced temperatures above a value \(\lvert \theta \rvert \sim z^{2/3}\) and times up to \(t \sim z^{-4/3}\).

The results in \reffig{fig:MzScaling128000h0p16}(a) show close quantitative agreement for $\sigma$, without any fitting parameters, over many decades of $t$. This agreement extends from the earliest times shown, up to at least \(t \simeq 10^3\) for \(T < T\sub{K}\) and up to around \(10^2\) for \(T = \qty{0.55}{\kelvin}\) (\(\theta = 0.19\)). For $\rho_1$, in \reffig{fig:MonScaling128000h0p16}(a), agreement is observed over a narrower time window, though still more than one decade for temperatures close to and below \(T\sub{K}\), and also for a more limited range of $T$.

At very short times, where scaling is not expected to apply, both \(\sigma\) and \(\rho_1\) increase linearly with time \cite{Pal2024}. This (coincidentally) agrees with the scaling form in \refeq{eq:ScalingForms1} for \(\sigma\),\footnote{As seen in \reffig{FigScalingFunction}, both \(\Psi(u)\) and \(\Phi(u)\) have finite limits as \(u \rightarrow 0\).} and so \reffig{fig:MzScaling128000h0p16}(a) shows agreement even at very small \(t\). This does not apply for \(\rho_1\), where the simulation results deviate significantly from the scaling function for \(t \lesssim 10^1\).

Figures~\ref{fig:MzScaling128000h0p16}(b) and \ref{fig:MonScaling128000h0p16}(b) show the same data plotted with \(u = \theta t^{1/2}\) on the horizontal axis and \(\theta^2 \sigma/z^2\) and \(\lvert \theta \rvert \rho_1/z^2\), respectively, on the vertical axes. If the small-\(z\) scaling forms in \refeq{eq:ScalingForms1} applied exactly, all data would collapse onto the black solid lines, which show \(u^2 \Phi(u)\) and \(u\Psi(u)\), where the one-parameter functions \(\Phi\) and \(\Psi\) are defined in \refeqand{eq:Phi}{eq:Psi}. In fact, while the data collapse quite well for small \(\lvert u\rvert\), they deviate substantially when \(\lvert u\rvert\) becomes larger. These deviations from the small-\(z\) scaling forms occur sooner for \(T\) closer to \(T\sub{K}\), and hence for smaller \(\lvert \theta\rvert\), consistent with the general scaling arguments above.

Finally, Figures~\ref{fig:MzScaling128000h0p16}(c) and \ref{fig:MonScaling128000h0p16}(c) show data collapse based on generalized scaling. In each case, the data are plotted in the same way as in panel (b), but for data with different \(h\), chosen to give the same four values of \(v = z\lvert \theta \rvert^{-3/2}\). As predicted by the generalized scaling forms in \refeq{eq:ScalingForms2}, data with the same \(v\) collapse onto the same curve.

Additional data for other system sizes and \(h\) values are shown in the appendix. For increasing \(h\), deviations from the single-string results become larger, as expected due to the larger value of \(z = e^{-\Delta/T}\) at the transition. We see only relatively small finite-size effects in the magnetization and monopole density, indicating that corrections due to the relevant scaling variables \(L_z^{-1}\) and \(L_\perp^{-1}\) \cite{Powell2013} are less important than those due to \(z\) for these parameters.


\subsection{Strings and clusters}
\label{sec:res_dstrings}

\begin{figure}
\centering
\includegraphics[width=0.9\linewidth]{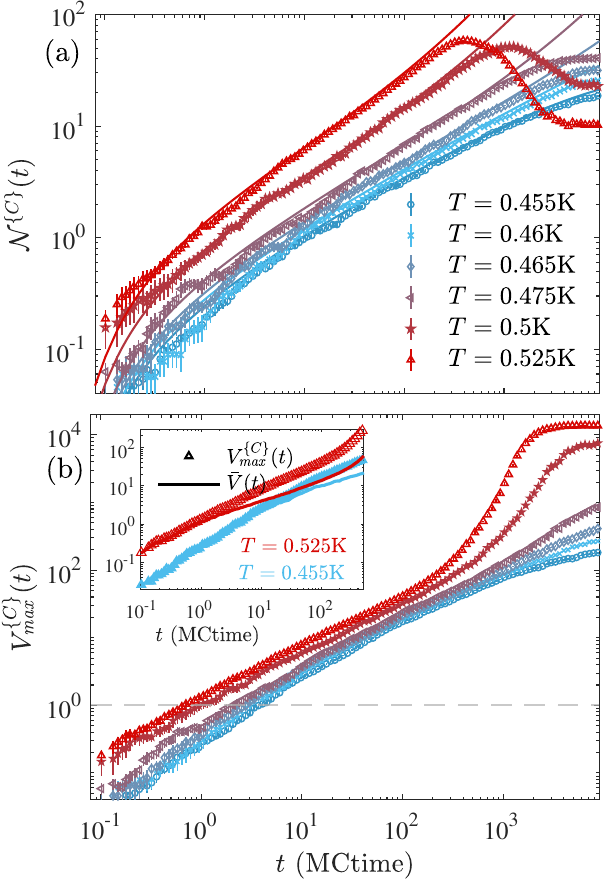}
\caption{Cluster dynamics following a field quench near the Kasteleyn transition. (a) Total number of clusters, $\mathcal{N}^{\{C\}}(t)$, as a function of time for various temperatures $T$. Symbols show MC results while the solid lines represent the analytical result for the string-length distribution, \(N(\ell,t)\), from the single-string model, summed over all string lengths \(\ell\). (b) Number of spins in the largest cluster, $V_{\rm max}^{\{ C\}}$, as a fraction of the total number of spins \(N\sub{s}\), versus \(t\). At early times, this can be interpreted as the length of the longest string. (Inset) Mean cluster size $\bar{V}(t)$ (solid lines) compared to maximum cluster size $V_{\rm max}^{\{C\}}$. At early time, $\bar{V} \approx V_{\rm max}^{\{ C\}}$ indicates a single-string dominant regime. Subsequently, formation of multiple strings of varying lengths is captured by the behavior $\bar{V} < V_{\rm max}^{\{C\}}$. This transition highlights a generic feature of the dynamics, and is irrespective of $T$. The simulation has been performed for a system of $N_s =128000$ and $h=\qty{0.16}{\kelvin}$ ($T_{\rm K}=\qty{0.4617}{\kelvin}$) and averaged over 200 independent runs.}
\label{fig:stringCount}
\end{figure}

In this section, we describe our simulation results for the growth of strings near the Kasteleyn transition and the scaling behavior of their distribution. First, in Fig.~\ref{fig:stringCount}, we show some general characteristics of the dynamics. As noted in \refcite{Pal2024}, once a significant fraction of the spins have been flipped downward, i.e., for \(\sigma\) of order \(1\), strings can no longer be uniquely identified, and one should instead consider clusters (defined as contiguous sets) of flipped spins.

The total number $\mathcal{N}^{\{C\}}$ of such clusters is shown in Fig.~\ref{fig:stringCount}(a) as a function of time for several values of $T$ near $T\sub{K}$. Its growth displays similar qualitative features to the dynamics in zero field \cite{Pal2024}, including non-monotonic behavior in time, which is visible in the plot for $T\gtrsim \qty{0.5}{\kelvin}$. The increase at early time is due to the creation of strings, and agrees quite well with the prediction \(\sum_\ell N(\ell,t)\) of the single-string model, while the subsequent decrease can be understood in terms of a consolidation process of strings into clusters.

Figure~\ref{fig:stringCount}(b) shows the volume $V_{\rm max}^{\{ C\}}$ of (i.e., number of spins in) the largest cluster. This saturates, at a value of order \(N\sub{s}\), at time $t\approx 2\times 10^3$, with $\mathcal{N}^{\{C\}}$ also reaching a steady state, indicating the approach to equilibrium. The inset shows the corresponding growth of the mean cluster size
\begin{equation}
\bar{V}(t) =  \frac{\sum_{\ell}\ell \mathcal{N}(\ell,t)}{\mathcal{N}^{\{C\}}(t)}
\punc,
\end{equation}
on both sides of the transition. It displays the same qualitative features as $V_{\rm max}^{\{ C\}}$. Although initially both quantities grow equally, at a later time, $\bar{V}(t) < V_{\rm max}^{\{C\}}(t)$, confirming the tendency of the system toward the formation of a single large cluster.


\begin{figure*}[p] 
\centering
\includegraphics[width=0.925\linewidth]{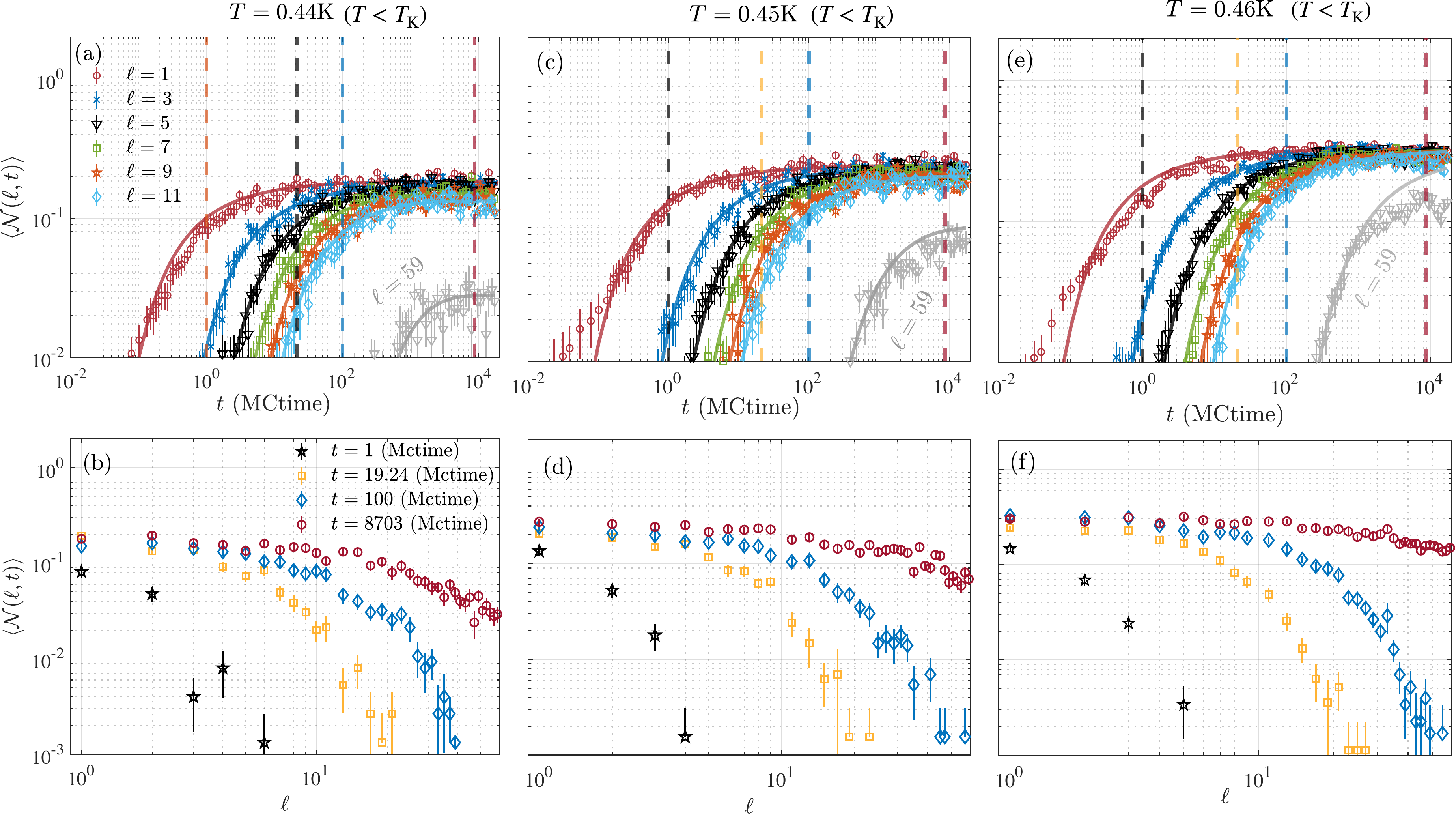}
\caption{Time evolution of the string length distribution for $T<T_{\rm K}$ for a system of $N_s =128000$ spins with $h=\qty{0.16}{\kelvin}$. For this field, $T_{\rm K}=\qty{0.4617}{\kelvin}$, and so, from left to right, the system is approaching criticality from below. (a) Time evolution of the average number of strings for six different string lengths, $\ell = 1, 3, 5, 7, 9, 11$. Symbols of different colors represent MC data which are averaged over 500 different trajectories. Solid lines are the plot of the analytical expression, $N(\ell,t)=N_s\varphi(\ell,t)$, where $\varphi = z^2\Chi(\ell t^{-1/2}, \theta t^{1/2})$, and $\Chi$ is given by Eq.~(\ref{eq:Chi}). (b) String length distributions are shown for four different times selected from (a) indicated by the similar colored vertical dashed lines. The middle column (c) \& (d) and the right column (e) \& (f) show the same quantities for $T=\qty{0.45}{\kelvin}$ and $T=\qty{0.46}{\kelvin}$, respectively.}
\label{Fig:stDistTltTK128000}
\end{figure*}
\begin{figure*}[p] 
\centering
\includegraphics[width=0.925\linewidth]{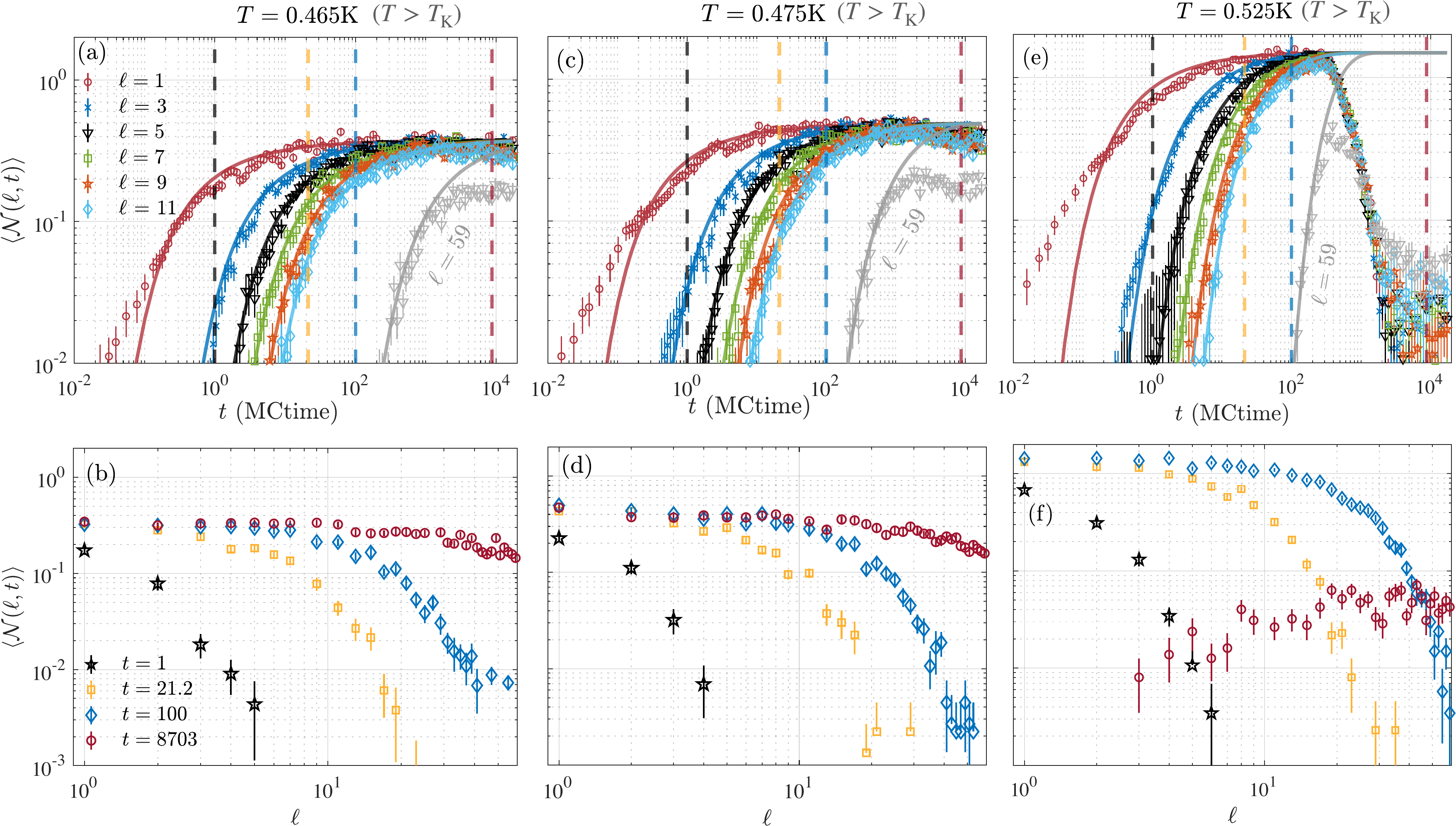}
\caption{Time evolution of the string length distribution, as in Fig.~\ref{Fig:stDistTltTK128000} but for $T>T_{\rm K}$. In this case, the system approaches criticality from above going from right to left. The left, middle and right columns show the same quantities as in Fig.~\ref{Fig:stDistTltTK128000}, for \(T = \qty{0.465}{\kelvin}\), $\qty{0.475}{\kelvin}$, and $T=\qty{0.525}{\kelvin}$ respectively. For all the cases each data point is averaged over 400 independent runs.}
\label{Fig:stDistTgtTK128000}
\end{figure*}

Next, we turn to the distribution of string lengths, \(N(\ell,t)\) (which includes clusters of size \(\ell\) as well as strings in the simulations). The time evolutions of a few individual lengths from \(\ell = 1\) to \(11\) are shown in the top panels [(a), (c), and (e)] of Figs.~\ref{Fig:stDistTltTK128000} (where \(T < T\sub{K}\)) and \ref{Fig:stDistTgtTK128000} (\(T > T\sub{K}\)), along with the analytical form from \refeq{eq:Chi}, which applies in the small-\(z\) limit. We see reasonable agreement in all cases, with deviations clearly apparent at very early and late times and becoming more prominent for \(T\) well above \(T\sub{K}\), as with thermodynamic quantities studied in \refsec{sec:res_Dscaling}. The quantitative agreement indicates that clusters of the sizes shown are predominantly strings of length \(\ell\), as expected for small \(z\).

For the largest temperature, \(T = \qty{0.525}{\kelvin}\), shown in \reffig{Fig:stDistTgtTK128000}(e), there is a clear qualitative change in the behavior starting around \(t \simeq 10^2\). We understand this change as due to cluster formation, as discussed above. For this case, data for \(\ell = 59\) are also shown, confirming that a significant population of large strings and clusters are produced in this regime.

The lower panels [(b), (d), and (f)] of Figs.~\ref{Fig:stDistTltTK128000} and \ref{Fig:stDistTgtTK128000} show the distribution \(N(\ell,t)\) versus \(\ell\) at four different times \(t\), indicated in each case by the vertical dashed lines in the top panels. While it should be noted that larger \(\ell\) values include significant contributions from clusters rather than simple strings, various qualitative features of the single-string model are visible in these results. First, very short strings dominate the distribution at the earliest times for all temperatures. At later times for \(T < T\sub{K}\), the distribution remains peaked at small \(\ell\) but with a width that increases with \(t\), consistent with the scaling function plotted in \reffig{FigXplot}. (Note that the latter uses a linear scale for \(s = \ell t^{-1/2}\), whereas \reffig{Fig:stDistTltTK128000} uses a logarithmic scale.) Finally, for \(T > T\sub{K}\), the distribution is essentially flat for small \(\ell\) before decreasing at a value of \(\ell\) that increases with \(t\).

Additional data for the string-length distributions are shown in the appendix. As with magnetization and monopole density, deviations from the single-string model are more significant for large \(h\). For smaller system sizes \(N\sub{s}\), there are significant finite-size effects when \(\ell\) is comparable to the linear system size.

\begin{figure}[!]
\centering
\includegraphics[width=1.0\linewidth]{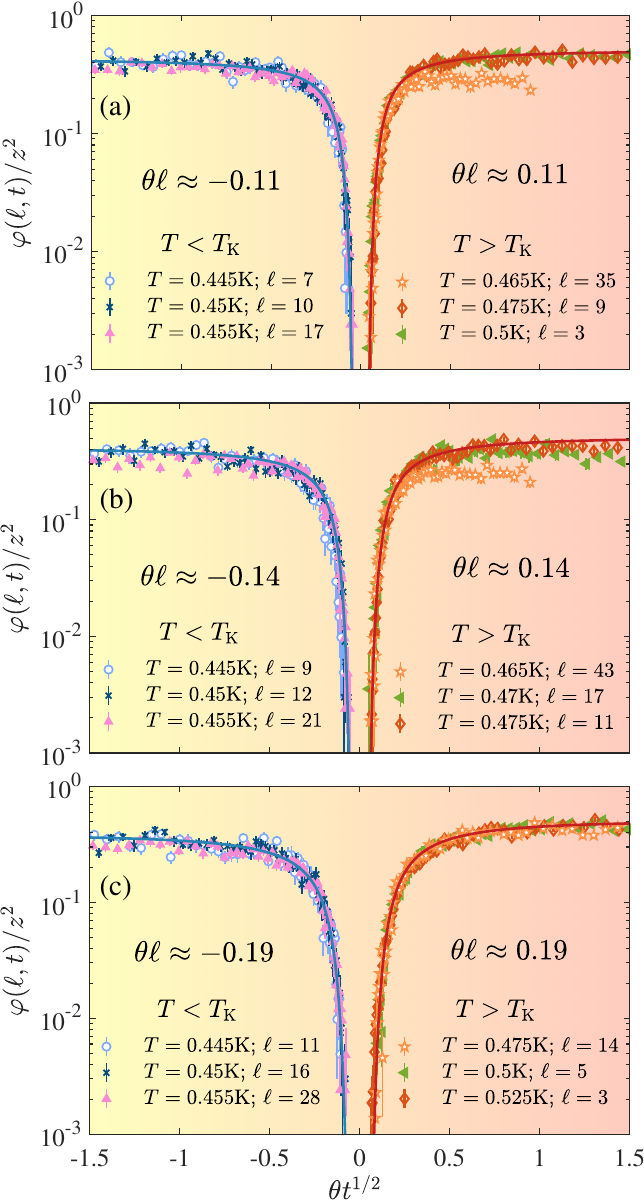}
\caption{Dynamic scaling for string length distribution \(\varphi(\ell,t)\) for $N\sub{s}=128000$ and $h=\qty{0.16}{\kelvin}$. In each panel, combinations of \(T\) and \(\ell\) are chosen so that \(\lvert\theta\rvert\ell\) takes approximately the same value: (a) $\theta\ell \simeq \pm0.11$, (b) $\theta\ell \simeq \pm0.14$, and (c) $\theta\ell \simeq \pm0.19$. For each, the solid line shows the scaling function \(\varphi(\ell, t) = z^2\Chi(\ell t^{-1/2}, \theta t^{1/2})\), where $\Chi$ is given in Eq.~(\ref{eq:Chi}). Each data point is averaged over 500 independent runs.}
\label{Fig:scalingThetaL}
\end{figure}

Finally, in Fig.~\ref{Fig:scalingThetaL} we compare the string-length distributions from our MC simulations with the scaling relation \(\varphi(\ell, t) = z^2\Chi(\ell t^{-1/2}, \theta t^{1/2})\) and the function \(\Chi\) given in \refeq{eq:Chi}. To test this scaling form, we show data for different combinations of \(T\) and \(\ell\) such that \(\lvert\theta\rvert\ell\) is approximately equal in each panel, and find reasonable collapse onto the analytic curve. Deviations from the predicted scaling behavior are most evident for larger \(\ell\) and hence smaller \(\theta\), where the effects are nonzero \(z\) are expected to be most noticeable, as with the comparable plots in Figs.~\ref{fig:MzScaling128000h0p16}(b) and \ref{fig:MonScaling128000h0p16}(b).

\section{Conclusions and Outlook}
\label{conclu}

In this work, we have used Monte Carlo simulations and dynamic scaling theory to study the out-of-equilibrium dynamics of classical spin ice following a sudden quench of a \([100]\) magnetic field to close to the Kasteleyn transition. As established previously \cite{Pal2024}, the early-time dynamics is governed by the formation and growth of strings of flipped spins, terminated at each end by magnetic monopoles. Here, we used a stochastic model for a single string to derive dynamic scaling relations for magnetization and monopole density and incorporated the effects of string interactions using a mean-field approach. We find good agreement between our MC results for these quantities and for the string-length distribution with explicit functions valid in the limit of low-string density as well as with generalized scaling forms that apply more broadly.

As seen in \reffig{fig:MzScaling128000h0p16}(a) and \reffig{fig:MonScaling128000h0p16}(a), we find reasonable agreement between our MC simulations, necessarily performed at nonzero \(z\), and the results of the single-string model for \(z \rightarrow 0\). This result may be somewhat surprising, since strings in reality interact by excluded-volume effects, whereas the model treats them as independent. Comparing the magnetization density \(1 - \sigma(t)\) and monopole density \(\rho_1(t)\), one sees significantly better agreement for the former. To understand this, note that using the large-\(t'\) behavior of \(e^{W t'}\) in \refeq{eq:Nellt} assumes that the (finite) contribution from small \(t'\) (i.e., recently created strings) is negligible. This is true as long as the rest of the integral grows without bound as \(t\) gets larger. The mean string lifetime grows at least as \(\sim \lvert \theta \rvert^{-1}\), and so this assumption is valid for small \(\theta\) even for \(n=0\), but for \(n=1\) the contribution of longer (and hence typically older) strings has more weight in the sum.

The generalized scaling forms derived in \refsec{SecGeneralizedScaling} give significantly better agreement with the simulations, applying for a broader range of \(z\) values. This generalized scaling picture neglects the spatial structure of the strings and their relative positions, and so is a mean-field theory. (It can be considered analogous to the well-mixed approximation for chemical reactions \cite{Schnoerr2017}.) The theory nonetheless keeps track of the full population \(N(\ell,t)\) of strings of all lengths \(\ell\). A simpler model that describes the dynamics only in terms of the number of strings and their mean length, or equivalently in terms of \(\rho_1\) and \(\sigma\), cannot give the correct scaling behavior.

Using the dependence of the scaling forms, \refeq{eq:ScalingForms2}, on \(u=\theta t^{1/2}\) we infer \(\nu \mathfrak{z} = 2\), where \(\nu=\frac{1}{2}\) is the correlation-length exponent \cite{Powell2013} and \(\mathfrak{z}\) the dynamic critical exponent \cite{Hohenberg1977}, and hence \(\mathfrak{z} = 4\). This value is implied by the scaling forms in the single-string limit, \refeq{eq:ScalingForms1}, and so does not rely on the mean-field treatment used for generalized scaling.

Our quantitative results for the distribution of string lengths also provide a clear understanding of the breakdown of the scaling picture at \(T\) further from \(T\sub{K}\), particularly on the high-temperature side. This is most clearly evident in \reffig{Fig:stDistTgtTK128000}(e), where a decrease in the population of short strings beyond \(t \sim 10^2\) indicates the merger of strings into larger clusters of flipped spins. The dynamic scaling forms that we have proposed here are governed by the Kasteleyn critical point at \(z = 0\) and \(\theta = 0\), and so require that \(z\) and \(\theta\) are sufficiently small that one is within its scaling regime. They therefore cannot apply when the system is dominated by large clusters, in which case the late-time behavior is instead similar to a quench to zero field \cite{Pal2024}.

In experiment, the clearest signatures of our predictions for dynamic scaling are likely to be in the relaxation of the magnetization, which is directly measurable in experiment \cite{Paulsen2014}. Neutron scattering experiments have shown evidence for string excitations in spin ice \cite{Morris2009} and could in principle give access to the string-length distributions, though the required time resolution may be challenging. Further work is needed to assess potential signatures for local probes such as muon spin relaxation \cite{Lago2007,Kirschner2018}. Quantitative comparison with experiment is also likely to require treatment of magnetothermal heating and resulting magnetization avalanches \cite{Jackson2014}, which we have not attempted to include here.

A similar analysis could be applied to spin ice with the field along other directions \cite{Potts2022}, including to the (two-dimensional) Kasteleyn transition with the field at a small angle to the [111] direction \cite{Moessner2003}. It can potentially also be extended to quenches in artificial spin ice \cite{Libal2020}, and the same physics may be accessible with analogue simulators \cite{Cuarda2025,Shah2025,Surace2020}.

An additional consideration for comparison with experiments on classical spin ice materials would be the effect of dipolar interactions between spins. As noted in \refsec{Sec:Hamiltonian}, these lead to magnetic Coulomb interactions between monopoles \cite{Castelnovo2008}, which would modify the effective rates for string growth and shrinkage for short strings \cite{Pal2024}. Since the scaling behavior is dominated by long strings and low monopole density, we do not expect it to be qualitatively changed by these effects.

On the theoretical side, our results should be compared with those of \refcite{Oakes2016}, which showed that dynamics of classical dimers on a square lattice could also be understood in terms of strings. In contrast to the situation here, the dynamics of dimers was treated as driven by plaquette flips, which correspond to ring exchange in spin ice \cite{Castelnovo2012}. Relaxation is then governed by the movement of a fixed density of strings that span the system, rather than by their formation and growth.

\acknowledgments
This work was supported by the Engineering and Physical Sciences Research Council grant number EP/T021691/1. The numerical simulations used resources provided by the University of Nottingham Augusta and Ada HPC services. S. Pal also acknowledges the computational support from the University of Trento (UniTn) HPC cluster, where the last stage of the calculations was performed. S. Pal also  highly acknowledges the hospitality and support from Lancaster University, where the manuscript was finalized for submission.


\appendix*

\section{Additional simulation results}

\subsection{Dynamic scaling of $m_z$}

\begin{figure}[H]
 \centering
 \includegraphics[width=0.885\linewidth]{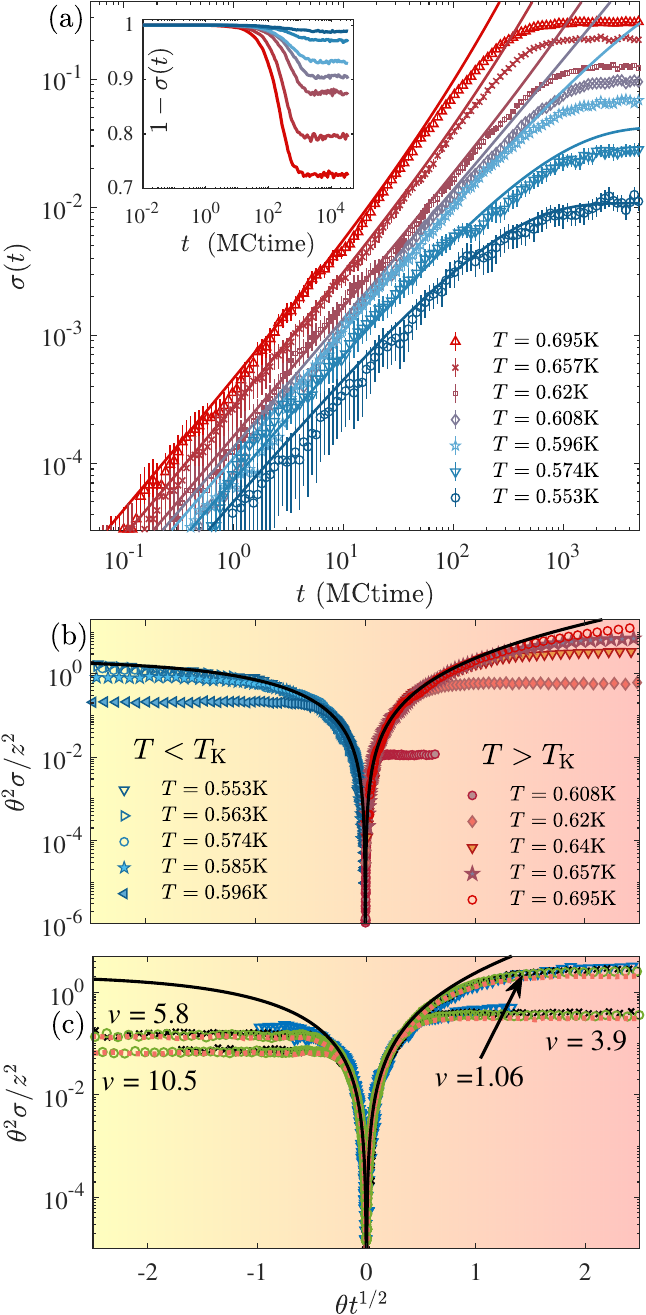}
 \caption{Dynamic scaling of magnetization for $N\sub{s} = 8192$ spins. As in \reffig{fig:MzScaling128000h0p16}, panel (a) shows MC data at $h=\qty{0.21}{\kelvin}$ for a set of temperatures close to $T\sub{K} = \qty{0.606}{\kelvin}$ compared with the scaling function in \refeq{eq:Phi} and panel (b) shows the same data plotted in terms of the dynamic scaling variables. Panel (c) shows data collapse using the generalized scaling behavior given in Eq.~(\ref{eq:ScalingForms2}), for \(v = 10.5\), \(5.8\), \(3.9\) and \(1.06\), with $h = 0.16$K (blue triangle), $h = 0.21$K (black cross), $h = 0.25$K (green open circle) and $h = 0.3$K (orange dots). Data are averaged over 50 independent runs.}
 \label{fig:MzScaling8192h0p21}
 \end{figure}

\begin{figure}[H]
\centering
\includegraphics[width=0.9\linewidth]{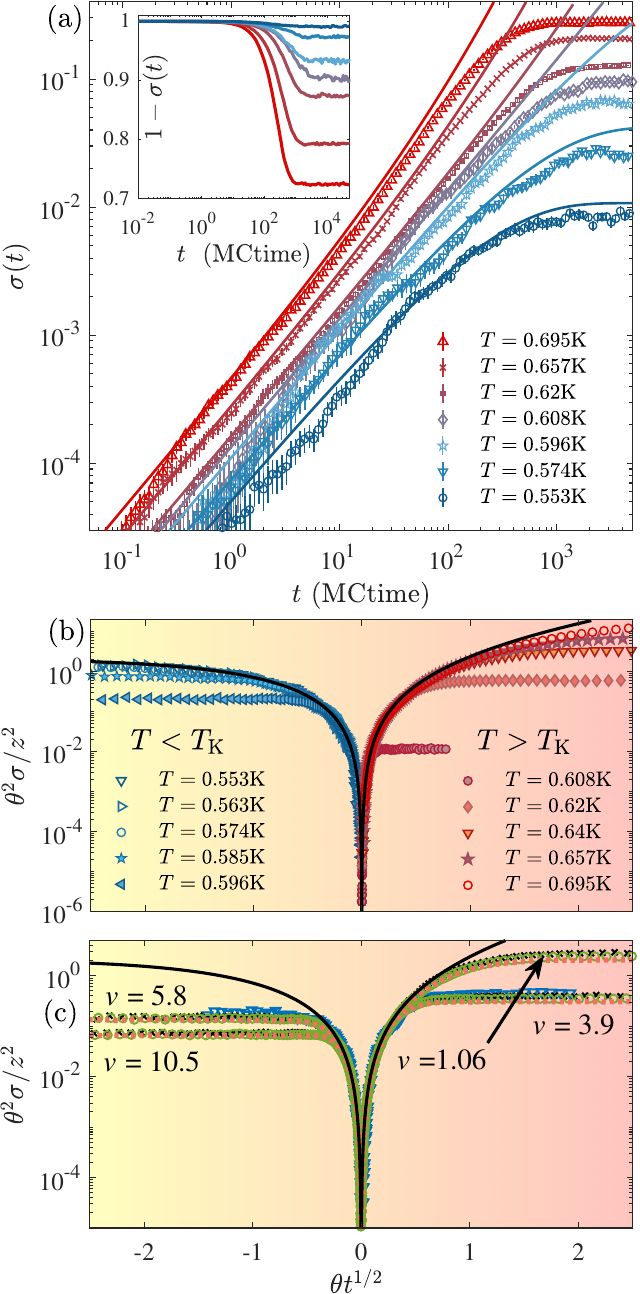}
\caption{Dynamic scaling of magnetization for $N\sub{s} = 27648$ spins. As in \reffig{fig:MzScaling128000h0p16}, panel (a) shows MC data at $h=\qty{0.21}{\kelvin}$ for a set of temperatures close to $T\sub{K} = \qty{0.606}{\kelvin}$ compared with the scaling function in \refeq{eq:Phi} and panel (b) shows the same data plotted in terms of the dynamic scaling variables. Panel (c) shows data collapse using the generalized scaling behavior given in Eq.~(\ref{eq:ScalingForms2}), with the same parameters as in Fig.~\ref{fig:MzScaling8192h0p21}. Data are averaged over 40 independent runs.}
\label{fig:MzScaling27648h0p21}
\end{figure}

First, we examine the role of system size on the scaling of the magnetization at fixed applied magnetic field $h=\qty{0.21}{\kelvin}$, in Figs.~\ref{fig:MzScaling8192h0p21} (number of spins $N\sub s = 8192$), \ref{fig:MzScaling27648h0p21} ($N\sub s = 27648$), and \ref{fig:MzScaling128000h0p21} ($N\sub s = 128000$). Finite-size effects are small for all parameter values shown, with deviations somewhat more noticeable close to the transition temperature \(T\sub{K} = \qty{0.606}{\kelvin}\). This can be understood by considering the inverse system size as an additional scaling variable, whose effects are more important when the other scaling variables (here, \(t^{-1}\), \(\theta\), and \(z\)) are smaller \cite{Powell2013}.

\subsubsection{MC results Set I: $N_s = 8192$, $h = \qty{0.21}{\kelvin}$ and $T\sub{K} = \qty{0.606}{\kelvin}$ }

 \subsubsection{MC results Set II: $N_s = 27648$, $h = \qty{0.21}{\kelvin}$ and $T\sub{K} = \qty{0.606}{\kelvin}$ }

 \subsubsection{MC results Set III: $N_s = 128000$, $h = \qty{0.21}{\kelvin}$ and $T\sub{K} = \qty{0.606}{\kelvin}$ }
\begin{figure}[H]
\centering
\includegraphics[width=1.0\linewidth]{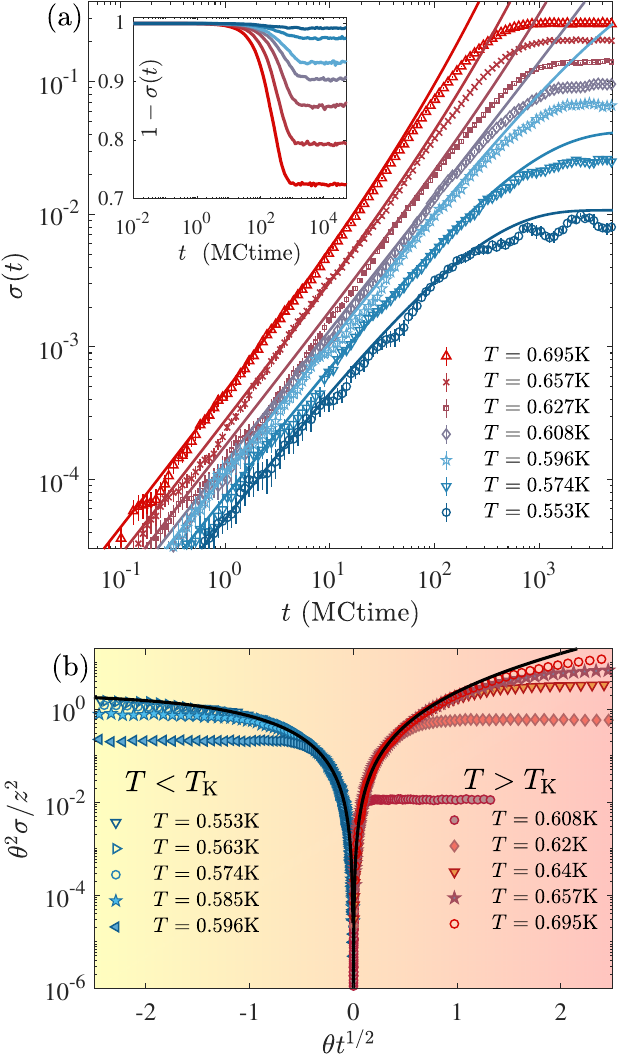}
\caption{Dynamic scaling of magnetization at $h=\qty{0.21}{\kelvin}$ for $N\sub{s} = 128000$ spins. As in \reffig{fig:MzScaling128000h0p16}, panel (a) shows MC data for a set of temperatures close to $T\sub{K} = \qty{0.606}{\kelvin}$ compared with the scaling function in \refeq{eq:Phi} and panel (b) shows the same data plotted in terms of the dynamic scaling variables. Data are averaged over 20 independent runs.}
\label{fig:MzScaling128000h0p21}
\end{figure}

\begin{figure}[H]
\centering
\includegraphics[width=1.0\linewidth]{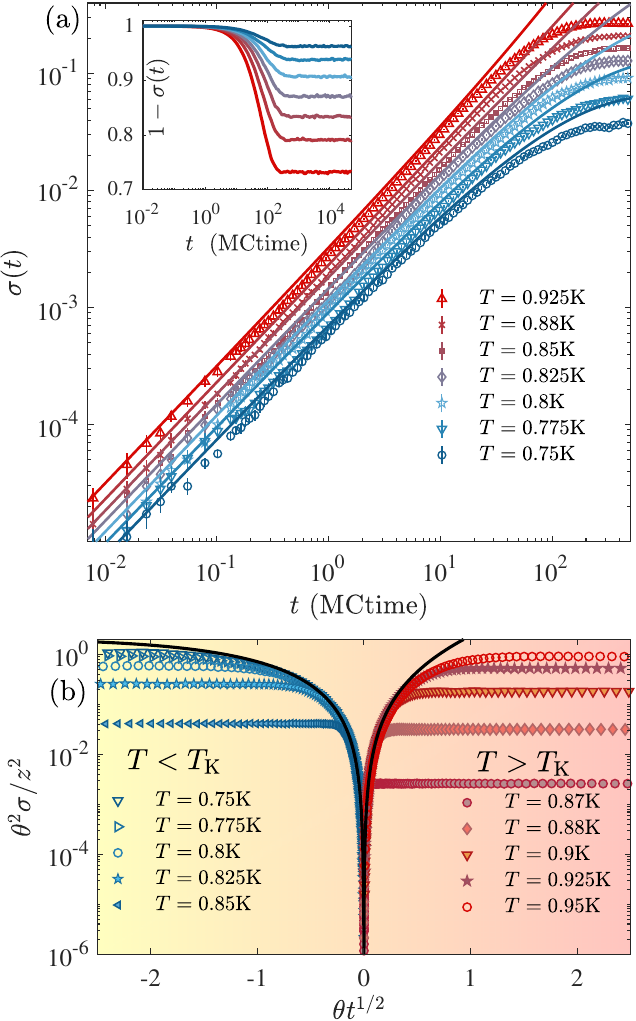}
\caption{Dynamic scaling of magnetization at $h=\qty{0.3}{\kelvin}$ for $N\sub{s} = 128000$ spins. As in \reffig{fig:MzScaling128000h0p16}, panel (a) shows MC data for a set of temperatures close to $T\sub{K} = \qty{0.866}{\kelvin}$ compared with the scaling function in \refeq{eq:Phi} and panel (b) shows the same data plotted in terms of the dynamic scaling variables. Data are averaged over 10 independent runs.}
\label{fig:MzScaling128000h0p3}
\end{figure}

To clarify the influence of the magnetic field, we investigate its variation within the largest system size (fixed $N\sub s$ = 128000). Comparing the results shown previously in Fig.~\ref{fig:MzScaling128000h0p16} (\(h = \qty{0.16}{\kelvin}\)) with  Figs.~\ref{fig:MzScaling128000h0p21} (\(h = \qty{0.21}{\kelvin}\)) and \ref{fig:MzScaling128000h0p3} (\(h = \qty{0.3}{\kelvin}\)), we note that the scaling relation holds best at the lowest magnetic field, where the value of \(z = e^{-\Delta/T}\) is smallest at the critical temperature \(T\sub K\). For higher fields, and particularly for \(h = \qty{0.3}{\kelvin}\), where \(T\sub{K} = \qty{0.866}{\kelvin}\), the agreement between simulation results and the single-string scaling functions is noticeably poorer.


\subsubsection{MC results Set IV: $N_s = 128000$, $h = \qty{0.3}{\kelvin}$ and $T\sub{K} = \qty{0.866}{\kelvin}$ }

\begin{figure}[H]
\centering
\includegraphics[width=1.0\linewidth]{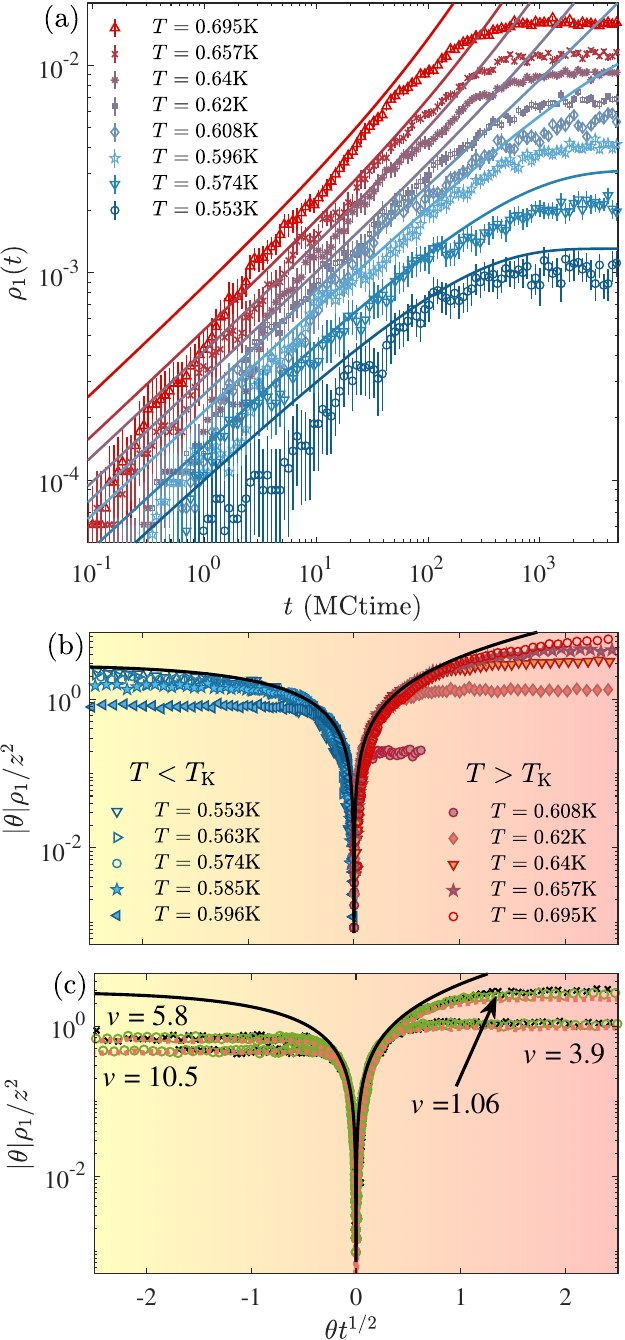}
\caption{Dynamic scaling of monopole density for $N\sub{s} = 8192$ spins. As in \reffig{fig:MonScaling128000h0p16}, panel (a) shows MC data at $h=\qty{0.21}{\kelvin}$ for a set of temperatures close to $T\sub{K} = \qty{0.606}{\kelvin}$ compared with the scaling function in \refeq{eq:Psi} and panel (b) shows the same data plotted in terms of the dynamic scaling variables. Panel (c) shows data collapse using the generalized scaling behavior given in Eq.~(\ref{eq:ScalingForms2}), for \(v = 10.5\), \(5.8\), \(3.9\) and \(1.06\), with $h = 0.16$K (blue triangle), $h = 0.21$K (black cross), $h = 0.25$K (green open circle) and $h = 0.3$K (orange dots). Data are averaged over 50 independent runs.}
\label{fig:MonScaling8192h0p21}
\end{figure}

\begin{figure}[H]
 \centering
 \includegraphics[width=1.0\linewidth]{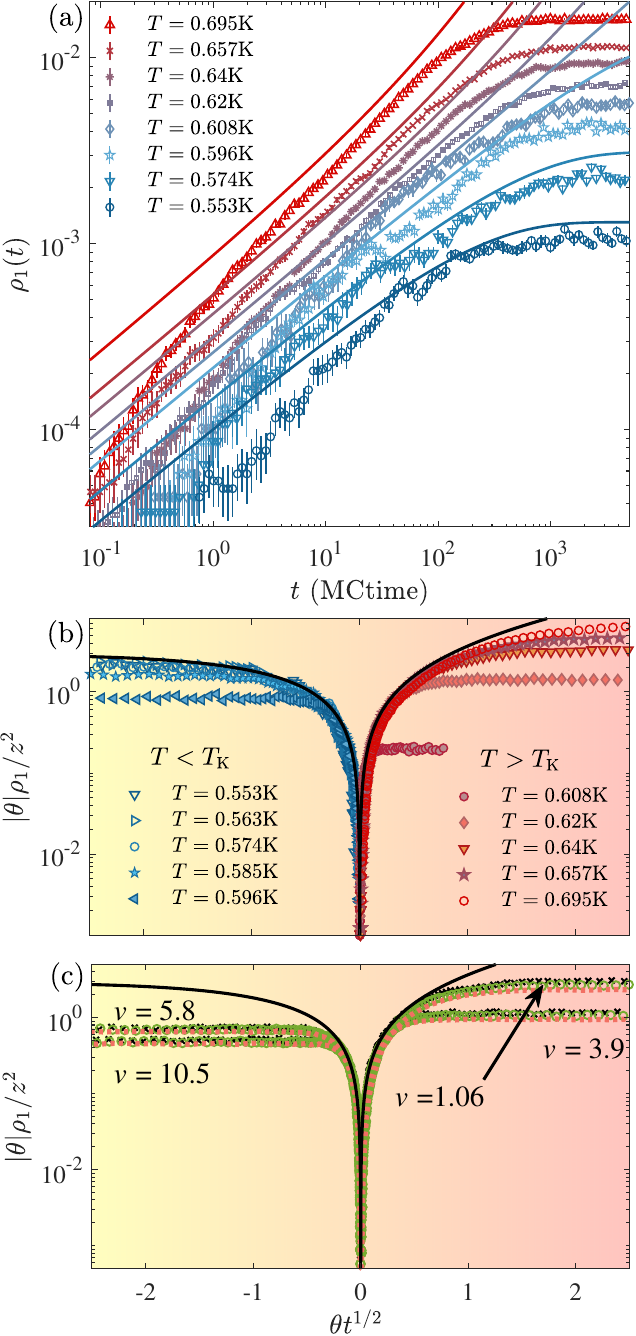}
 \caption{Dynamic scaling of monopole density for $N\sub{s} = 27648$ spins. As in \reffig{fig:MonScaling128000h0p16}, panel (a) shows MC data at $h=\qty{0.21}{\kelvin}$ for a set of temperatures close to $T\sub{K} = \qty{0.606}{\kelvin}$ compared with the scaling function in \refeq{eq:Psi} and panel (b) shows the same data plotted in terms of the dynamic scaling variables. Panel (c) shows data collapse using the generalized scaling behavior given in Eq.~(\ref{eq:ScalingForms2}), with the same parameters as in Fig.~\ref{fig:MzScaling8192h0p21}. Data are averaged over 40 independent runs.
 } 
 \label{fig:MonScaling27648h0p21}
 \end{figure}

\begin{figure}[H]
\centering
\includegraphics[width=0.85\linewidth]{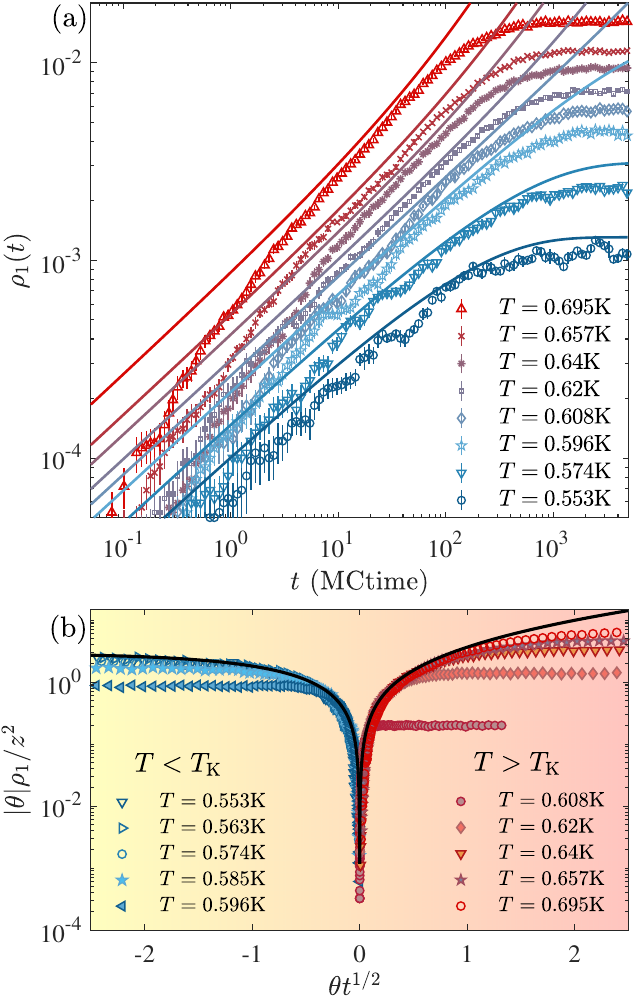}
\caption{Dynamic scaling of monopole density at $h=\qty{0.21}{\kelvin}$ for $N\sub{s} = 128000$ spins. As in \reffig{fig:MonScaling128000h0p16}, panel (a) shows MC data for a set of temperatures close to $T\sub{K} = \qty{0.606}{\kelvin}$ compared with the scaling function in \refeq{eq:Phi} and panel (b) shows the same data plotted in terms of the dynamic scaling variables. Data are averaged over 20 independent runs.}
\label{fig:MonScaling12800h0p21}
\end{figure}

\begin{figure}[H]
\centering
\includegraphics[width=0.85\linewidth]{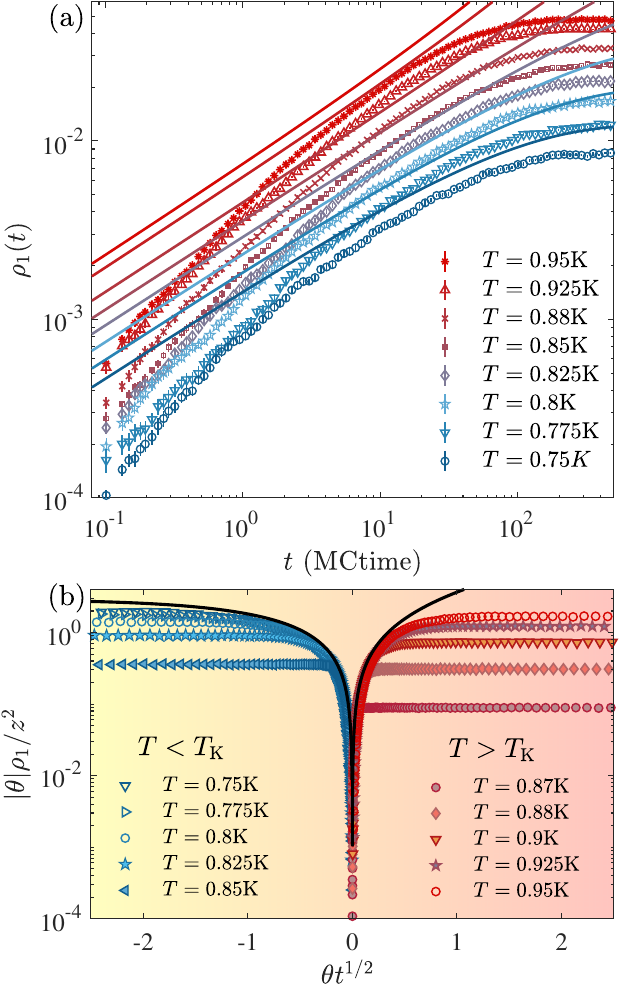}
\caption{Dynamic scaling of monopole density at $h=\qty{0.3}{\kelvin}$ for $N\sub{s} = 128000$ spins. As in \reffig{fig:MonScaling128000h0p16}, panel (a) shows MC data for a set of temperatures close to $T\sub{K} = \qty{0.866}{\kelvin}$ compared with the scaling function in \refeq{eq:Phi} and panel (b) shows the same data plotted in terms of the dynamic scaling variables. Data are averaged over 10 independent runs.}
\label{fig:MonScaling12800h0p3}
\end{figure}

Corresponding results for scaling of the monopole density are shown for the same set of system sizes in Figs.~\ref{fig:MonScaling8192h0p21}--\ref{fig:MonScaling12800h0p21} and for the same fields in Figs.~\ref{fig:MonScaling128000h0p16}, \ref{fig:MonScaling12800h0p21}, and \ref{fig:MonScaling12800h0p3}. As noted in the main text, agreement with the scaling forms is generally somewhat poorer for the monopole density than for the magnetization, and this is equally true at smaller system size and with larger applied field.


\subsection{Dynamic scaling of monopole density}
\subsubsection{MC results Set I: $N_s = 8192$, $h = \qty{0.21}{\kelvin}$ and $T\sub{K} = \qty{0.606}{\kelvin}$ }

\subsubsection{MC results Set II: $N_s = 27648$, $h = \qty{0.21}{\kelvin}$ and $T\sub{K} = \qty{0.606}{\kelvin}$ }
 \subsubsection{MC results Set III: $N_s = 128000$, $h=\qty{0.21}{\kelvin}$ and $T\sub{K} = \qty{0.606}{\kelvin}$}

\subsubsection{MC results Set IV: $N_s = 128000$, $h=\qty{0.3}{\kelvin}$ and $T\sub{K} = \qty{0.866}{\kelvin}$}

\begin{widetext}

\subsection{String-length distributions}

In this section, we show string length distributions from simulations with various values of the system size $N\sub{s}$ and applied field $h$.

In Figs.~\ref{fig:stDist@8192h0p21}, \ref{fig:stDist@27648h0p21}, and \ref{fig:stDist@128000h0p21}, we show the distributions at fixed $h=\qty{0.21}{\kelvin}$ for \(N\sub{s}=8192\), \(27468\), and \(128000\), respectively. These values of \(N\sub{s}\) correspond to cubic systems of linear size \(L_z = 32\), \(48\), and \(80\), in the same units as the string length. Finite-size effects are again relatively modest at \(t \lesssim 10^2\), with the simulation results following the single-string predictions quite closely, even for the smallest size. On the other hand, there is significant size dependence once strings comparable to or greater than the linear dimension begin to appear. (Strings with \(\ell > L_z\) are possible because of the periodic boundary conditions.) As discussed in \refsec{sec:res_dstrings}, cluster formation at very late times causes a decrease in the number of short strings. This is clearly visible in \reffig{fig:stDist@128000h0p21} at the latest time, \(t = 8703\), but is much less pronounced at the same \(t\) for smaller system sizes. To understand this, note that the typical transverse size of a system-spanning string is \(\sim \sqrt{L_z}\) and so the probability of two strings meeting and forming a cluster is lower for small system size.

The effect of the changing the field \(h\) on the string distribution can be seen by comparing Figs.~\ref{Fig:stDistTltTK128000} and \ref{Fig:stDistTgtTK128000} for \(h = \qty{0.16}{\kelvin}\) with Figs.~\ref{fig:stDist@128000h0p21} ($h=\qty{0.21}{\kelvin}$) and \ref{fig:stDist@128000h0p3} ($h=\qty{0.3}{\kelvin}$), all of which have \(N\sub{s}=128000\). As with the magnetization and monopole density, the quality of the match to the single-string model decreases with increasing \(h\). The onset of cluster formation, signaled by a rapid decrease in the number of short strings, is particularly prominent for \(h = \qty{0.3}{\kelvin}\).

\subsubsection{MC results Set I: $N_s = 8192$, $h=\qty{0.21}{\kelvin}$ and $T\sub{K} = \qty{0.606}{\kelvin}$}

\begin{figure}[H]
\centering
\includegraphics[width=0.86\linewidth]{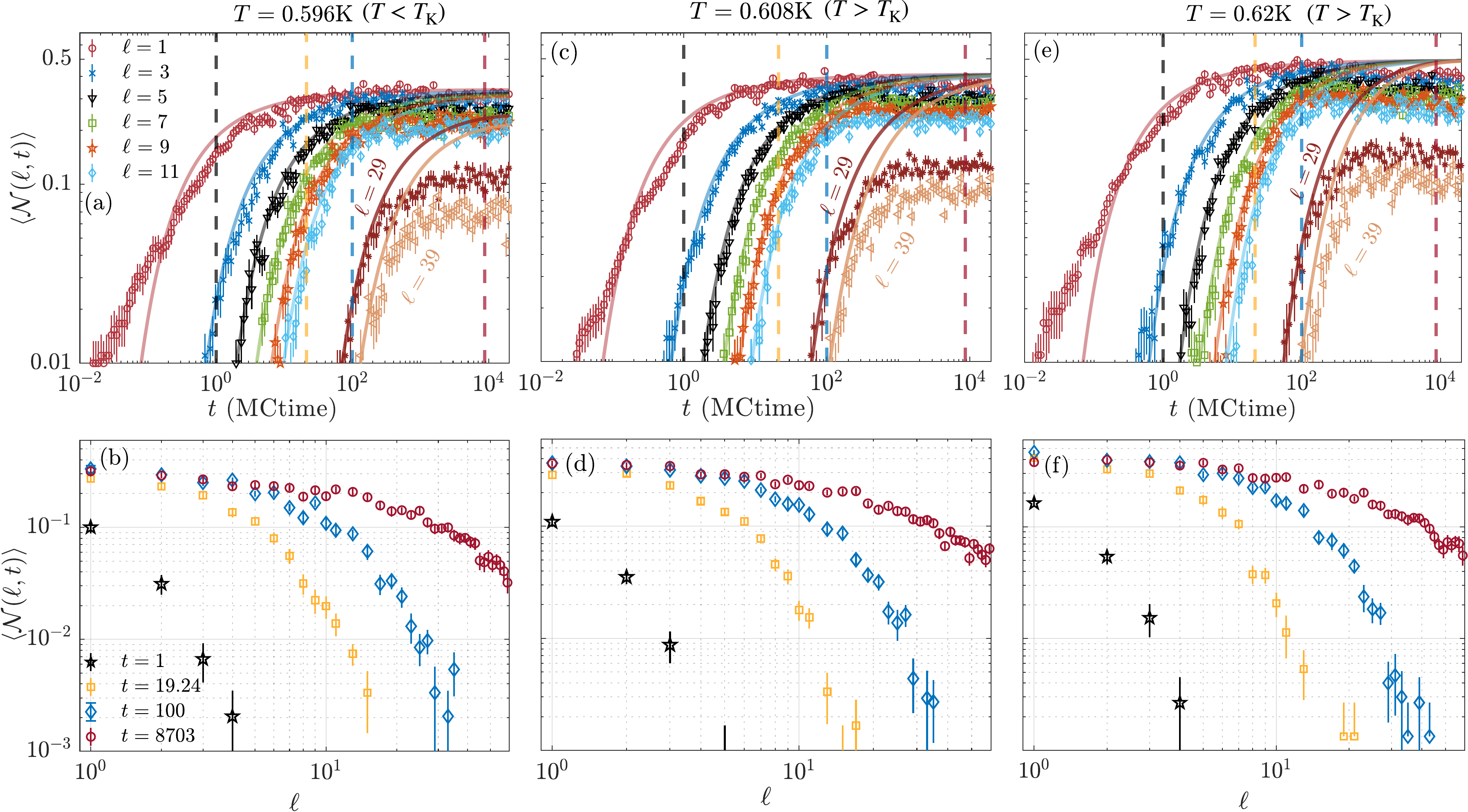}
\caption{String length distribution \(\langle \mathcal{N}(\ell,t)\rangle\) at \(h = \qty{0.21}{\kelvin}\) (\(T\sub{K} = \qty{0.606}{\kelvin}\)) for \(N\sub{s}=8192\) spins, plotted versus time \(t\) (top row) and length \(\ell\) (bottom row). Each data point is averaged over 500 independent runs. Symbols and lines have the same meaning as in Fig.~\ref{Fig:stDistTltTK128000}.}
\label{fig:stDist@8192h0p21}
\end{figure} 

\subsubsection{MC results Set II: $N_s = 27648$, $h=\qty{0.21}{\kelvin}$ and $T\sub{K} = \qty{0.606}{\kelvin}$}

\begin{figure}[H]
\centering
\includegraphics[width=0.86\linewidth]{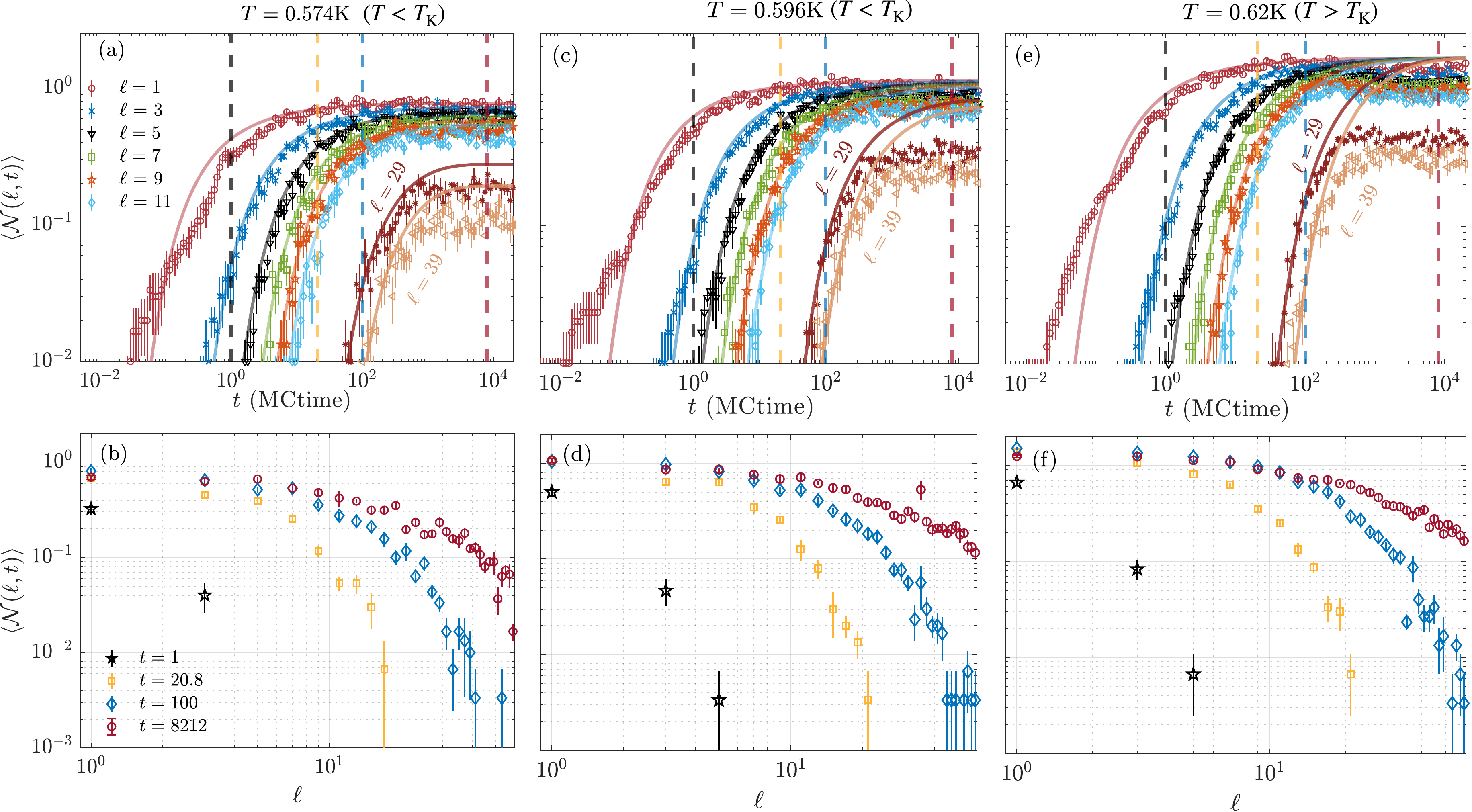}
\caption{String length distribution \(\langle \mathcal{N}(\ell,t)\rangle\) at \(h = \qty{0.21}{\kelvin}\) (\(T\sub{K} = \qty{0.606}{\kelvin}\)) for \(N\sub{s}=27648\) spins, plotted versus time \(t\) (top row) and length \(\ell\) (bottom row), as in Fig.~\ref{Fig:stDistTltTK128000}. Results are averaged over 300 independent runs.}
\label{fig:stDist@27648h0p21}
\end{figure} 

\subsubsection{MC results Set III: $N_s = 128000$, $h=\qty{0.21}{\kelvin}$ and $T\sub{K} = \qty{0.606}{\kelvin}$}

\begin{figure}[H]
\centering
\includegraphics[width=0.85\linewidth]{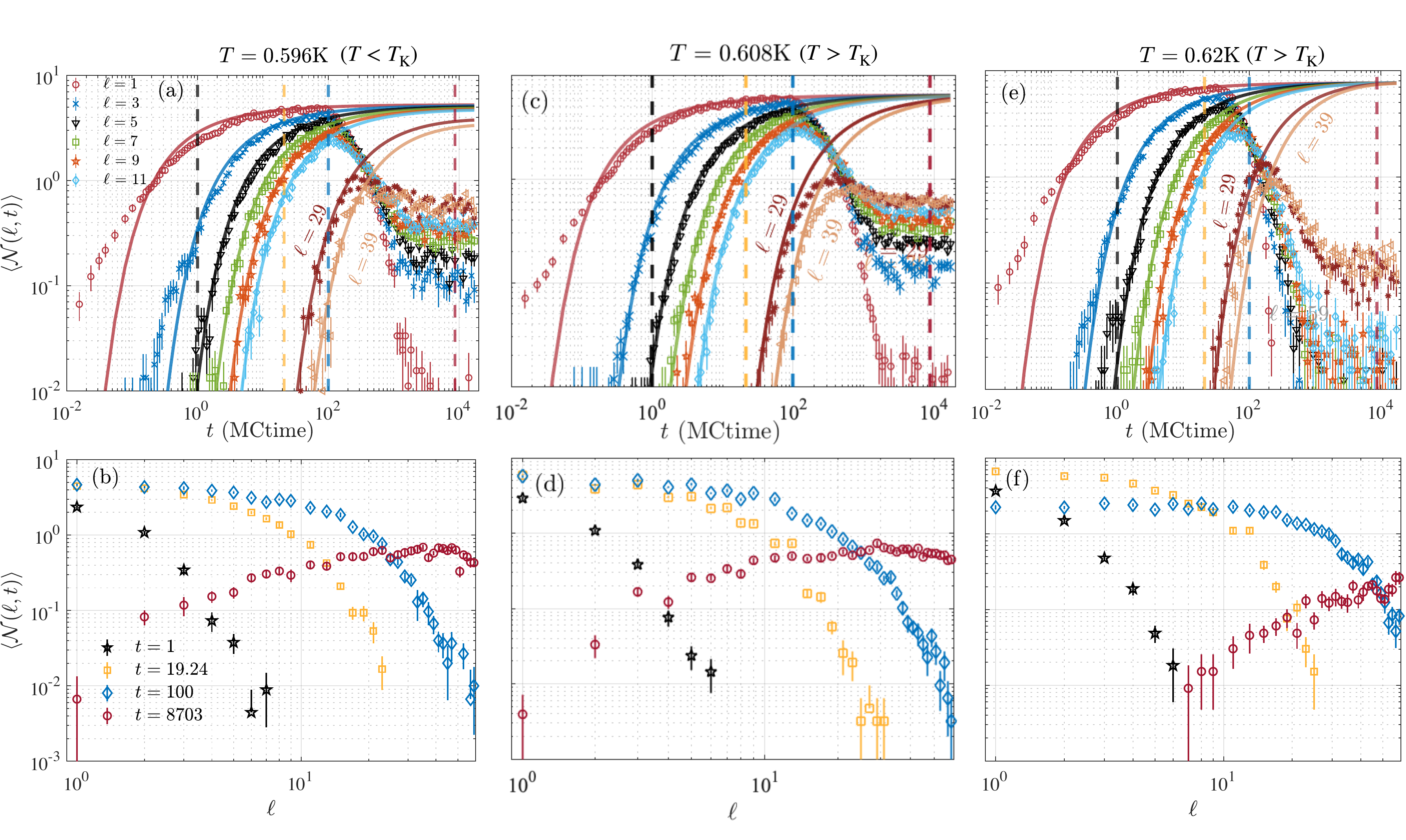}
\caption{String length distribution \(\langle \mathcal{N}(\ell,t)\rangle\) at \(h = \qty{0.21}{\kelvin}\) (\(T\sub{K} = \qty{0.606}{\kelvin}\)) for \(N\sub{s}=128000\) spins, plotted versus time \(t\) (top row) and length \(\ell\) (bottom row), as in Fig.~\ref{Fig:stDistTltTK128000}. Results are averaged over 30 independent runs.}
\label{fig:stDist@128000h0p21}
\end{figure} 

\subsubsection{MC results Set IV: $N_s = 128000$, $h=\qty{0.3}{\kelvin}$ and $T\sub{K} = \qty{0.866}{\kelvin}$}

\begin{figure}[H]
\centering
\includegraphics[width=0.85\linewidth]{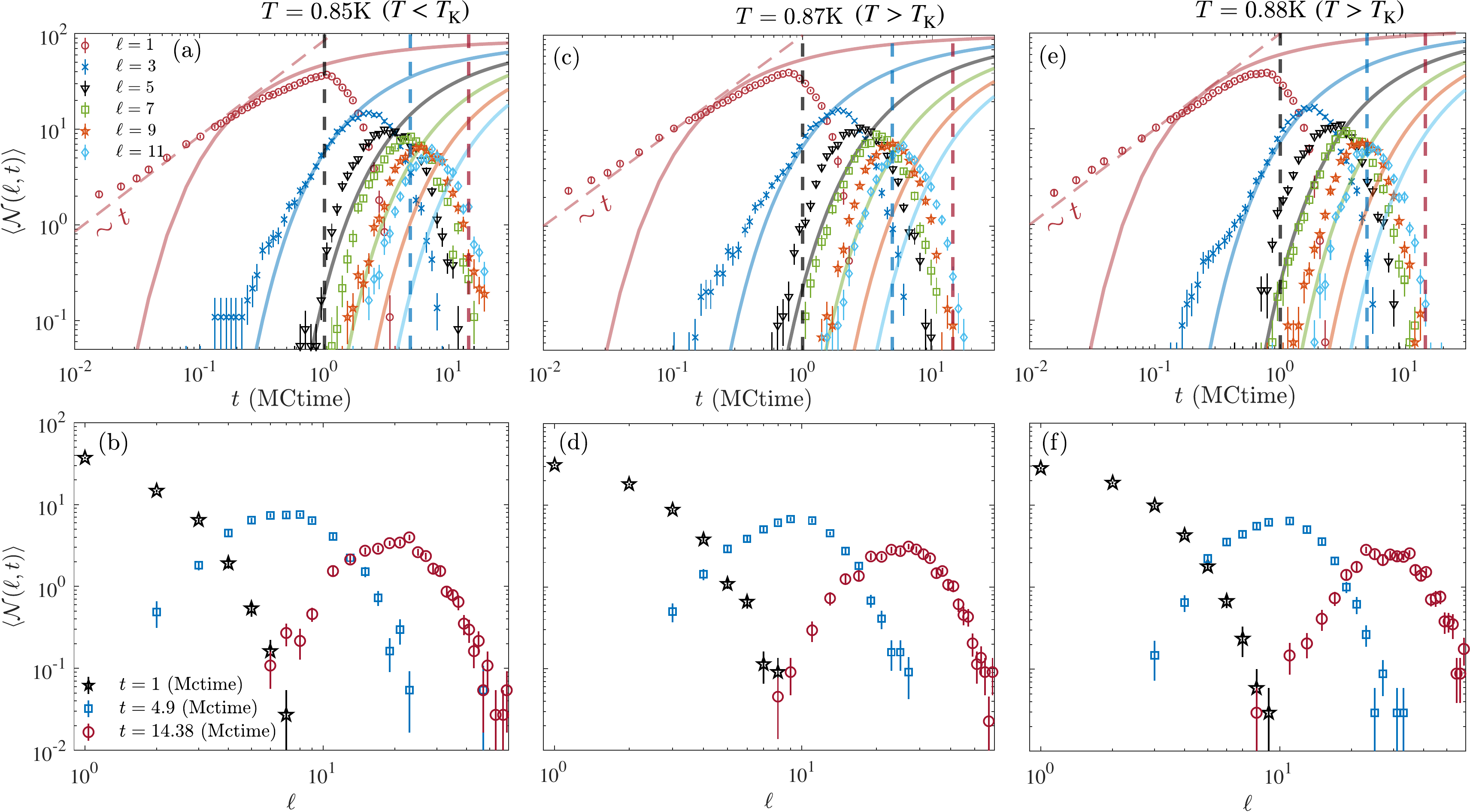}
\caption{String length distribution \(\langle \mathcal{N}(\ell,t)\rangle\) at \(h = \qty{0.3}{\kelvin}\) (\(T\sub{K} = \qty{0.866}{\kelvin}\)) for \(N\sub{s}=128000\) spins, plotted versus time \(t\) (top row) and length \(\ell\) (bottom row), as in Fig.~\ref{Fig:stDistTltTK128000}. Each data point is averaged over 50 independent runs.}
\label{fig:stDist@128000h0p3}
\end{figure} 

\end{widetext}




\bibliography{spinice}{}

\begin{thebibliography}{86}%
\makeatletter
\providecommand \@ifxundefined [1]{%
 \@ifx{#1\undefined}
}%
\providecommand \@ifnum [1]{%
 \ifnum #1\expandafter \@firstoftwo
 \else \expandafter \@secondoftwo
 \fi
}%
\providecommand \@ifx [1]{%
 \ifx #1\expandafter \@firstoftwo
 \else \expandafter \@secondoftwo
 \fi
}%
\providecommand \natexlab [1]{#1}%
\providecommand \enquote  [1]{``#1''}%
\providecommand \bibnamefont  [1]{#1}%
\providecommand \bibfnamefont [1]{#1}%
\providecommand \citenamefont [1]{#1}%
\providecommand \href@noop [0]{\@secondoftwo}%
\providecommand \href [0]{\begingroup \@sanitize@url \@href}%
\providecommand \@href[1]{\@@startlink{#1}\@@href}%
\providecommand \@@href[1]{\endgroup#1\@@endlink}%
\providecommand \@sanitize@url [0]{\catcode `\\12\catcode `\$12\catcode
  `\&12\catcode `\#12\catcode `\^12\catcode `\_12\catcode `\%12\relax}%
\providecommand \@@startlink[1]{}%
\providecommand \@@endlink[0]{}%
\providecommand \url  [0]{\begingroup\@sanitize@url \@url }%
\providecommand \@url [1]{\endgroup\@href {#1}{\urlprefix }}%
\providecommand \urlprefix  [0]{URL }%
\providecommand \Eprint [0]{\href }%
\providecommand \doibase [0]{https://doi.org/}%
\providecommand \selectlanguage [0]{\@gobble}%
\providecommand \bibinfo  [0]{\@secondoftwo}%
\providecommand \bibfield  [0]{\@secondoftwo}%
\providecommand \translation [1]{[#1]}%
\providecommand \BibitemOpen [0]{}%
\providecommand \bibitemStop [0]{}%
\providecommand \bibitemNoStop [0]{.\EOS\space}%
\providecommand \EOS [0]{\spacefactor3000\relax}%
\providecommand \BibitemShut  [1]{\csname bibitem#1\endcsname}%
\let\auto@bib@innerbib\@empty
\bibitem [{\citenamefont {Moessner}\ and\ \citenamefont
  {Ramirez}(2006)}]{Moessner2006}%
  \BibitemOpen
  \bibfield  {author} {\bibinfo {author} {\bibfnamefont {R.}~\bibnamefont
  {Moessner}}\ and\ \bibinfo {author} {\bibfnamefont {A.~P.}\ \bibnamefont
  {Ramirez}},\ }\bibfield  {title} {\bibinfo {title} {{Geometrical
  frustration}},\ }\href {https://doi.org/10.1063/1.2186278} {\bibfield
  {journal} {\bibinfo  {journal} {Physics Today}\ }\textbf {\bibinfo {volume}
  {59}},\ \bibinfo {pages} {24} (\bibinfo {year} {2006})}\BibitemShut {NoStop}%
\bibitem [{\citenamefont {Anderson}(1956)}]{Anderson1956}%
  \BibitemOpen
  \bibfield  {author} {\bibinfo {author} {\bibfnamefont {P.~W.}\ \bibnamefont
  {Anderson}},\ }\bibfield  {title} {\bibinfo {title} {Ordering and
  antiferromagnetism in ferrites},\ }\href
  {https://doi.org/10.1103/PhysRev.102.1008} {\bibfield  {journal} {\bibinfo
  {journal} {Phys. Rev.}\ }\textbf {\bibinfo {volume} {102}},\ \bibinfo {pages}
  {1008} (\bibinfo {year} {1956})}\BibitemShut {NoStop}%
\bibitem [{\citenamefont {Pauling}(1935)}]{Pauling1935}%
  \BibitemOpen
  \bibfield  {author} {\bibinfo {author} {\bibfnamefont {L.}~\bibnamefont
  {Pauling}},\ }\bibfield  {title} {\bibinfo {title} {The structure and entropy
  of ice and of other crystals with some randomness of atomic arrangement},\
  }\href {https://doi.org/10.1021/ja01315a102} {\bibfield  {journal} {\bibinfo
  {journal} {Journal of the American Chemical Society}\ }\textbf {\bibinfo
  {volume} {57}},\ \bibinfo {pages} {2680} (\bibinfo {year}
  {1935})}\BibitemShut {NoStop}%
\bibitem [{\citenamefont {Fowler}\ and\ \citenamefont
  {Rushbrooke}(1937)}]{Fowler1937}%
  \BibitemOpen
  \bibfield  {author} {\bibinfo {author} {\bibfnamefont {R.~H.}\ \bibnamefont
  {Fowler}}\ and\ \bibinfo {author} {\bibfnamefont {G.~S.}\ \bibnamefont
  {Rushbrooke}},\ }\bibfield  {title} {\bibinfo {title} {An attempt to extend
  the statistical theory of perfect solutions},\ }\href
  {https://doi.org/10.1039/TF9373301272} {\bibfield  {journal} {\bibinfo
  {journal} {Trans. Faraday Soc.}\ }\textbf {\bibinfo {volume} {33}},\ \bibinfo
  {pages} {1272} (\bibinfo {year} {1937})}\BibitemShut {NoStop}%
\bibitem [{\citenamefont {Glaetzle}\ \emph {et~al.}(2014)\citenamefont
  {Glaetzle}, \citenamefont {Dalmonte}, \citenamefont {Nath}, \citenamefont
  {Rousochatzakis}, \citenamefont {Moessner},\ and\ \citenamefont
  {Zoller}}]{Glaetzle2014}%
  \BibitemOpen
  \bibfield  {author} {\bibinfo {author} {\bibfnamefont {A.~W.}\ \bibnamefont
  {Glaetzle}}, \bibinfo {author} {\bibfnamefont {M.}~\bibnamefont {Dalmonte}},
  \bibinfo {author} {\bibfnamefont {R.}~\bibnamefont {Nath}}, \bibinfo {author}
  {\bibfnamefont {I.}~\bibnamefont {Rousochatzakis}}, \bibinfo {author}
  {\bibfnamefont {R.}~\bibnamefont {Moessner}},\ and\ \bibinfo {author}
  {\bibfnamefont {P.}~\bibnamefont {Zoller}},\ }\bibfield  {title} {\bibinfo
  {title} {Quantum spin-ice and dimer models with {R}ydberg atoms},\ }\href
  {https://doi.org/10.1103/PhysRevX.4.041037} {\bibfield  {journal} {\bibinfo
  {journal} {Phys. Rev. X}\ }\textbf {\bibinfo {volume} {4}},\ \bibinfo {pages}
  {041037} (\bibinfo {year} {2014})}\BibitemShut {NoStop}%
\bibitem [{\citenamefont {Zhu}\ \emph {et~al.}(2016)\citenamefont {Zhu},
  \citenamefont {Koch},\ and\ \citenamefont {Martin}}]{Zhu2016}%
  \BibitemOpen
  \bibfield  {author} {\bibinfo {author} {\bibfnamefont {G.}~\bibnamefont
  {Zhu}}, \bibinfo {author} {\bibfnamefont {J.}~\bibnamefont {Koch}},\ and\
  \bibinfo {author} {\bibfnamefont {I.}~\bibnamefont {Martin}},\ }\bibfield
  {title} {\bibinfo {title} {Nematic quantum liquid crystals of bosons in
  frustrated lattices},\ }\href {https://doi.org/10.1103/PhysRevB.93.144508}
  {\bibfield  {journal} {\bibinfo  {journal} {Phys. Rev. B}\ }\textbf {\bibinfo
  {volume} {93}},\ \bibinfo {pages} {144508} (\bibinfo {year}
  {2016})}\BibitemShut {NoStop}%
\bibitem [{\citenamefont {Menu}\ \emph {et~al.}(2024)\citenamefont {Menu},
  \citenamefont {Malo}, \citenamefont {Vuleti\ifmmode~\acute{c}\else
  \'{c}\fi{}}, \citenamefont {Chiofalo},\ and\ \citenamefont
  {Morigi}}]{Menu2024}%
  \BibitemOpen
  \bibfield  {author} {\bibinfo {author} {\bibfnamefont {R.}~\bibnamefont
  {Menu}}, \bibinfo {author} {\bibfnamefont {J.~Y.}\ \bibnamefont {Malo}},
  \bibinfo {author} {\bibfnamefont {V.}~\bibnamefont
  {Vuleti\ifmmode~\acute{c}\else \'{c}\fi{}}}, \bibinfo {author} {\bibfnamefont
  {M.~L.}\ \bibnamefont {Chiofalo}},\ and\ \bibinfo {author} {\bibfnamefont
  {G.}~\bibnamefont {Morigi}},\ }\bibfield  {title} {\bibinfo {title} {Quantum
  frustrated {W}igner chains},\ }\href
  {https://doi.org/10.1103/PhysRevB.110.155121} {\bibfield  {journal} {\bibinfo
   {journal} {Phys. Rev. B}\ }\textbf {\bibinfo {volume} {110}},\ \bibinfo
  {pages} {155121} (\bibinfo {year} {2024})}\BibitemShut {NoStop}%
\bibitem [{\citenamefont {Araki}\ \emph {et~al.}(2013)\citenamefont {Araki},
  \citenamefont {Serra},\ and\ \citenamefont {Tanaka}}]{Takeaki2013}%
  \BibitemOpen
  \bibfield  {author} {\bibinfo {author} {\bibfnamefont {T.}~\bibnamefont
  {Araki}}, \bibinfo {author} {\bibfnamefont {F.}~\bibnamefont {Serra}},\ and\
  \bibinfo {author} {\bibfnamefont {H.}~\bibnamefont {Tanaka}},\ }\bibfield
  {title} {\bibinfo {title} {Defect science and engineering of liquid crystals
  under geometrical frustration},\ }\href {https://doi.org/10.1039/C3SM50468A}
  {\bibfield  {journal} {\bibinfo  {journal} {Soft Matter}\ }\textbf {\bibinfo
  {volume} {9}},\ \bibinfo {pages} {8107} (\bibinfo {year} {2013})}\BibitemShut
  {NoStop}%
\bibitem [{\citenamefont {Jacqmin}\ \emph {et~al.}(2014)\citenamefont
  {Jacqmin}, \citenamefont {Carusotto}, \citenamefont {Sagnes}, \citenamefont
  {Abbarchi}, \citenamefont {Solnyshkov}, \citenamefont {Malpuech},
  \citenamefont {Galopin}, \citenamefont {Lema\^{\i}tre}, \citenamefont
  {Bloch},\ and\ \citenamefont {Amo}}]{Jacqmin2014}%
  \BibitemOpen
  \bibfield  {author} {\bibinfo {author} {\bibfnamefont {T.}~\bibnamefont
  {Jacqmin}}, \bibinfo {author} {\bibfnamefont {I.}~\bibnamefont {Carusotto}},
  \bibinfo {author} {\bibfnamefont {I.}~\bibnamefont {Sagnes}}, \bibinfo
  {author} {\bibfnamefont {M.}~\bibnamefont {Abbarchi}}, \bibinfo {author}
  {\bibfnamefont {D.~D.}\ \bibnamefont {Solnyshkov}}, \bibinfo {author}
  {\bibfnamefont {G.}~\bibnamefont {Malpuech}}, \bibinfo {author}
  {\bibfnamefont {E.}~\bibnamefont {Galopin}}, \bibinfo {author} {\bibfnamefont
  {A.}~\bibnamefont {Lema\^{\i}tre}}, \bibinfo {author} {\bibfnamefont
  {J.}~\bibnamefont {Bloch}},\ and\ \bibinfo {author} {\bibfnamefont
  {A.}~\bibnamefont {Amo}},\ }\bibfield  {title} {\bibinfo {title} {Direct
  observation of {D}irac cones and a flatband in a honeycomb lattice for
  polaritons},\ }\href {https://doi.org/10.1103/PhysRevLett.112.116402}
  {\bibfield  {journal} {\bibinfo  {journal} {Phys. Rev. Lett.}\ }\textbf
  {\bibinfo {volume} {112}},\ \bibinfo {pages} {116402} (\bibinfo {year}
  {2014})}\BibitemShut {NoStop}%
\bibitem [{\citenamefont {Schmidt}(2016)}]{Schmidt2016}%
  \BibitemOpen
  \bibfield  {author} {\bibinfo {author} {\bibfnamefont {S.}~\bibnamefont
  {Schmidt}},\ }\bibfield  {title} {\bibinfo {title} {Frustrated polaritons},\
  }\href {https://doi.org/10.1088/0031-8949/91/7/073006} {\bibfield  {journal}
  {\bibinfo  {journal} {Physica Scripta}\ }\textbf {\bibinfo {volume} {91}},\
  \bibinfo {pages} {073006} (\bibinfo {year} {2016})}\BibitemShut {NoStop}%
\bibitem [{\citenamefont {Koraltan}\ \emph {et~al.}(2021)\citenamefont
  {Koraltan}, \citenamefont {Slanovc}, \citenamefont {Bruckner}, \citenamefont
  {Nisoli}, \citenamefont {Chumak}, \citenamefont {Dobrovolskiy}, \citenamefont
  {Abert},\ and\ \citenamefont {Suess}}]{Koraltan2021}%
  \BibitemOpen
  \bibfield  {author} {\bibinfo {author} {\bibfnamefont {S.}~\bibnamefont
  {Koraltan}}, \bibinfo {author} {\bibfnamefont {F.}~\bibnamefont {Slanovc}},
  \bibinfo {author} {\bibfnamefont {F.}~\bibnamefont {Bruckner}}, \bibinfo
  {author} {\bibfnamefont {C.}~\bibnamefont {Nisoli}}, \bibinfo {author}
  {\bibfnamefont {A.~V.}\ \bibnamefont {Chumak}}, \bibinfo {author}
  {\bibfnamefont {O.~V.}\ \bibnamefont {Dobrovolskiy}}, \bibinfo {author}
  {\bibfnamefont {C.}~\bibnamefont {Abert}},\ and\ \bibinfo {author}
  {\bibfnamefont {D.}~\bibnamefont {Suess}},\ }\bibfield  {title} {\bibinfo
  {title} {Tension-free {D}irac strings and steered magnetic charges in {3D}
  artificial spin ice},\ }\href {https://doi.org/10.1038/s41524-021-00593-7}
  {\bibfield  {journal} {\bibinfo  {journal} {npj Computational Materials}\
  }\textbf {\bibinfo {volume} {7}},\ \bibinfo {pages} {125} (\bibinfo {year}
  {2021})}\BibitemShut {NoStop}%
\bibitem [{\citenamefont {Petrescu}\ \emph {et~al.}(2012)\citenamefont
  {Petrescu}, \citenamefont {Houck},\ and\ \citenamefont
  {Le~Hur}}]{Petrescu2012}%
  \BibitemOpen
  \bibfield  {author} {\bibinfo {author} {\bibfnamefont {A.}~\bibnamefont
  {Petrescu}}, \bibinfo {author} {\bibfnamefont {A.~A.}\ \bibnamefont
  {Houck}},\ and\ \bibinfo {author} {\bibfnamefont {K.}~\bibnamefont
  {Le~Hur}},\ }\bibfield  {title} {\bibinfo {title} {Anomalous {H}all effects
  of light and chiral edge modes on the kagom{\'e} lattice},\ }\href
  {https://doi.org/10.1103/PhysRevA.86.053804} {\bibfield  {journal} {\bibinfo
  {journal} {Phys. Rev. A}\ }\textbf {\bibinfo {volume} {86}},\ \bibinfo
  {pages} {053804} (\bibinfo {year} {2012})}\BibitemShut {NoStop}%
\bibitem [{\citenamefont {Biondi}\ \emph {et~al.}(2015)\citenamefont {Biondi},
  \citenamefont {van Nieuwenburg}, \citenamefont {Blatter}, \citenamefont
  {Huber},\ and\ \citenamefont {Schmidt}}]{Biondi2015}%
  \BibitemOpen
  \bibfield  {author} {\bibinfo {author} {\bibfnamefont {M.}~\bibnamefont
  {Biondi}}, \bibinfo {author} {\bibfnamefont {E.~P.~L.}\ \bibnamefont {van
  Nieuwenburg}}, \bibinfo {author} {\bibfnamefont {G.}~\bibnamefont {Blatter}},
  \bibinfo {author} {\bibfnamefont {S.~D.}\ \bibnamefont {Huber}},\ and\
  \bibinfo {author} {\bibfnamefont {S.}~\bibnamefont {Schmidt}},\ }\bibfield
  {title} {\bibinfo {title} {Incompressible polaritons in a flat band},\ }\href
  {https://doi.org/10.1103/PhysRevLett.115.143601} {\bibfield  {journal}
  {\bibinfo  {journal} {Phys. Rev. Lett.}\ }\textbf {\bibinfo {volume} {115}},\
  \bibinfo {pages} {143601} (\bibinfo {year} {2015})}\BibitemShut {NoStop}%
\bibitem [{\citenamefont {Ma}(2018)}]{Ma2018}%
  \BibitemOpen
  \bibfield  {author} {\bibinfo {author} {\bibfnamefont {S.}~\bibnamefont
  {Ma}},\ }\href {https://books.google.co.nz/books?id=R0haDwAAQBAJ} {\emph
  {\bibinfo {title} {Modern Theory Of Critical Phenomena}}}\ (\bibinfo
  {publisher} {Taylor \& Francis},\ \bibinfo {year} {2018})\BibitemShut
  {NoStop}%
\bibitem [{\citenamefont {Wilson}\ and\ \citenamefont
  {Kogut}(1974)}]{Wilson1974}%
  \BibitemOpen
  \bibfield  {author} {\bibinfo {author} {\bibfnamefont {K.~G.}\ \bibnamefont
  {Wilson}}\ and\ \bibinfo {author} {\bibfnamefont {J.}~\bibnamefont {Kogut}},\
  }\bibfield  {title} {\bibinfo {title} {The renormalization group and the
  $\epsilon$ expansion},\ }\href
  {https://doi.org/https://doi.org/10.1016/0370-1573(74)90023-4} {\bibfield
  {journal} {\bibinfo  {journal} {Physics Reports}\ }\textbf {\bibinfo {volume}
  {12}},\ \bibinfo {pages} {75} (\bibinfo {year} {1974})}\BibitemShut {NoStop}%
\bibitem [{\citenamefont {Senthil}\ \emph
  {et~al.}(2004{\natexlab{a}})\citenamefont {Senthil}, \citenamefont
  {Vishwanath}, \citenamefont {Balents}, \citenamefont {Sachdev},\ and\
  \citenamefont {Fisher}}]{Senthil2004a}%
  \BibitemOpen
  \bibfield  {author} {\bibinfo {author} {\bibfnamefont {T.}~\bibnamefont
  {Senthil}}, \bibinfo {author} {\bibfnamefont {A.}~\bibnamefont {Vishwanath}},
  \bibinfo {author} {\bibfnamefont {L.}~\bibnamefont {Balents}}, \bibinfo
  {author} {\bibfnamefont {S.}~\bibnamefont {Sachdev}},\ and\ \bibinfo {author}
  {\bibfnamefont {M.~P.~A.}\ \bibnamefont {Fisher}},\ }\bibfield  {title}
  {\bibinfo {title} {Deconfined quantum critical points},\ }\href
  {https://doi.org/10.1126/science.1091806} {\bibfield  {journal} {\bibinfo
  {journal} {Science}\ }\textbf {\bibinfo {volume} {303}},\ \bibinfo {pages}
  {1490 } (\bibinfo {year} {2004}{\natexlab{a}})}\BibitemShut {NoStop}%
\bibitem [{\citenamefont {Senthil}\ \emph
  {et~al.}(2004{\natexlab{b}})\citenamefont {Senthil}, \citenamefont {Balents},
  \citenamefont {Sachdev}, \citenamefont {Vishwanath},\ and\ \citenamefont
  {Fisher}}]{Senthil2004b}%
  \BibitemOpen
  \bibfield  {author} {\bibinfo {author} {\bibfnamefont {T.}~\bibnamefont
  {Senthil}}, \bibinfo {author} {\bibfnamefont {L.}~\bibnamefont {Balents}},
  \bibinfo {author} {\bibfnamefont {S.}~\bibnamefont {Sachdev}}, \bibinfo
  {author} {\bibfnamefont {A.}~\bibnamefont {Vishwanath}},\ and\ \bibinfo
  {author} {\bibfnamefont {M.~P.~A.}\ \bibnamefont {Fisher}},\ }\bibfield
  {title} {\bibinfo {title} {Quantum criticality beyond the
  {L}andau-{G}inzburg-{W}ilson paradigm},\ }\href
  {https://doi.org/10.1103/PhysRevB.70.144407} {\bibfield  {journal} {\bibinfo
  {journal} {Phys. Rev. B}\ }\textbf {\bibinfo {volume} {70}},\ \bibinfo
  {pages} {144407} (\bibinfo {year} {2004}{\natexlab{b}})}\BibitemShut
  {NoStop}%
\bibitem [{\citenamefont {Senthil}\ and\ \citenamefont
  {Fisher}(2006)}]{Senthil2006}%
  \BibitemOpen
  \bibfield  {author} {\bibinfo {author} {\bibfnamefont {T.}~\bibnamefont
  {Senthil}}\ and\ \bibinfo {author} {\bibfnamefont {M.~P.~A.}\ \bibnamefont
  {Fisher}},\ }\bibfield  {title} {\bibinfo {title} {Competing orders,
  nonlinear sigma models, and topological terms in quantum magnets},\ }\href
  {https://doi.org/10.1103/PhysRevB.74.064405} {\bibfield  {journal} {\bibinfo
  {journal} {Phys. Rev. B}\ }\textbf {\bibinfo {volume} {74}},\ \bibinfo
  {pages} {064405} (\bibinfo {year} {2006})}\BibitemShut {NoStop}%
\bibitem [{\citenamefont {Metlitski}\ and\ \citenamefont
  {Thorngren}(2018)}]{Metlitski2018}%
  \BibitemOpen
  \bibfield  {author} {\bibinfo {author} {\bibfnamefont {M.~A.}\ \bibnamefont
  {Metlitski}}\ and\ \bibinfo {author} {\bibfnamefont {R.}~\bibnamefont
  {Thorngren}},\ }\bibfield  {title} {\bibinfo {title} {Intrinsic and emergent
  anomalies at deconfined critical points},\ }\href
  {https://doi.org/10.1103/PhysRevB.98.085140} {\bibfield  {journal} {\bibinfo
  {journal} {Phys. Rev. B}\ }\textbf {\bibinfo {volume} {98}},\ \bibinfo
  {pages} {085140} (\bibinfo {year} {2018})}\BibitemShut {NoStop}%
\bibitem [{\citenamefont {Ferioli}\ \emph {et~al.}(2023)\citenamefont
  {Ferioli}, \citenamefont {Glicenstein}, \citenamefont {Ferrier-Barbut},\ and\
  \citenamefont {Browaeys}}]{Ferioli2023}%
  \BibitemOpen
  \bibfield  {author} {\bibinfo {author} {\bibfnamefont {G.}~\bibnamefont
  {Ferioli}}, \bibinfo {author} {\bibfnamefont {A.}~\bibnamefont
  {Glicenstein}}, \bibinfo {author} {\bibfnamefont {I.}~\bibnamefont
  {Ferrier-Barbut}},\ and\ \bibinfo {author} {\bibfnamefont {A.}~\bibnamefont
  {Browaeys}},\ }\bibfield  {title} {\bibinfo {title} {A non-equilibrium
  superradiant phase transition in free space},\ }\href
  {https://doi.org/10.1038/s41567-023-02064-w} {\bibfield  {journal} {\bibinfo
  {journal} {Nature Physics}\ }\textbf {\bibinfo {volume} {19}},\ \bibinfo
  {pages} {1345} (\bibinfo {year} {2023})}\BibitemShut {NoStop}%
\bibitem [{\citenamefont {Song}\ \emph {et~al.}(2024)\citenamefont {Song},
  \citenamefont {Zhang},\ and\ \citenamefont {Senthil}}]{Song2024}%
  \BibitemOpen
  \bibfield  {author} {\bibinfo {author} {\bibfnamefont {X.-Y.}\ \bibnamefont
  {Song}}, \bibinfo {author} {\bibfnamefont {Y.-H.}\ \bibnamefont {Zhang}},\
  and\ \bibinfo {author} {\bibfnamefont {T.}~\bibnamefont {Senthil}},\
  }\bibfield  {title} {\bibinfo {title} {Phase transitions out of quantum
  {H}all states in moir\'e materials},\ }\href
  {https://doi.org/10.1103/PhysRevB.109.085143} {\bibfield  {journal} {\bibinfo
   {journal} {Phys. Rev. B}\ }\textbf {\bibinfo {volume} {109}},\ \bibinfo
  {pages} {085143} (\bibinfo {year} {2024})}\BibitemShut {NoStop}%
\bibitem [{\citenamefont {Jaubert}\ \emph {et~al.}(2008)\citenamefont
  {Jaubert}, \citenamefont {Chalker}, \citenamefont {Holdsworth},\ and\
  \citenamefont {Moessner}}]{Jaubert2008}%
  \BibitemOpen
  \bibfield  {author} {\bibinfo {author} {\bibfnamefont {L.~D.~C.}\
  \bibnamefont {Jaubert}}, \bibinfo {author} {\bibfnamefont {J.~T.}\
  \bibnamefont {Chalker}}, \bibinfo {author} {\bibfnamefont {P.~C.~W.}\
  \bibnamefont {Holdsworth}},\ and\ \bibinfo {author} {\bibfnamefont
  {R.}~\bibnamefont {Moessner}},\ }\bibfield  {title} {\bibinfo {title}
  {Three-dimensional {K}asteleyn transition: Spin ice in a [100] field},\
  }\href {https://doi.org/10.1103/PhysRevLett.100.067207} {\bibfield  {journal}
  {\bibinfo  {journal} {Phys. Rev. Lett.}\ }\textbf {\bibinfo {volume} {100}},\
  \bibinfo {pages} {067207} (\bibinfo {year} {2008})}\BibitemShut {NoStop}%
\bibitem [{\citenamefont {Powell}(2015)}]{Powell2015}%
  \BibitemOpen
  \bibfield  {author} {\bibinfo {author} {\bibfnamefont {S.}~\bibnamefont
  {Powell}},\ }\bibfield  {title} {\bibinfo {title} {Ferromagnetic {C}oulomb
  phase in classical spin ice},\ }\href
  {https://doi.org/10.1103/PhysRevB.91.094431} {\bibfield  {journal} {\bibinfo
  {journal} {Phys. Rev. B}\ }\textbf {\bibinfo {volume} {91}},\ \bibinfo
  {pages} {094431} (\bibinfo {year} {2015})}\BibitemShut {NoStop}%
\bibitem [{\citenamefont {Szab\'o}\ \emph {et~al.}(2025)\citenamefont
  {Szab\'o}, \citenamefont {Grigera}, \citenamefont {Holdsworth}, \citenamefont
  {Jaubert}, \citenamefont {Moessner}, \citenamefont {Slobinsky}, \citenamefont
  {Sturla},\ and\ \citenamefont {Borzi}}]{Szabo2025}%
  \BibitemOpen
  \bibfield  {author} {\bibinfo {author} {\bibfnamefont {A.}~\bibnamefont
  {Szab\'o}}, \bibinfo {author} {\bibfnamefont {S.~A.}\ \bibnamefont
  {Grigera}}, \bibinfo {author} {\bibfnamefont {P.~C.~W.}\ \bibnamefont
  {Holdsworth}}, \bibinfo {author} {\bibfnamefont {L.~D.~C.}\ \bibnamefont
  {Jaubert}}, \bibinfo {author} {\bibfnamefont {R.}~\bibnamefont {Moessner}},
  \bibinfo {author} {\bibfnamefont {D.~G.}\ \bibnamefont {Slobinsky}}, \bibinfo
  {author} {\bibfnamefont {M.}~\bibnamefont {Sturla}},\ and\ \bibinfo {author}
  {\bibfnamefont {R.~A.}\ \bibnamefont {Borzi}},\ }\bibfield  {title} {\bibinfo
  {title} {Hidden order and {$\mathbb{Z}_{2}$} confinement transition in a
  fully packed monopole liquid},\ }\href {https://doi.org/10.1103/m7ck-5bz2}
  {\bibfield  {journal} {\bibinfo  {journal} {Phys. Rev. B}\ }\textbf {\bibinfo
  {volume} {112}},\ \bibinfo {pages} {014452} (\bibinfo {year}
  {2025})}\BibitemShut {NoStop}%
\bibitem [{\citenamefont {Alet}\ \emph {et~al.}(2006)\citenamefont {Alet},
  \citenamefont {Walczak},\ and\ \citenamefont {Fisher}}]{Alet2006}%
  \BibitemOpen
  \bibfield  {author} {\bibinfo {author} {\bibfnamefont {F.}~\bibnamefont
  {Alet}}, \bibinfo {author} {\bibfnamefont {A.~M.}\ \bibnamefont {Walczak}},\
  and\ \bibinfo {author} {\bibfnamefont {M.~P.}\ \bibnamefont {Fisher}},\
  }\bibfield  {title} {\bibinfo {title} {Exotic quantum phases and phase
  transitions in correlated matter},\ }\href
  {https://doi.org/https://doi.org/10.1016/j.physa.2006.04.003} {\bibfield
  {journal} {\bibinfo  {journal} {Physica A: Statistical Mechanics and its
  Applications}\ }\textbf {\bibinfo {volume} {369}},\ \bibinfo {pages} {122}
  (\bibinfo {year} {2006})}\BibitemShut {NoStop}%
\bibitem [{\citenamefont {Wilkins}\ and\ \citenamefont
  {Powell}(2019)}]{Wilkins2019}%
  \BibitemOpen
  \bibfield  {author} {\bibinfo {author} {\bibfnamefont {N.}~\bibnamefont
  {Wilkins}}\ and\ \bibinfo {author} {\bibfnamefont {S.}~\bibnamefont
  {Powell}},\ }\bibfield  {title} {\bibinfo {title} {Synchronization transition
  in the double dimer model on the cubic lattice},\ }\href
  {https://doi.org/10.1103/PhysRevB.99.144403} {\bibfield  {journal} {\bibinfo
  {journal} {Phys. Rev. B}\ }\textbf {\bibinfo {volume} {99}},\ \bibinfo
  {pages} {144403} (\bibinfo {year} {2019})}\BibitemShut {NoStop}%
\bibitem [{\citenamefont {Desai}\ \emph {et~al.}(2021)\citenamefont {Desai},
  \citenamefont {Pujari},\ and\ \citenamefont {Damle}}]{Desai2021}%
  \BibitemOpen
  \bibfield  {author} {\bibinfo {author} {\bibfnamefont {N.}~\bibnamefont
  {Desai}}, \bibinfo {author} {\bibfnamefont {S.}~\bibnamefont {Pujari}},\ and\
  \bibinfo {author} {\bibfnamefont {K.}~\bibnamefont {Damle}},\ }\bibfield
  {title} {\bibinfo {title} {Bilayer {C}oulomb phase of two-dimensional dimer
  models: Absence of power-law columnar order},\ }\href
  {https://doi.org/10.1103/PhysRevE.103.042136} {\bibfield  {journal} {\bibinfo
   {journal} {Phys. Rev. E}\ }\textbf {\bibinfo {volume} {103}},\ \bibinfo
  {pages} {042136} (\bibinfo {year} {2021})}\BibitemShut {NoStop}%
\bibitem [{\citenamefont {Hohenberg}\ and\ \citenamefont
  {Halperin}(1977)}]{Hohenberg1977}%
  \BibitemOpen
  \bibfield  {author} {\bibinfo {author} {\bibfnamefont {P.~C.}\ \bibnamefont
  {Hohenberg}}\ and\ \bibinfo {author} {\bibfnamefont {B.~I.}\ \bibnamefont
  {Halperin}},\ }\bibfield  {title} {\bibinfo {title} {Theory of dynamic
  critical phenomena},\ }\href {https://doi.org/10.1103/RevModPhys.49.435}
  {\bibfield  {journal} {\bibinfo  {journal} {Rev. Mod. Phys.}\ }\textbf
  {\bibinfo {volume} {49}},\ \bibinfo {pages} {435} (\bibinfo {year}
  {1977})}\BibitemShut {NoStop}%
\bibitem [{\citenamefont {Polkovnikov}(2005)}]{Polkovnikov2005}%
  \BibitemOpen
  \bibfield  {author} {\bibinfo {author} {\bibfnamefont {A.}~\bibnamefont
  {Polkovnikov}},\ }\bibfield  {title} {\bibinfo {title} {Universal adiabatic
  dynamics in the vicinity of a quantum critical point},\ }\href
  {https://doi.org/10.1103/PhysRevB.72.161201} {\bibfield  {journal} {\bibinfo
  {journal} {Phys. Rev. B}\ }\textbf {\bibinfo {volume} {72}},\ \bibinfo
  {pages} {161201} (\bibinfo {year} {2005})}\BibitemShut {NoStop}%
\bibitem [{\citenamefont {Zurek}\ \emph {et~al.}(2005)\citenamefont {Zurek},
  \citenamefont {Dorner},\ and\ \citenamefont {Zoller}}]{Zurek2005}%
  \BibitemOpen
  \bibfield  {author} {\bibinfo {author} {\bibfnamefont {W.~H.}\ \bibnamefont
  {Zurek}}, \bibinfo {author} {\bibfnamefont {U.}~\bibnamefont {Dorner}},\ and\
  \bibinfo {author} {\bibfnamefont {P.}~\bibnamefont {Zoller}},\ }\bibfield
  {title} {\bibinfo {title} {Dynamics of a quantum phase transition},\ }\href
  {https://doi.org/10.1103/PhysRevLett.95.105701} {\bibfield  {journal}
  {\bibinfo  {journal} {Phys. Rev. Lett.}\ }\textbf {\bibinfo {volume} {95}},\
  \bibinfo {pages} {105701} (\bibinfo {year} {2005})}\BibitemShut {NoStop}%
\bibitem [{\citenamefont {Polkovnikov}\ \emph {et~al.}(2011)\citenamefont
  {Polkovnikov}, \citenamefont {Sengupta}, \citenamefont {Silva},\ and\
  \citenamefont {Vengalattore}}]{Polkovnikov2011}%
  \BibitemOpen
  \bibfield  {author} {\bibinfo {author} {\bibfnamefont {A.}~\bibnamefont
  {Polkovnikov}}, \bibinfo {author} {\bibfnamefont {K.}~\bibnamefont
  {Sengupta}}, \bibinfo {author} {\bibfnamefont {A.}~\bibnamefont {Silva}},\
  and\ \bibinfo {author} {\bibfnamefont {M.}~\bibnamefont {Vengalattore}},\
  }\bibfield  {title} {\bibinfo {title} {Colloquium: Nonequilibrium dynamics of
  closed interacting quantum systems},\ }\href
  {https://doi.org/10.1103/RevModPhys.83.863} {\bibfield  {journal} {\bibinfo
  {journal} {Rev. Mod. Phys.}\ }\textbf {\bibinfo {volume} {83}},\ \bibinfo
  {pages} {863} (\bibinfo {year} {2011})}\BibitemShut {NoStop}%
\bibitem [{\citenamefont {Kolodrubetz}\ \emph {et~al.}(2012)\citenamefont
  {Kolodrubetz}, \citenamefont {Clark},\ and\ \citenamefont
  {Huse}}]{Kolodrubetz2012}%
  \BibitemOpen
  \bibfield  {author} {\bibinfo {author} {\bibfnamefont {M.}~\bibnamefont
  {Kolodrubetz}}, \bibinfo {author} {\bibfnamefont {B.~K.}\ \bibnamefont
  {Clark}},\ and\ \bibinfo {author} {\bibfnamefont {D.~A.}\ \bibnamefont
  {Huse}},\ }\bibfield  {title} {\bibinfo {title} {Nonequilibrium dynamic
  critical scaling of the quantum {I}sing chain},\ }\href
  {https://doi.org/10.1103/PhysRevLett.109.015701} {\bibfield  {journal}
  {\bibinfo  {journal} {Phys. Rev. Lett.}\ }\textbf {\bibinfo {volume} {109}},\
  \bibinfo {pages} {015701} (\bibinfo {year} {2012})}\BibitemShut {NoStop}%
\bibitem [{\citenamefont {Liu}\ \emph {et~al.}(2014)\citenamefont {Liu},
  \citenamefont {Polkovnikov},\ and\ \citenamefont {Sandvik}}]{Liu2014}%
  \BibitemOpen
  \bibfield  {author} {\bibinfo {author} {\bibfnamefont {C.-W.}\ \bibnamefont
  {Liu}}, \bibinfo {author} {\bibfnamefont {A.}~\bibnamefont {Polkovnikov}},\
  and\ \bibinfo {author} {\bibfnamefont {A.~W.}\ \bibnamefont {Sandvik}},\
  }\bibfield  {title} {\bibinfo {title} {Dynamic scaling at classical phase
  transitions approached through nonequilibrium quenching},\ }\href
  {https://doi.org/10.1103/PhysRevB.89.054307} {\bibfield  {journal} {\bibinfo
  {journal} {Phys. Rev. B}\ }\textbf {\bibinfo {volume} {89}},\ \bibinfo
  {pages} {054307} (\bibinfo {year} {2014})}\BibitemShut {NoStop}%
\bibitem [{\citenamefont {Kibble}(1980)}]{Kibble1980}%
  \BibitemOpen
  \bibfield  {author} {\bibinfo {author} {\bibfnamefont {T.}~\bibnamefont
  {Kibble}},\ }\bibfield  {title} {\bibinfo {title} {Some implications of a
  cosmological phase transition},\ }\href
  {https://doi.org/https://doi.org/10.1016/0370-1573(80)90091-5} {\bibfield
  {journal} {\bibinfo  {journal} {Physics Reports}\ }\textbf {\bibinfo {volume}
  {67}},\ \bibinfo {pages} {183} (\bibinfo {year} {1980})}\BibitemShut
  {NoStop}%
\bibitem [{\citenamefont {Zurek}(1985)}]{Zurek1985}%
  \BibitemOpen
  \bibfield  {author} {\bibinfo {author} {\bibfnamefont {W.~H.}\ \bibnamefont
  {Zurek}},\ }\bibfield  {title} {\bibinfo {title} {Cosmological experiments in
  superfluid helium?},\ }\href {https://doi.org/10.1038/317505a0} {\bibfield
  {journal} {\bibinfo  {journal} {Nature}\ }\textbf {\bibinfo {volume} {317}},\
  \bibinfo {pages} {505} (\bibinfo {year} {1985})}\BibitemShut {NoStop}%
\bibitem [{\citenamefont {Hindmarsh}\ and\ \citenamefont
  {Rajantie}(2000)}]{Hindmarsh2000}%
  \BibitemOpen
  \bibfield  {author} {\bibinfo {author} {\bibfnamefont {M.}~\bibnamefont
  {Hindmarsh}}\ and\ \bibinfo {author} {\bibfnamefont {A.}~\bibnamefont
  {Rajantie}},\ }\bibfield  {title} {\bibinfo {title} {Defect formation and
  local gauge invariance},\ }\href
  {https://doi.org/10.1103/PhysRevLett.85.4660} {\bibfield  {journal} {\bibinfo
   {journal} {Phys. Rev. Lett.}\ }\textbf {\bibinfo {volume} {85}},\ \bibinfo
  {pages} {4660} (\bibinfo {year} {2000})}\BibitemShut {NoStop}%
\bibitem [{\citenamefont {Chandran}\ \emph {et~al.}(2012)\citenamefont
  {Chandran}, \citenamefont {Erez}, \citenamefont {Gubser},\ and\ \citenamefont
  {Sondhi}}]{Chandran2012}%
  \BibitemOpen
  \bibfield  {author} {\bibinfo {author} {\bibfnamefont {A.}~\bibnamefont
  {Chandran}}, \bibinfo {author} {\bibfnamefont {A.}~\bibnamefont {Erez}},
  \bibinfo {author} {\bibfnamefont {S.~S.}\ \bibnamefont {Gubser}},\ and\
  \bibinfo {author} {\bibfnamefont {S.~L.}\ \bibnamefont {Sondhi}},\ }\bibfield
   {title} {\bibinfo {title} {Kibble-{Z}urek problem: Universality and the
  scaling limit},\ }\href {https://doi.org/10.1103/PhysRevB.86.064304}
  {\bibfield  {journal} {\bibinfo  {journal} {Phys. Rev. B}\ }\textbf {\bibinfo
  {volume} {86}},\ \bibinfo {pages} {064304} (\bibinfo {year}
  {2012})}\BibitemShut {NoStop}%
\bibitem [{\citenamefont {Lacroix}\ \emph {et~al.}(2011)\citenamefont
  {Lacroix}, \citenamefont {Mendels},\ and\ \citenamefont
  {Mila}}]{FmagSpringer2011}%
  \BibitemOpen
  \bibinfo {editor} {\bibfnamefont {C.}~\bibnamefont {Lacroix}}, \bibinfo
  {editor} {\bibfnamefont {P.}~\bibnamefont {Mendels}},\ and\ \bibinfo {editor}
  {\bibfnamefont {F.}~\bibnamefont {Mila}},\ eds.,\ \href
  {https://doi.org/https://doi.org/10.1007/978-3-642-10589-0} {\emph {\bibinfo
  {title} {Introduction to Frustrated Magnetism}}},\ \bibinfo {edition} {1st}\
  ed.\ (\bibinfo  {publisher} {Springer Berlin, Heidelberg},\ \bibinfo {year}
  {2011})\BibitemShut {NoStop}%
\bibitem [{\citenamefont {Sakakibara}\ \emph {et~al.}(2003)\citenamefont
  {Sakakibara}, \citenamefont {Tayama}, \citenamefont {Hiroi}, \citenamefont
  {Matsuhira},\ and\ \citenamefont {Takagi}}]{Sakakibara2003}%
  \BibitemOpen
  \bibfield  {author} {\bibinfo {author} {\bibfnamefont {T.}~\bibnamefont
  {Sakakibara}}, \bibinfo {author} {\bibfnamefont {T.}~\bibnamefont {Tayama}},
  \bibinfo {author} {\bibfnamefont {Z.}~\bibnamefont {Hiroi}}, \bibinfo
  {author} {\bibfnamefont {K.}~\bibnamefont {Matsuhira}},\ and\ \bibinfo
  {author} {\bibfnamefont {S.}~\bibnamefont {Takagi}},\ }\bibfield  {title}
  {\bibinfo {title} {Observation of a liquid-gas-type transition in the
  pyrochlore spin ice compound {D}y$_2${T}i$_2${O}$_7$ in a magnetic field},\
  }\href {https://doi.org/10.1103/PhysRevLett.90.207205} {\bibfield  {journal}
  {\bibinfo  {journal} {Phys. Rev. Lett.}\ }\textbf {\bibinfo {volume} {90}},\
  \bibinfo {pages} {207205} (\bibinfo {year} {2003})}\BibitemShut {NoStop}%
\bibitem [{\citenamefont {Castelnovo}\ \emph {et~al.}(2008)\citenamefont
  {Castelnovo}, \citenamefont {Moessner},\ and\ \citenamefont
  {Sondhi}}]{Castelnovo2008}%
  \BibitemOpen
  \bibfield  {author} {\bibinfo {author} {\bibfnamefont {C.}~\bibnamefont
  {Castelnovo}}, \bibinfo {author} {\bibfnamefont {R.}~\bibnamefont
  {Moessner}},\ and\ \bibinfo {author} {\bibfnamefont {S.~L.}\ \bibnamefont
  {Sondhi}},\ }\bibfield  {title} {\bibinfo {title} {Magnetic monopoles in spin
  ice},\ }\href {https://doi.org/10.1038/nature06433} {\bibfield  {journal}
  {\bibinfo  {journal} {Nature}\ }\textbf {\bibinfo {volume} {451}},\ \bibinfo
  {pages} {42} (\bibinfo {year} {2008})}\BibitemShut {NoStop}%
\bibitem [{\citenamefont {Henley}(2010)}]{Henley2010}%
  \BibitemOpen
  \bibfield  {author} {\bibinfo {author} {\bibfnamefont {C.~L.}\ \bibnamefont
  {Henley}},\ }\bibfield  {title} {\bibinfo {title} {The
  {\textquotedblleft}{C}oulomb phase{\textquotedblright} in frustrated
  systems},\ }\href {https://doi.org/10.1146/annurev-conmatphys-070909-104138}
  {\bibfield  {journal} {\bibinfo  {journal} {Annual Review of Condensed Matter
  Physics}\ }\textbf {\bibinfo {volume} {1}},\ \bibinfo {pages} {179} (\bibinfo
  {year} {2010})}\BibitemShut {NoStop}%
\bibitem [{\citenamefont {Udagawa}\ and\ \citenamefont
  {Jaubert}(2021)}]{SIceSpringer2011}%
  \BibitemOpen
  \bibinfo {editor} {\bibfnamefont {M.}~\bibnamefont {Udagawa}}\ and\ \bibinfo
  {editor} {\bibfnamefont {L.}~\bibnamefont {Jaubert}},\ eds.,\ \href
  {https://doi.org/https://doi.org/10.1007/978-3-030-70860-3} {\emph {\bibinfo
  {title} {Spin Ice}}},\ \bibinfo {edition} {1st}\ ed.\ (\bibinfo  {publisher}
  {Springer Cham},\ \bibinfo {year} {2021})\BibitemShut {NoStop}%
\bibitem [{\citenamefont {Gingras}\ and\ \citenamefont
  {McClarty}(2014)}]{Gingras2014}%
  \BibitemOpen
  \bibfield  {author} {\bibinfo {author} {\bibfnamefont {M.~J.~P.}\
  \bibnamefont {Gingras}}\ and\ \bibinfo {author} {\bibfnamefont {P.~A.}\
  \bibnamefont {McClarty}},\ }\bibfield  {title} {\bibinfo {title} {Quantum
  spin ice: a search for gapless quantum spin liquids in pyrochlore magnets},\
  }\href {https://doi.org/10.1088/0034-4885/77/5/056501} {\bibfield  {journal}
  {\bibinfo  {journal} {Reports on Progress in Physics}\ }\textbf {\bibinfo
  {volume} {77}},\ \bibinfo {pages} {056501} (\bibinfo {year}
  {2014})}\BibitemShut {NoStop}%
\bibitem [{\citenamefont {Savary}\ and\ \citenamefont
  {Balents}(2016)}]{Savary2016}%
  \BibitemOpen
  \bibfield  {author} {\bibinfo {author} {\bibfnamefont {L.}~\bibnamefont
  {Savary}}\ and\ \bibinfo {author} {\bibfnamefont {L.}~\bibnamefont
  {Balents}},\ }\bibfield  {title} {\bibinfo {title} {Quantum spin liquids: a
  review},\ }\href {https://doi.org/10.1088/0034-4885/80/1/016502} {\bibfield
  {journal} {\bibinfo  {journal} {Reports on Progress in Physics}\ }\textbf
  {\bibinfo {volume} {80}},\ \bibinfo {pages} {016502} (\bibinfo {year}
  {2016})}\BibitemShut {NoStop}%
\bibitem [{\citenamefont {Powell}(2020)}]{BPowell2020}%
  \BibitemOpen
  \bibfield  {author} {\bibinfo {author} {\bibfnamefont {B.~J.}\ \bibnamefont
  {Powell}},\ }\bibfield  {title} {\bibinfo {title} {Emergent particles and
  gauge fields in quantum matter},\ }\href
  {https://doi.org/10.1080/00107514.2020.1832350} {\bibfield  {journal}
  {\bibinfo  {journal} {Contemporary Physics}\ }\textbf {\bibinfo {volume}
  {61}},\ \bibinfo {pages} {96} (\bibinfo {year} {2020})}\BibitemShut {NoStop}%
\bibitem [{\citenamefont {Ramirez}\ \emph {et~al.}(1999)\citenamefont
  {Ramirez}, \citenamefont {Hayashi}, \citenamefont {Cava}, \citenamefont
  {Siddharthan},\ and\ \citenamefont {Shastry}}]{Ramirez1999}%
  \BibitemOpen
  \bibfield  {author} {\bibinfo {author} {\bibfnamefont {A.~P.}\ \bibnamefont
  {Ramirez}}, \bibinfo {author} {\bibfnamefont {A.}~\bibnamefont {Hayashi}},
  \bibinfo {author} {\bibfnamefont {R.~J.}\ \bibnamefont {Cava}}, \bibinfo
  {author} {\bibfnamefont {R.}~\bibnamefont {Siddharthan}},\ and\ \bibinfo
  {author} {\bibfnamefont {B.~S.}\ \bibnamefont {Shastry}},\ }\bibfield
  {title} {\bibinfo {title} {Zero-point entropy in `spin ice'},\ }\href
  {https://doi.org/10.1038/20619} {\bibfield  {journal} {\bibinfo  {journal}
  {Nature}\ }\textbf {\bibinfo {volume} {399}},\ \bibinfo {pages} {333}
  (\bibinfo {year} {1999})}\BibitemShut {NoStop}%
\bibitem [{\citenamefont {Paulsen}\ \emph {et~al.}(2014)\citenamefont
  {Paulsen}, \citenamefont {Jackson}, \citenamefont {Lhotel}, \citenamefont
  {Canals}, \citenamefont {Prabhakaran}, \citenamefont {Matsuhira},
  \citenamefont {Giblin},\ and\ \citenamefont {Bramwell}}]{Paulsen2014}%
  \BibitemOpen
  \bibfield  {author} {\bibinfo {author} {\bibfnamefont {C.}~\bibnamefont
  {Paulsen}}, \bibinfo {author} {\bibfnamefont {M.~J.}\ \bibnamefont
  {Jackson}}, \bibinfo {author} {\bibfnamefont {E.}~\bibnamefont {Lhotel}},
  \bibinfo {author} {\bibfnamefont {B.}~\bibnamefont {Canals}}, \bibinfo
  {author} {\bibfnamefont {D.}~\bibnamefont {Prabhakaran}}, \bibinfo {author}
  {\bibfnamefont {K.}~\bibnamefont {Matsuhira}}, \bibinfo {author}
  {\bibfnamefont {S.~R.}\ \bibnamefont {Giblin}},\ and\ \bibinfo {author}
  {\bibfnamefont {S.~T.}\ \bibnamefont {Bramwell}},\ }\bibfield  {title}
  {\bibinfo {title} {Far-from-equilibrium monopole dynamics in spin ice},\
  }\href {https://doi.org/10.1038/nphys2847} {\bibfield  {journal} {\bibinfo
  {journal} {Nature Physics}\ }\textbf {\bibinfo {volume} {10}},\ \bibinfo
  {pages} {135} (\bibinfo {year} {2014})}\BibitemShut {NoStop}%
\bibitem [{\citenamefont {Raban}\ \emph {et~al.}(2022)\citenamefont {Raban},
  \citenamefont {Berthier},\ and\ \citenamefont {Holdsworth}}]{Raban2022}%
  \BibitemOpen
  \bibfield  {author} {\bibinfo {author} {\bibfnamefont {V.}~\bibnamefont
  {Raban}}, \bibinfo {author} {\bibfnamefont {L.}~\bibnamefont {Berthier}},\
  and\ \bibinfo {author} {\bibfnamefont {P.~C.~W.}\ \bibnamefont
  {Holdsworth}},\ }\bibfield  {title} {\bibinfo {title} {Violation of the
  fluctuation-dissipation theorem and effective temperatures in spin ice},\
  }\href {https://doi.org/10.1103/PhysRevB.105.134431} {\bibfield  {journal}
  {\bibinfo  {journal} {Phys. Rev. B}\ }\textbf {\bibinfo {volume} {105}},\
  \bibinfo {pages} {134431} (\bibinfo {year} {2022})}\BibitemShut {NoStop}%
\bibitem [{\citenamefont {Hall\'en}\ \emph {et~al.}(2022)\citenamefont
  {Hall\'en}, \citenamefont {Grigera}, \citenamefont {Tennant}, \citenamefont
  {Castelnovo},\ and\ \citenamefont {Moessner}}]{Hallen2022}%
  \BibitemOpen
  \bibfield  {author} {\bibinfo {author} {\bibfnamefont {J.~N.}\ \bibnamefont
  {Hall\'en}}, \bibinfo {author} {\bibfnamefont {S.~A.}\ \bibnamefont
  {Grigera}}, \bibinfo {author} {\bibfnamefont {D.~A.}\ \bibnamefont
  {Tennant}}, \bibinfo {author} {\bibfnamefont {C.}~\bibnamefont
  {Castelnovo}},\ and\ \bibinfo {author} {\bibfnamefont {R.}~\bibnamefont
  {Moessner}},\ }\bibfield  {title} {\bibinfo {title} {Dynamical fractal and
  anomalous noise in a clean magnetic crystal},\ }\href
  {https://doi.org/10.1126/science.add1644} {\bibfield  {journal} {\bibinfo
  {journal} {Science}\ }\textbf {\bibinfo {volume} {378}},\ \bibinfo {pages}
  {1218} (\bibinfo {year} {2022})}\BibitemShut {NoStop}%
\bibitem [{\citenamefont {Ehlers}\ \emph {et~al.}(2004)\citenamefont {Ehlers},
  \citenamefont {Cornelius}, \citenamefont {Fennell}, \citenamefont {Koza},
  \citenamefont {Bramwell},\ and\ \citenamefont {Gardner}}]{Ehlers2004}%
  \BibitemOpen
  \bibfield  {author} {\bibinfo {author} {\bibfnamefont {G.}~\bibnamefont
  {Ehlers}}, \bibinfo {author} {\bibfnamefont {A.~L.}\ \bibnamefont
  {Cornelius}}, \bibinfo {author} {\bibfnamefont {T.}~\bibnamefont {Fennell}},
  \bibinfo {author} {\bibfnamefont {M.}~\bibnamefont {Koza}}, \bibinfo {author}
  {\bibfnamefont {S.~T.}\ \bibnamefont {Bramwell}},\ and\ \bibinfo {author}
  {\bibfnamefont {J.~S.}\ \bibnamefont {Gardner}},\ }\bibfield  {title}
  {\bibinfo {title} {Evidence for two distinct spin relaxation mechanisms in
  `hot' spin ice {H}o$_{2}${T}i$_{2}${O}$_{7}$},\ }\href
  {https://doi.org/10.1088/0953-8984/16/11/010} {\bibfield  {journal} {\bibinfo
   {journal} {Journal of Physics: Condensed Matter}\ }\textbf {\bibinfo
  {volume} {16}},\ \bibinfo {pages} {S635} (\bibinfo {year}
  {2004})}\BibitemShut {NoStop}%
\bibitem [{\citenamefont {Fennell}\ \emph {et~al.}(2009)\citenamefont
  {Fennell}, \citenamefont {Deen}, \citenamefont {Wildes}, \citenamefont
  {Schmalzl}, \citenamefont {Prabhakaran}, \citenamefont {Boothroyd},
  \citenamefont {Aldus}, \citenamefont {McMorrow},\ and\ \citenamefont
  {Bramwell}}]{Fenell2009}%
  \BibitemOpen
  \bibfield  {author} {\bibinfo {author} {\bibfnamefont {T.}~\bibnamefont
  {Fennell}}, \bibinfo {author} {\bibfnamefont {P.~P.}\ \bibnamefont {Deen}},
  \bibinfo {author} {\bibfnamefont {A.~R.}\ \bibnamefont {Wildes}}, \bibinfo
  {author} {\bibfnamefont {K.}~\bibnamefont {Schmalzl}}, \bibinfo {author}
  {\bibfnamefont {D.}~\bibnamefont {Prabhakaran}}, \bibinfo {author}
  {\bibfnamefont {A.~T.}\ \bibnamefont {Boothroyd}}, \bibinfo {author}
  {\bibfnamefont {R.~J.}\ \bibnamefont {Aldus}}, \bibinfo {author}
  {\bibfnamefont {D.~F.}\ \bibnamefont {McMorrow}},\ and\ \bibinfo {author}
  {\bibfnamefont {S.~T.}\ \bibnamefont {Bramwell}},\ }\bibfield  {title}
  {\bibinfo {title} {Magnetic {C}oulomb phase in the spin ice
  {H}o$_{2}${T}i$_{2}${O}$_{7}$},\ }\href
  {https://doi.org/10.1126/science.1177582} {\bibfield  {journal} {\bibinfo
  {journal} {Science}\ }\textbf {\bibinfo {volume} {326}},\ \bibinfo {pages}
  {415} (\bibinfo {year} {2009})}\BibitemShut {NoStop}%
\bibitem [{\citenamefont {Bramwell}\ and\ \citenamefont
  {Gingras}(2001)}]{Bramwell2001}%
  \BibitemOpen
  \bibfield  {author} {\bibinfo {author} {\bibfnamefont {S.~T.}\ \bibnamefont
  {Bramwell}}\ and\ \bibinfo {author} {\bibfnamefont {M.~J.~P.}\ \bibnamefont
  {Gingras}},\ }\bibfield  {title} {\bibinfo {title} {Spin ice state in
  frustrated magnetic pyrochlore materials},\ }\href
  {https://doi.org/10.1126/science.1064761} {\bibfield  {journal} {\bibinfo
  {journal} {Science}\ }\textbf {\bibinfo {volume} {294}},\ \bibinfo {pages}
  {1495} (\bibinfo {year} {2001})}\BibitemShut {NoStop}%
\bibitem [{\citenamefont {Morris}\ \emph {et~al.}(2009)\citenamefont {Morris},
  \citenamefont {Tennant}, \citenamefont {Grigera}, \citenamefont {Klemke},
  \citenamefont {Castelnovo}, \citenamefont {Moessner}, \citenamefont
  {Czternasty}, \citenamefont {Meissner}, \citenamefont {Rule}, \citenamefont
  {Hoffmann}, \citenamefont {Kiefer}, \citenamefont {Gerischer}, \citenamefont
  {Slobinsky},\ and\ \citenamefont {Perry}}]{Morris2009}%
  \BibitemOpen
  \bibfield  {author} {\bibinfo {author} {\bibfnamefont {D.~J.~P.}\
  \bibnamefont {Morris}}, \bibinfo {author} {\bibfnamefont {D.~A.}\
  \bibnamefont {Tennant}}, \bibinfo {author} {\bibfnamefont {S.~A.}\
  \bibnamefont {Grigera}}, \bibinfo {author} {\bibfnamefont {B.}~\bibnamefont
  {Klemke}}, \bibinfo {author} {\bibfnamefont {C.}~\bibnamefont {Castelnovo}},
  \bibinfo {author} {\bibfnamefont {R.}~\bibnamefont {Moessner}}, \bibinfo
  {author} {\bibfnamefont {C.}~\bibnamefont {Czternasty}}, \bibinfo {author}
  {\bibfnamefont {M.}~\bibnamefont {Meissner}}, \bibinfo {author}
  {\bibfnamefont {K.~C.}\ \bibnamefont {Rule}}, \bibinfo {author}
  {\bibfnamefont {J.-U.}\ \bibnamefont {Hoffmann}}, \bibinfo {author}
  {\bibfnamefont {K.}~\bibnamefont {Kiefer}}, \bibinfo {author} {\bibfnamefont
  {S.}~\bibnamefont {Gerischer}}, \bibinfo {author} {\bibfnamefont
  {D.}~\bibnamefont {Slobinsky}},\ and\ \bibinfo {author} {\bibfnamefont
  {R.~S.}\ \bibnamefont {Perry}},\ }\bibfield  {title} {\bibinfo {title} {Dirac
  strings and magnetic monopoles in the spin ice
  {D}y$_{2}${T}i$_{2}${O}$_{7}$},\ }\href
  {https://doi.org/10.1126/science.1178868} {\bibfield  {journal} {\bibinfo
  {journal} {Science}\ }\textbf {\bibinfo {volume} {326}},\ \bibinfo {pages}
  {411} (\bibinfo {year} {2009})}\BibitemShut {NoStop}%
\bibitem [{\citenamefont {Jaubert}\ and\ \citenamefont
  {Holdsworth}(2009)}]{Jaubert2009}%
  \BibitemOpen
  \bibfield  {author} {\bibinfo {author} {\bibfnamefont {L.~D.~C.}\
  \bibnamefont {Jaubert}}\ and\ \bibinfo {author} {\bibfnamefont {P.~C.~W.}\
  \bibnamefont {Holdsworth}},\ }\bibfield  {title} {\bibinfo {title} {Signature
  of magnetic monopole and {D}irac string dynamics in spin ice},\ }\href
  {https://doi.org/10.1038/nphys1227} {\bibfield  {journal} {\bibinfo
  {journal} {Nature Physics}\ }\textbf {\bibinfo {volume} {5}},\ \bibinfo
  {pages} {258} (\bibinfo {year} {2009})}\BibitemShut {NoStop}%
\bibitem [{\citenamefont {Castelnovo}\ \emph {et~al.}(2010)\citenamefont
  {Castelnovo}, \citenamefont {Moessner},\ and\ \citenamefont
  {Sondhi}}]{Castelnovo2010}%
  \BibitemOpen
  \bibfield  {author} {\bibinfo {author} {\bibfnamefont {C.}~\bibnamefont
  {Castelnovo}}, \bibinfo {author} {\bibfnamefont {R.}~\bibnamefont
  {Moessner}},\ and\ \bibinfo {author} {\bibfnamefont {S.~L.}\ \bibnamefont
  {Sondhi}},\ }\bibfield  {title} {\bibinfo {title} {Thermal quenches in spin
  ice},\ }\href {https://doi.org/10.1103/PhysRevLett.104.107201} {\bibfield
  {journal} {\bibinfo  {journal} {Phys. Rev. Lett.}\ }\textbf {\bibinfo
  {volume} {104}},\ \bibinfo {pages} {107201} (\bibinfo {year}
  {2010})}\BibitemShut {NoStop}%
\bibitem [{\citenamefont {Jaubert}\ and\ \citenamefont
  {Holdsworth}(2011)}]{Jaubert2011}%
  \BibitemOpen
  \bibfield  {author} {\bibinfo {author} {\bibfnamefont {L.~D.~C.}\
  \bibnamefont {Jaubert}}\ and\ \bibinfo {author} {\bibfnamefont {P.~C.~W.}\
  \bibnamefont {Holdsworth}},\ }\bibfield  {title} {\bibinfo {title} {Magnetic
  monopole dynamics in spin ice},\ }\href
  {https://doi.org/10.1088/0953-8984/23/16/164222} {\bibfield  {journal}
  {\bibinfo  {journal} {Journal of Physics: Condensed Matter}\ }\textbf
  {\bibinfo {volume} {23}},\ \bibinfo {pages} {164222} (\bibinfo {year}
  {2011})}\BibitemShut {NoStop}%
\bibitem [{\citenamefont {Mostame}\ \emph {et~al.}(2014)\citenamefont
  {Mostame}, \citenamefont {Castelnovo}, \citenamefont {Moessner},\ and\
  \citenamefont {Sondhi}}]{Mostame2014b}%
  \BibitemOpen
  \bibfield  {author} {\bibinfo {author} {\bibfnamefont {S.}~\bibnamefont
  {Mostame}}, \bibinfo {author} {\bibfnamefont {C.}~\bibnamefont {Castelnovo}},
  \bibinfo {author} {\bibfnamefont {R.}~\bibnamefont {Moessner}},\ and\
  \bibinfo {author} {\bibfnamefont {S.~L.}\ \bibnamefont {Sondhi}},\ }\bibfield
   {title} {\bibinfo {title} {Tunable nonequilibrium dynamics of field quenches
  in spin ice},\ }\href {https://doi.org/10.1073/pnas.1317631111} {\bibfield
  {journal} {\bibinfo  {journal} {Proceedings of the National Academy of
  Sciences}\ }\textbf {\bibinfo {volume} {111}},\ \bibinfo {pages} {640}
  (\bibinfo {year} {2014})}\BibitemShut {NoStop}%
\bibitem [{\citenamefont {Grams}\ \emph {et~al.}(2014)\citenamefont {Grams},
  \citenamefont {Valldor}, \citenamefont {Garst},\ and\ \citenamefont
  {Hemberger}}]{Grams2014}%
  \BibitemOpen
  \bibfield  {author} {\bibinfo {author} {\bibfnamefont {C.~P.}\ \bibnamefont
  {Grams}}, \bibinfo {author} {\bibfnamefont {M.}~\bibnamefont {Valldor}},
  \bibinfo {author} {\bibfnamefont {M.}~\bibnamefont {Garst}},\ and\ \bibinfo
  {author} {\bibfnamefont {J.}~\bibnamefont {Hemberger}},\ }\bibfield  {title}
  {\bibinfo {title} {Critical speeding-up in the magnetoelectric response of
  spin-ice near its monopole liquid--gas transition},\ }\href
  {https://doi.org/10.1038/ncomms5853} {\bibfield  {journal} {\bibinfo
  {journal} {Nature Communications}\ }\textbf {\bibinfo {volume} {5}},\
  \bibinfo {pages} {4853} (\bibinfo {year} {2014})}\BibitemShut {NoStop}%
\bibitem [{\citenamefont {Jackson}\ \emph {et~al.}(2014)\citenamefont
  {Jackson}, \citenamefont {Lhotel}, \citenamefont {Giblin}, \citenamefont
  {Bramwell}, \citenamefont {Prabhakaran}, \citenamefont {Matsuhira},
  \citenamefont {Hiroi}, \citenamefont {Yu},\ and\ \citenamefont
  {Paulsen}}]{Jackson2014}%
  \BibitemOpen
  \bibfield  {author} {\bibinfo {author} {\bibfnamefont {M.~J.}\ \bibnamefont
  {Jackson}}, \bibinfo {author} {\bibfnamefont {E.}~\bibnamefont {Lhotel}},
  \bibinfo {author} {\bibfnamefont {S.~R.}\ \bibnamefont {Giblin}}, \bibinfo
  {author} {\bibfnamefont {S.~T.}\ \bibnamefont {Bramwell}}, \bibinfo {author}
  {\bibfnamefont {D.}~\bibnamefont {Prabhakaran}}, \bibinfo {author}
  {\bibfnamefont {K.}~\bibnamefont {Matsuhira}}, \bibinfo {author}
  {\bibfnamefont {Z.}~\bibnamefont {Hiroi}}, \bibinfo {author} {\bibfnamefont
  {Q.}~\bibnamefont {Yu}},\ and\ \bibinfo {author} {\bibfnamefont
  {C.}~\bibnamefont {Paulsen}},\ }\bibfield  {title} {\bibinfo {title} {Dynamic
  behavior of magnetic avalanches in the spin-ice compound
  {${\mathrm{Dy}}_{2}{\mathrm{Ti}}_{2}{\mathrm{O}}_{7}$}},\ }\href
  {https://doi.org/10.1103/PhysRevB.90.064427} {\bibfield  {journal} {\bibinfo
  {journal} {Phys. Rev. B}\ }\textbf {\bibinfo {volume} {90}},\ \bibinfo
  {pages} {064427} (\bibinfo {year} {2014})}\BibitemShut {NoStop}%
\bibitem [{\citenamefont {Tang}\ \emph {et~al.}(2025)\citenamefont {Tang},
  \citenamefont {Glamsch}, \citenamefont {Aqeel}, \citenamefont
  {Scheuchenpflug}, \citenamefont {Schulze}, \citenamefont {Liebald},
  \citenamefont {Rytz}, \citenamefont {Guguschev}, \citenamefont {Albrecht},\
  and\ \citenamefont {Gegenwart}}]{Tang2025}%
  \BibitemOpen
  \bibfield  {author} {\bibinfo {author} {\bibfnamefont {N.}~\bibnamefont
  {Tang}}, \bibinfo {author} {\bibfnamefont {S.}~\bibnamefont {Glamsch}},
  \bibinfo {author} {\bibfnamefont {A.}~\bibnamefont {Aqeel}}, \bibinfo
  {author} {\bibfnamefont {L.}~\bibnamefont {Scheuchenpflug}}, \bibinfo
  {author} {\bibfnamefont {M.}~\bibnamefont {Schulze}}, \bibinfo {author}
  {\bibfnamefont {C.}~\bibnamefont {Liebald}}, \bibinfo {author} {\bibfnamefont
  {D.}~\bibnamefont {Rytz}}, \bibinfo {author} {\bibfnamefont {C.}~\bibnamefont
  {Guguschev}}, \bibinfo {author} {\bibfnamefont {M.}~\bibnamefont
  {Albrecht}},\ and\ \bibinfo {author} {\bibfnamefont {P.}~\bibnamefont
  {Gegenwart}},\ }\href {https://doi.org/10.48550/arXiv.2509.18422} {\bibinfo
  {title} {Observation via spin {S}eebeck effect of macroscopic magnetic
  transport from emergent magnetic monopoles}} (\bibinfo {year} {2025}),\
  \Eprint {https://arxiv.org/abs/2509.18422} {arXiv:2509.18422
  [cond-mat.str-el]} \BibitemShut {NoStop}%
\bibitem [{\citenamefont {Powell}\ and\ \citenamefont
  {Chalker}(2008)}]{Powell2008}%
  \BibitemOpen
  \bibfield  {author} {\bibinfo {author} {\bibfnamefont {S.}~\bibnamefont
  {Powell}}\ and\ \bibinfo {author} {\bibfnamefont {J.~T.}\ \bibnamefont
  {Chalker}},\ }\bibfield  {title} {\bibinfo {title} {Classical to quantum
  mappings for geometrically frustrated systems: Spin-ice in a [100] field},\
  }\href {https://doi.org/10.1103/PhysRevB.78.024422} {\bibfield  {journal}
  {\bibinfo  {journal} {Phys. Rev. B}\ }\textbf {\bibinfo {volume} {78}},\
  \bibinfo {pages} {024422} (\bibinfo {year} {2008})}\BibitemShut {NoStop}%
\bibitem [{\citenamefont {Powell}(2013)}]{Powell2013}%
  \BibitemOpen
  \bibfield  {author} {\bibinfo {author} {\bibfnamefont {S.}~\bibnamefont
  {Powell}},\ }\bibfield  {title} {\bibinfo {title} {Confinement of monopoles
  and scaling theory near unconventional critical points},\ }\href
  {https://doi.org/10.1103/PhysRevB.87.064414} {\bibfield  {journal} {\bibinfo
  {journal} {Phys. Rev. B}\ }\textbf {\bibinfo {volume} {87}},\ \bibinfo
  {pages} {064414} (\bibinfo {year} {2013})}\BibitemShut {NoStop}%
\bibitem [{\citenamefont {Kasteleyn}(1963)}]{Kasteleyn1963}%
  \BibitemOpen
  \bibfield  {author} {\bibinfo {author} {\bibfnamefont {P.~W.}\ \bibnamefont
  {Kasteleyn}},\ }\bibfield  {title} {\bibinfo {title} {Dimer statistics and
  phase transitions},\ }\href {https://doi.org/10.1063/1.1703953} {\bibfield
  {journal} {\bibinfo  {journal} {Journal of Mathematical Physics}\ }\textbf
  {\bibinfo {volume} {4}},\ \bibinfo {pages} {287} (\bibinfo {year}
  {1963})}\BibitemShut {NoStop}%
\bibitem [{\citenamefont {Grigera}\ and\ \citenamefont
  {Hooley}(2018)}]{Grigera2018}%
  \BibitemOpen
  \bibfield  {author} {\bibinfo {author} {\bibfnamefont {S.~A.}\ \bibnamefont
  {Grigera}}\ and\ \bibinfo {author} {\bibfnamefont {C.~A.}\ \bibnamefont
  {Hooley}},\ }\bibfield  {title} {\bibinfo {title} {Entropy as a function of
  magnetisation for a {2D} spin-ice model exhibiting a {K}asteleyn
  transition},\ }\href {https://doi.org/10.1088/2399-6528/aad492} {\bibfield
  {journal} {\bibinfo  {journal} {Journal of Physics Communications}\ }\textbf
  {\bibinfo {volume} {2}},\ \bibinfo {pages} {085004} (\bibinfo {year}
  {2018})}\BibitemShut {NoStop}%
\bibitem [{\citenamefont {Potts}\ and\ \citenamefont
  {Benton}(2022)}]{Potts2022}%
  \BibitemOpen
  \bibfield  {author} {\bibinfo {author} {\bibfnamefont {M.}~\bibnamefont
  {Potts}}\ and\ \bibinfo {author} {\bibfnamefont {O.}~\bibnamefont {Benton}},\
  }\bibfield  {title} {\bibinfo {title} {Spin ice in a general applied magnetic
  field: {K}asteleyn transition, magnetic torque, and rotational magnetocaloric
  effect},\ }\href {https://doi.org/10.1103/PhysRevB.106.054437} {\bibfield
  {journal} {\bibinfo  {journal} {Phys. Rev. B}\ }\textbf {\bibinfo {volume}
  {106}},\ \bibinfo {pages} {054437} (\bibinfo {year} {2022})}\BibitemShut
  {NoStop}%
\bibitem [{\citenamefont {Wan}\ and\ \citenamefont
  {Tchernyshyov}(2012)}]{Yuan2012}%
  \BibitemOpen
  \bibfield  {author} {\bibinfo {author} {\bibfnamefont {Y.}~\bibnamefont
  {Wan}}\ and\ \bibinfo {author} {\bibfnamefont {O.}~\bibnamefont
  {Tchernyshyov}},\ }\bibfield  {title} {\bibinfo {title} {Quantum strings in
  quantum spin ice},\ }\href {https://doi.org/10.1103/PhysRevLett.108.247210}
  {\bibfield  {journal} {\bibinfo  {journal} {Phys. Rev. Lett.}\ }\textbf
  {\bibinfo {volume} {108}},\ \bibinfo {pages} {247210} (\bibinfo {year}
  {2012})}\BibitemShut {NoStop}%
\bibitem [{\citenamefont {Powell}(2022)}]{Powell2022}%
  \BibitemOpen
  \bibfield  {author} {\bibinfo {author} {\bibfnamefont {S.}~\bibnamefont
  {Powell}},\ }\bibfield  {title} {\bibinfo {title} {Quantum {K}asteleyn
  transition},\ }\href {https://doi.org/10.1103/PhysRevB.105.064413} {\bibfield
   {journal} {\bibinfo  {journal} {Phys. Rev. B}\ }\textbf {\bibinfo {volume}
  {105}},\ \bibinfo {pages} {064413} (\bibinfo {year} {2022})}\BibitemShut
  {NoStop}%
\bibitem [{\citenamefont {Pal}\ and\ \citenamefont {Powell}(2024)}]{Pal2024}%
  \BibitemOpen
  \bibfield  {author} {\bibinfo {author} {\bibfnamefont {S.}~\bibnamefont
  {Pal}}\ and\ \bibinfo {author} {\bibfnamefont {S.}~\bibnamefont {Powell}},\
  }\bibfield  {title} {\bibinfo {title} {Crossover from string to cluster
  dynamics following a field quench in spin ice},\ }\href
  {https://doi.org/10.1103/PhysRevB.109.094427} {\bibfield  {journal} {\bibinfo
   {journal} {Phys. Rev. B}\ }\textbf {\bibinfo {volume} {109}},\ \bibinfo
  {pages} {094427} (\bibinfo {year} {2024})}\BibitemShut {NoStop}%
\bibitem [{\citenamefont {Huster~Zapke}\ and\ \citenamefont
  {Holdsworth}()}]{Zapke2025}%
  \BibitemOpen
  \bibfield  {author} {\bibinfo {author} {\bibfnamefont {A.}~\bibnamefont
  {Huster~Zapke}}\ and\ \bibinfo {author} {\bibfnamefont {P.~C.~W.}\
  \bibnamefont {Holdsworth}},\ }\href
  {https://doi.org/10.48550/arXiv.2505.14811} {\bibinfo {title} {Monopoles,
  {D}irac strings and magnetic noise in model spin ice}},\ \Eprint
  {https://arxiv.org/abs/2505.14811} {arXiv:2505.14811} \BibitemShut {NoStop}%
\bibitem [{\citenamefont {Powell}\ and\ \citenamefont
  {Pal}(2025)}]{Powell2025}%
  \BibitemOpen
  \bibfield  {author} {\bibinfo {author} {\bibfnamefont {S.}~\bibnamefont
  {Powell}}\ and\ \bibinfo {author} {\bibfnamefont {S.}~\bibnamefont {Pal}},\
  }\bibfield  {title} {\bibinfo {title} {Dynamic scaling theory for a field
  quench near the {K}asteleyn transition in spin ice},\ }\href
  {https://doi.org/10.1103/b9gd-y4y6} {\bibfield  {journal} {\bibinfo
  {journal} {Phys. Rev. Lett.}\ }\textbf {\bibinfo {volume} {134}},\ \bibinfo
  {pages} {256701} (\bibinfo {year} {2025})}\BibitemShut {NoStop}%
\bibitem [{\citenamefont {Gardner}\ \emph {et~al.}(2010)\citenamefont
  {Gardner}, \citenamefont {Gingras},\ and\ \citenamefont
  {Greedan}}]{Gardiner2010}%
  \BibitemOpen
  \bibfield  {author} {\bibinfo {author} {\bibfnamefont {J.~S.}\ \bibnamefont
  {Gardner}}, \bibinfo {author} {\bibfnamefont {M.~J.~P.}\ \bibnamefont
  {Gingras}},\ and\ \bibinfo {author} {\bibfnamefont {J.~E.}\ \bibnamefont
  {Greedan}},\ }\bibfield  {title} {\bibinfo {title} {Magnetic pyrochlore
  oxides},\ }\href {https://doi.org/10.1103/RevModPhys.82.53} {\bibfield
  {journal} {\bibinfo  {journal} {Rev. Mod. Phys.}\ }\textbf {\bibinfo {volume}
  {82}},\ \bibinfo {pages} {53} (\bibinfo {year} {2010})}\BibitemShut {NoStop}%
\bibitem [{\citenamefont {Ryzhkin}(2005)}]{Ryzhkin2005}%
  \BibitemOpen
  \bibfield  {author} {\bibinfo {author} {\bibfnamefont {I.~A.}\ \bibnamefont
  {Ryzhkin}},\ }\bibfield  {title} {\bibinfo {title} {Magnetic relaxation in
  rare-earth oxide pyrochlores},\ }\href@noop {} {\bibfield  {journal}
  {\bibinfo  {journal} {Journal of Experimental and Theoretical Physics}\
  }\textbf {\bibinfo {volume} {101}},\ \bibinfo {pages} {481} (\bibinfo {year}
  {2005})}\BibitemShut {NoStop}%
\bibitem [{\citenamefont {Glauber}(1963)}]{Glauber1963}%
  \BibitemOpen
  \bibfield  {author} {\bibinfo {author} {\bibfnamefont {R.~J.}\ \bibnamefont
  {Glauber}},\ }\bibfield  {title} {\bibinfo {title} {Time-dependent statistics
  of the {I}sing model},\ }\href {https://doi.org/10.1063/1.1703954} {\bibfield
   {journal} {\bibinfo  {journal} {Journal of Mathematical Physics}\ }\textbf
  {\bibinfo {volume} {4}},\ \bibinfo {pages} {294} (\bibinfo {year}
  {1963})}\BibitemShut {NoStop}%
\bibitem [{\citenamefont {Cardy}(1996)}]{Cardy1996}%
  \BibitemOpen
  \bibfield  {author} {\bibinfo {author} {\bibfnamefont {J.}~\bibnamefont
  {Cardy}},\ }\href@noop {} {\emph {\bibinfo {title} {Scaling and
  Renormalization in Statistical Physics}}},\ Cambridge Lecture Notes in
  Physics\ (\bibinfo  {publisher} {Cambridge University Press},\ \bibinfo
  {address} {Cambridge, UK},\ \bibinfo {year} {1996})\BibitemShut {NoStop}%
\bibitem [{\citenamefont {Snyder}\ \emph {et~al.}(2004)\citenamefont {Snyder},
  \citenamefont {Ueland}, \citenamefont {Slusky}, \citenamefont {Karunadasa},
  \citenamefont {Cava},\ and\ \citenamefont {Schiffer}}]{Snyder2004}%
  \BibitemOpen
  \bibfield  {author} {\bibinfo {author} {\bibfnamefont {J.}~\bibnamefont
  {Snyder}}, \bibinfo {author} {\bibfnamefont {B.~G.}\ \bibnamefont {Ueland}},
  \bibinfo {author} {\bibfnamefont {J.~S.}\ \bibnamefont {Slusky}}, \bibinfo
  {author} {\bibfnamefont {H.}~\bibnamefont {Karunadasa}}, \bibinfo {author}
  {\bibfnamefont {R.~J.}\ \bibnamefont {Cava}},\ and\ \bibinfo {author}
  {\bibfnamefont {P.}~\bibnamefont {Schiffer}},\ }\bibfield  {title} {\bibinfo
  {title} {Low-temperature spin freezing in the
  {${\mathrm{Dy}}_{2}{\mathrm{Ti}}_{2}{\mathrm{O}}_{7}$} spin ice},\ }\href
  {https://doi.org/10.1103/PhysRevB.69.064414} {\bibfield  {journal} {\bibinfo
  {journal} {Phys. Rev. B}\ }\textbf {\bibinfo {volume} {69}},\ \bibinfo
  {pages} {064414} (\bibinfo {year} {2004})}\BibitemShut {NoStop}%
\bibitem [{\citenamefont {Goldenfeld}(1992)}]{Goldenfeld}%
  \BibitemOpen
  \bibfield  {author} {\bibinfo {author} {\bibfnamefont {N.}~\bibnamefont
  {Goldenfeld}},\ }\href {https://doi.org/10.1201/9780429493492} {\emph
  {\bibinfo {title} {Lectures On Phase Transitions And The Renormalization
  Group}}},\ \bibinfo {edition} {1st}\ ed.\ (\bibinfo  {publisher} {CRC Press,
  Boca Raton},\ \bibinfo {year} {1992})\BibitemShut {NoStop}%
\bibitem [{\citenamefont {Schnoerr}\ \emph {et~al.}(2017)\citenamefont
  {Schnoerr}, \citenamefont {Sanguinetti},\ and\ \citenamefont
  {Grima}}]{Schnoerr2017}%
  \BibitemOpen
  \bibfield  {author} {\bibinfo {author} {\bibfnamefont {D.}~\bibnamefont
  {Schnoerr}}, \bibinfo {author} {\bibfnamefont {G.}~\bibnamefont
  {Sanguinetti}},\ and\ \bibinfo {author} {\bibfnamefont {R.}~\bibnamefont
  {Grima}},\ }\bibfield  {title} {\bibinfo {title} {Approximation and inference
  methods for stochastic biochemical kinetics---a tutorial review},\ }\href
  {https://doi.org/10.1088/1751-8121/aa54d9} {\bibfield  {journal} {\bibinfo
  {journal} {Journal of Physics A: Mathematical and Theoretical}\ }\textbf
  {\bibinfo {volume} {50}},\ \bibinfo {pages} {093001} (\bibinfo {year}
  {2017})}\BibitemShut {NoStop}%
\bibitem [{\citenamefont {Lago}\ \emph {et~al.}(2007)\citenamefont {Lago},
  \citenamefont {Blundell},\ and\ \citenamefont {Baines}}]{Lago2007}%
  \BibitemOpen
  \bibfield  {author} {\bibinfo {author} {\bibfnamefont {J.}~\bibnamefont
  {Lago}}, \bibinfo {author} {\bibfnamefont {S.~J.}\ \bibnamefont {Blundell}},\
  and\ \bibinfo {author} {\bibfnamefont {C.}~\bibnamefont {Baines}},\
  }\bibfield  {title} {\bibinfo {title} {{\(\mu\)SR} investigation of spin
  dynamics in the spin-ice material
  {${\mathrm{Dy}}_{2}{\mathrm{Ti}}_{2}{\mathrm{O}}_{7}$}},\ }\href
  {https://doi.org/10.1088/0953-8984/19/32/326210} {\bibfield  {journal}
  {\bibinfo  {journal} {Journal of Physics: Condensed Matter}\ }\textbf
  {\bibinfo {volume} {19}},\ \bibinfo {pages} {326210} (\bibinfo {year}
  {2007})}\BibitemShut {NoStop}%
\bibitem [{\citenamefont {Kirschner}\ \emph {et~al.}(2018)\citenamefont
  {Kirschner}, \citenamefont {Flicker}, \citenamefont {Yacoby}, \citenamefont
  {Yao},\ and\ \citenamefont {Blundell}}]{Kirschner2018}%
  \BibitemOpen
  \bibfield  {author} {\bibinfo {author} {\bibfnamefont {F.~K.~K.}\
  \bibnamefont {Kirschner}}, \bibinfo {author} {\bibfnamefont {F.}~\bibnamefont
  {Flicker}}, \bibinfo {author} {\bibfnamefont {A.}~\bibnamefont {Yacoby}},
  \bibinfo {author} {\bibfnamefont {N.~Y.}\ \bibnamefont {Yao}},\ and\ \bibinfo
  {author} {\bibfnamefont {S.~J.}\ \bibnamefont {Blundell}},\ }\bibfield
  {title} {\bibinfo {title} {Proposal for the detection of magnetic monopoles
  in spin ice via nanoscale magnetometry},\ }\href
  {https://doi.org/10.1103/PhysRevB.97.140402} {\bibfield  {journal} {\bibinfo
  {journal} {Phys. Rev. B}\ }\textbf {\bibinfo {volume} {97}},\ \bibinfo
  {pages} {140402} (\bibinfo {year} {2018})}\BibitemShut {NoStop}%
\bibitem [{\citenamefont {Moessner}\ and\ \citenamefont
  {Sondhi}(2003)}]{Moessner2003}%
  \BibitemOpen
  \bibfield  {author} {\bibinfo {author} {\bibfnamefont {R.}~\bibnamefont
  {Moessner}}\ and\ \bibinfo {author} {\bibfnamefont {S.~L.}\ \bibnamefont
  {Sondhi}},\ }\bibfield  {title} {\bibinfo {title} {Theory of the [111]
  magnetization plateau in spin ice},\ }\href
  {https://doi.org/10.1103/PhysRevB.68.064411} {\bibfield  {journal} {\bibinfo
  {journal} {Phys. Rev. B}\ }\textbf {\bibinfo {volume} {68}},\ \bibinfo
  {pages} {064411} (\bibinfo {year} {2003})}\BibitemShut {NoStop}%
\bibitem [{\citenamefont {Lib\'al}\ \emph {et~al.}(2020)\citenamefont
  {Lib\'al}, \citenamefont {del Campo}, \citenamefont {Nisoli}, \citenamefont
  {Reichhardt},\ and\ \citenamefont {Reichhardt}}]{Libal2020}%
  \BibitemOpen
  \bibfield  {author} {\bibinfo {author} {\bibfnamefont {A.}~\bibnamefont
  {Lib\'al}}, \bibinfo {author} {\bibfnamefont {A.}~\bibnamefont {del Campo}},
  \bibinfo {author} {\bibfnamefont {C.}~\bibnamefont {Nisoli}}, \bibinfo
  {author} {\bibfnamefont {C.}~\bibnamefont {Reichhardt}},\ and\ \bibinfo
  {author} {\bibfnamefont {C.~J.~O.}\ \bibnamefont {Reichhardt}},\ }\bibfield
  {title} {\bibinfo {title} {Quenched dynamics of artificial colloidal spin
  ice},\ }\href {https://doi.org/10.1103/PhysRevResearch.2.033433} {\bibfield
  {journal} {\bibinfo  {journal} {Phys. Rev. Res.}\ }\textbf {\bibinfo {volume}
  {2}},\ \bibinfo {pages} {033433} (\bibinfo {year} {2020})}\BibitemShut
  {NoStop}%
\bibitem [{\citenamefont {Gonz\'alez-Cuadra}\ \emph {et~al.}(2025)\citenamefont
  {Gonz\'alez-Cuadra}, \citenamefont {Hamdan}, \citenamefont {Zache},
  \citenamefont {Braverman}, \citenamefont {Kornja\v{c}a}, \citenamefont
  {Lukin}, \citenamefont {Cant\'u}, \citenamefont {Liu}, \citenamefont {Wang},
  \citenamefont {Keesling}, \citenamefont {Lukin}, \citenamefont {Zoller},\
  and\ \citenamefont {Bylinskii}}]{Cuarda2025}%
  \BibitemOpen
  \bibfield  {author} {\bibinfo {author} {\bibfnamefont {D.}~\bibnamefont
  {Gonz\'alez-Cuadra}}, \bibinfo {author} {\bibfnamefont {M.}~\bibnamefont
  {Hamdan}}, \bibinfo {author} {\bibfnamefont {T.~V.}\ \bibnamefont {Zache}},
  \bibinfo {author} {\bibfnamefont {B.}~\bibnamefont {Braverman}}, \bibinfo
  {author} {\bibfnamefont {M.}~\bibnamefont {Kornja\v{c}a}}, \bibinfo {author}
  {\bibfnamefont {A.}~\bibnamefont {Lukin}}, \bibinfo {author} {\bibfnamefont
  {S.~H.}\ \bibnamefont {Cant\'u}}, \bibinfo {author} {\bibfnamefont
  {F.}~\bibnamefont {Liu}}, \bibinfo {author} {\bibfnamefont {S.-T.}\
  \bibnamefont {Wang}}, \bibinfo {author} {\bibfnamefont {A.}~\bibnamefont
  {Keesling}}, \bibinfo {author} {\bibfnamefont {M.~D.}\ \bibnamefont {Lukin}},
  \bibinfo {author} {\bibfnamefont {P.}~\bibnamefont {Zoller}},\ and\ \bibinfo
  {author} {\bibfnamefont {A.}~\bibnamefont {Bylinskii}},\ }\bibfield  {title}
  {\bibinfo {title} {Observation of string breaking on a {\((2 + 1)\)d}
  {R}ydberg quantum simulator},\ }\href
  {https://doi.org/10.1038/s41586-025-09051-6} {\bibfield  {journal} {\bibinfo
  {journal} {Nature}\ }\textbf {\bibinfo {volume} {642}},\ \bibinfo {pages}
  {321 } (\bibinfo {year} {2025})}\BibitemShut {NoStop}%
\bibitem [{\citenamefont {Shah}\ \emph {et~al.}(2025)\citenamefont {Shah},
  \citenamefont {Nambiar}, \citenamefont {Gorshkov},\ and\ \citenamefont
  {Galitski}}]{Shah2025}%
  \BibitemOpen
  \bibfield  {author} {\bibinfo {author} {\bibfnamefont {J.}~\bibnamefont
  {Shah}}, \bibinfo {author} {\bibfnamefont {G.}~\bibnamefont {Nambiar}},
  \bibinfo {author} {\bibfnamefont {A.~V.}\ \bibnamefont {Gorshkov}},\ and\
  \bibinfo {author} {\bibfnamefont {V.}~\bibnamefont {Galitski}},\ }\bibfield
  {title} {\bibinfo {title} {Quantum spin ice in three-dimensional {R}ydberg
  atom arrays},\ }\href {https://doi.org/10.1103/PhysRevX.15.011025} {\bibfield
   {journal} {\bibinfo  {journal} {Phys. Rev. X}\ }\textbf {\bibinfo {volume}
  {15}},\ \bibinfo {pages} {011025} (\bibinfo {year} {2025})}\BibitemShut
  {NoStop}%
\bibitem [{\citenamefont {Surace}\ \emph {et~al.}(2020)\citenamefont {Surace},
  \citenamefont {Mazza}, \citenamefont {Giudici}, \citenamefont {Lerose},
  \citenamefont {Gambassi},\ and\ \citenamefont {Dalmonte}}]{Surace2020}%
  \BibitemOpen
  \bibfield  {author} {\bibinfo {author} {\bibfnamefont {F.~M.}\ \bibnamefont
  {Surace}}, \bibinfo {author} {\bibfnamefont {P.~P.}\ \bibnamefont {Mazza}},
  \bibinfo {author} {\bibfnamefont {G.}~\bibnamefont {Giudici}}, \bibinfo
  {author} {\bibfnamefont {A.}~\bibnamefont {Lerose}}, \bibinfo {author}
  {\bibfnamefont {A.}~\bibnamefont {Gambassi}},\ and\ \bibinfo {author}
  {\bibfnamefont {M.}~\bibnamefont {Dalmonte}},\ }\bibfield  {title} {\bibinfo
  {title} {Lattice gauge theories and string dynamics in {R}ydberg atom quantum
  simulators},\ }\href {https://doi.org/10.1103/PhysRevX.10.021041} {\bibfield
  {journal} {\bibinfo  {journal} {Phys. Rev. X}\ }\textbf {\bibinfo {volume}
  {10}},\ \bibinfo {pages} {021041} (\bibinfo {year} {2020})}\BibitemShut
  {NoStop}%
\bibitem [{\citenamefont {Oakes}\ \emph {et~al.}(2016)\citenamefont {Oakes},
  \citenamefont {Garrahan},\ and\ \citenamefont {Powell}}]{Oakes2016}%
  \BibitemOpen
  \bibfield  {author} {\bibinfo {author} {\bibfnamefont {T.}~\bibnamefont
  {Oakes}}, \bibinfo {author} {\bibfnamefont {J.~P.}\ \bibnamefont
  {Garrahan}},\ and\ \bibinfo {author} {\bibfnamefont {S.}~\bibnamefont
  {Powell}},\ }\bibfield  {title} {\bibinfo {title} {Emergence of cooperative
  dynamics in fully packed classical dimers},\ }\href
  {https://doi.org/10.1103/PhysRevE.93.032129} {\bibfield  {journal} {\bibinfo
  {journal} {Phys. Rev. E}\ }\textbf {\bibinfo {volume} {93}},\ \bibinfo
  {pages} {032129} (\bibinfo {year} {2016})}\BibitemShut {NoStop}%
\bibitem [{\citenamefont {Castelnovo}\ \emph {et~al.}(2012)\citenamefont
  {Castelnovo}, \citenamefont {Moessner},\ and\ \citenamefont
  {Sondhi}}]{Castelnovo2012}%
  \BibitemOpen
  \bibfield  {author} {\bibinfo {author} {\bibfnamefont {C.}~\bibnamefont
  {Castelnovo}}, \bibinfo {author} {\bibfnamefont {R.}~\bibnamefont
  {Moessner}},\ and\ \bibinfo {author} {\bibfnamefont {S.}~\bibnamefont
  {Sondhi}},\ }\bibfield  {title} {\bibinfo {title} {Spin ice,
  fractionalization, and topological order},\ }\href
  {https://doi.org/10.1146/annurev-conmatphys-020911-125058} {\bibfield
  {journal} {\bibinfo  {journal} {Annual Review of Condensed Matter Physics}\
  }\textbf {\bibinfo {volume} {3}},\ \bibinfo {pages} {35} (\bibinfo {year}
  {2012})}\BibitemShut {NoStop}%
\end{thebibliography}%

\end{document}